\documentclass[a4paper, amsmath, amssymb, twocolumn]{quantumarticle}
\pdfoutput=1

\usepackage{graphicx}
\usepackage{bm}
\usepackage{hyperref}
\usepackage{upgreek}
\usepackage{float}
\usepackage{enumitem}
\usepackage{diagbox}
\usepackage{braket}
\usepackage{xcolor}
\usepackage{verbatim}
\setlist{noitemsep}
\usepackage{soul}
\usepackage[T1]{fontenc}
\usepackage{hhline} 
\usepackage{wasysym} 
\usepackage[numbers,sort&compress]{natbib}

\makeatletter
\def\frontmatter@maketitle{%
\@author@finish
\title@column\titleblock@produce
\suppressfloats[t]%
}%
\makeatother

\begin{document}

\title{The Magic Scroll: Leveraging biased noise to improve magic state cultivation in register-based architectures}

\author{I. D. Thorvaldson}
\author{J. Marshall}
\author{J. R. Craig}
\author{S. K. Gorman}
\author{C. D. Hill}
\author{M. Y. Simmons}

\affiliation{Silicon Quantum Computing Pty. Ltd., UNSW Sydney, 2052, Australia}
\affiliation{Centre of Excellence for Quantum Computation and Communication Technology,
School of Physics, UNSW Sydney, 2052, NSW Australia}

\date{August 10, 2026}

\maketitle

\section*{Abstract}

Multiple quantum computing platforms across neutral atoms~\cite{Bluvstein2024_neutral_atoms}, nitrogen vacancy centres~\cite{Waldherr2014, Bradley_10_qubit_NVs}, gate-defined dots~\cite{hrl_3d, amaha2025verticallycoupleddoublequantum, tidjani2025_3darray} and 14|15 phosphorus atom qubits~\cite{Thorvaldson2025, Edlbauer2025} in silicon are experimentally exploring the use of high connectivity qubits, beyond that of nearest-neighbour planar lattices. Theoretical works consider modifications to fault-tolerant codes to leverage this higher qubit connectivity, such as non-local LDPC codes~\cite{Bravyi2024_LDPC}, inspiring superconducting~\cite{yoder2025tourgrossmodularquantum} and photonic~\cite{litinski2022activevolume} platforms to also seek non-planar connectivity. In addition, separate theoretical works consider biased noise, where bit- and phase-flip errors are not equally likely. In this work, we present efficient methods for implementing 6.6.6 and 4.8.8 colour codes, as well as bilayer and folded surface codes, using two-qubit registers. We also leverage noise bias to avoid hook errors, demonstrating comparable performance to the surface code. By combining colour and bilayer codes, we show how magic state cultivation procedures~\cite{gidney_cultivation, Chamberland2020, vaknin2025magicstatecultivationsurface} can be improved, in a procedure we refer to as the Magic Scroll. Not only does the Magic Scroll escape to a standard surface code, it also improves cultivation volumes by $3\times$ and supports magic $\ket{T}$ state fidelities as low as $10^{-9}$. We show that this technique can be further leveraged to improve distillation performance, showing a $3 \times$ improvement to distillation volumes for error rates of $\sim10^{-15}$. Through these constructions, we demonstrate techniques to leverage noise bias and high qubit connectivity, showing how register-based architectures can improve error rates and reduce quantum volumes in fault-tolerant quantum computing.

\section{Introduction}

High qubit connectivity has been of particular interest for recent work in quantum information science, with many platforms demonstrating qubit couplings which go beyond planar nearest-neighbour interaction. Such work spans a variety of modalities, with all-to-all coupling demonstrated in neutral atom systems~\cite{Bluvstein2024_neutral_atoms}, as well as local all-to-all ``registers'' of qubits being demonstrated in nitrogen vacancy centres~\cite{Waldherr2014, Bradley_10_qubit_NVs}, trapped ions~\cite{ion_trap_registers}, as well as silicon-based quantum dots~\cite{hrl_3d, amaha2025verticallycoupleddoublequantum, tidjani2025_3darray}, T-centres~\cite{simmons2023scalablefaulttolerantquantumtechnologies} and phosphorus atom qubits (ie. the 14|15 platform)~\cite{Thorvaldson2025, Edlbauer2025}. The success of these highly connected qubits have motivated proposals for the introduction of long-range connectivity in architectures which do not natively support them, such as superconducting~\cite{yoder2025tourgrossmodularquantum} and photonic qubits~\cite{litinski2022activevolume}.\\

In tandem with these recent experimental results, theoretical studies have explored how higher connectivity could be used to improve the performance of logical qubits. These include LDPC codes built on highly non-local connectivity~\cite{Bravyi2024_LDPC}, error correction built on grids of qudits~\cite{moussa_bilayer}, as well as fusion-based fault tolerance with 3-dimensional lattice connectivity~\cite{Bartolucci2023_fusion} and surface codes defined in 3 dimensions~\cite{3d_surface_codes}, to name a few. These studies have shown that fault tolerant computation can be improved by using higher connectivity, either by improving code rates, logical qubit performance, or the available gateset. As such, recent focus has been to refine these high-connectivity qubit codes for application on near-term devices.\\

\begin{figure*}[]
\includegraphics[width=0.79\textwidth]{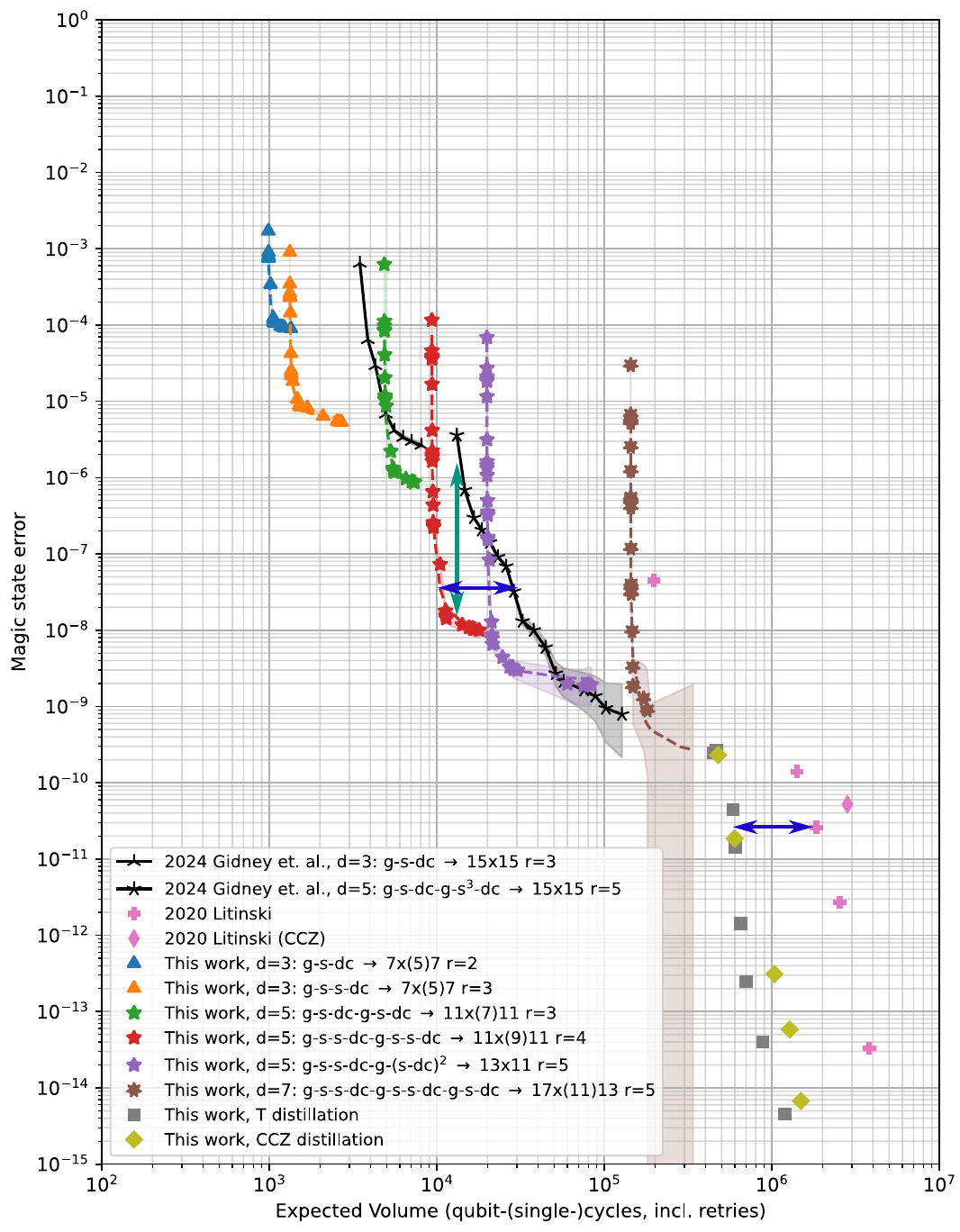}
\centering
\caption{\textbf{Comparison of simulated magic $\ket{T}$ state production using the Magic Scroll and Magic Scroll distillation.} Here, we plot the magic state error as a function of expected volume of the Magic Scroll cultivation and distillation techniques explored in this work, over different cultivation sequences, final code distances and escape repetitions. For cultivation, each set of same-coloured, same-shape points represents a series of different complementary gap thresholds, with the highest error point representing no complementary gap. Simulation settings are notated here in the way specified in the main text Eq.~\ref{eq:cultivation_label} - for example, ``d=5: g-s-dc-g-s-dc -> 13x(7)11 r=3'' indicates colour code distance-5, a particular ordering of grow (g), stabilise (s) and double-check (dc) for cultivation, and that we escape to a surface code of 13x7 over 3 cycles before finally ending in a 13x11 surface code. Shaded regions indicate error rates within a factor of 10 of maximum likelihood error. For comparison, the results reported by Gidney \textit{et. al.}~\cite{gidney_cultivation} are shown, modelled under unbiased noise. The magic scroll demonstrates improvements over past results, with the cyan arrow showing the volume where the Magic Scroll provides a $\sim100\times$ improvement over the error rates reported by Gidney \textit{et. al.}, and the blue arrow marking the error rate where the Magic Scroll shows $\sim 3\times$ improvement in volume over previous results. The boosted performance of the Magic Scroll in comparison to that reported by Gidney \textit{et. al.} is attributed to biased noise and register connectivity, improving magic state cultivation efficiency. Note that the $d=7$ cultivation results include an extrapolated fit (see Appendix~\ref{sec:supp_magic_scroll_extrapolation}), as an order of magnitude approximation. For distillation, we compare the performance of bilayer two-stage distillation using Magic Scrolls to the unbiased noise, single-layer surface code results shown by Litinski~\cite{Litinski_distillation}, with an additional blue arrow indicating where our work improves on the volume reported by Litinski by a factor of $\sim 3\times$. The exact distillation settings are shown in Table~\ref{tab:distillation_results}. All results cultivate $\ket{T}$ magic states, except where indicated that $\ket{CCZ}$ states are being synthillated.~\label{fig:gidney_style_figure}}
\end{figure*}

In parallel to optimising fault tolerant codes through high qubit connectivities, recent research has also shown improved logical qubit performance by leveraging systems with intrinsic noise bias. For example, bosonic-cat qubits~\cite{biased_cat_codes} or spin-cat qubits~\cite{kruckenhauser2025_spin_cats} are significantly more susceptible to phase errors than bit-flip errors, resulting in highly biased noise. Solid-state spin qubits also exhibit high intrinsic noise bias. Here, phosphorus-bound electron spins in particular exhibit extremely long relaxation (bit-flip) times of tens of seconds~\cite{silicon_electron_lifetimes}, but with dephasing (phase-flip) times of tens of microseconds~\cite{Thorvaldson2025} without refocusing. This results in $\sim 5$ orders of magnitude noise bias. Locally transformed error correcting codes can exploit such noise bias - for example, modifying surface codes by applying local Hadamard gates to every second data qubit yields the XZZX code~\cite{BonillaAtaides2021_XZZX}, and colour codes can be similarly modified with domain walls~\cite{Tiurev_domain_wall_colour_code}. Compared to systems with unbiased noise, systems with biased noise can utilise these codes can protect logical states more efficiently, and with a higher threshold - up to $50\%$ code-capacity threshold with maximum noise bias~\cite{BonillaAtaides2021_XZZX}.\\

In this work, we leverage both noise bias and high qubit connectivity to improve fault-tolerant operation. We base our architecture and noise model on multi-nuclear spin registers in the 14|15 platform in silicon, which exhibit both strong noise bias and the ability to operate highly-connected registers~\cite{Thorvaldson2025, Edlbauer2025}. First, we show that using a square grid of one- and two-qubit registers, we can implement 6.6.6 and 4.8.8 colour codes efficiently, with both demonstrating crossing thresholds of $\sim 0.55 \%$ using the Chromobius decoder~\cite{gidney2023_chromobius}. Secondly, we show that the same biased-noise register grid can support bilayer/folded surface codes, adding an extra layer to the surface code while maintaining a crossing threshold of $\sim 0.7 \%$, on par with the single-layer surface code.\\

We then combine the 4.8.8 colour code with the bilayer surface code to show that recent protocols for efficient cultivation of $\ket{T}$ magic states~\cite{gidney_cultivation, Chamberland2020, vaknin2025magicstatecultivationsurface, sahay2025foldtransversalsurfacecodecultivation, claes2025cultivatingtstatessurface, vaknin2025efficientmagicstatecultivation, hirano2025efficientmagicstatecultivation} can be improved under biased noise and two-qubit register connectivity for efficient magic state production. We also show how we can escape the magic state to a standard surface code with matchable errors by presenting a new protocol which we refer to as the Magic Scroll cultivation procedure, building on the recent results by Gidney \textit{et. al.}~\cite{gidney_cultivation} to enable a further $\sim 3\times$ improvement in expected quantum volumes. Finally, we show how the Magic Scroll can be used as input for further distillation, which itself can be optimised by leveraging biased noise and register connectivity along with complementary gapping techniques~\cite{gidney2023yokedsurfacecodes}. This improves on the results presented by Litinski~\cite{Litinski_distillation}, resulting in a further $3\times$ reduction in average distillation volumes. The results of $\ket{T}$ state cultivation and distillation protocols are summarised in Fig.~\ref{fig:gidney_style_figure}, with the various protocol parameter choices (ie. different colours) and corresponding performance discussed later in this work. Through the performance of these constructions, we show how high connectivity combined with biased errors has the potential to significantly improve the performance of fault-tolerant error correcting codes.

\subsection{Organisation}

The paper is organised as follows: In Section~\ref{sec:register_model}, we describe the local register connectivity and biased-noise error models used in this work, motivated by the performance of spin qubit register architectures. In Section~\ref{sec:color_codes}, we describe how these models can be applied to stabilisation of the 6.6.6 and 4.8.8 colour codes, and show the simulation results of the performance of these codes on par with state of the art colour code implementations. In Section~\ref{sec:bilayer_codes}, we further show how biased-noise register architectures can be used to implement a bilayer surface code, showing simulation results and the extended logical operations available to these codes while maintaining the usual surface code crossing threshold. In Section~\ref{sec:magic_scroll}, we show how these 4.8.8 colour codes and bilayer codes can be combined in the ``Magic Scroll'' protocol to improve the expected quantum volume of magic state cultivation in biased-noise register architectures by up to a factor of three, describing all steps of magic state cultivation and escape. Finally, in Section~\ref{sec:magic_scroll_distillation}, we show how Magic Scrolls can be combined with noise-biased asymmetric bilayer surface codes and used to improve magic state distillation procedures, again by a factor of three in expected quantum volume.

\section{Multi-Qubit Register Model~\label{sec:register_model}}

The error-correcting codes we consider in this work are based on two physically-motivated assumed characteristics that are not commonly assumed when studying fault tolerant quantum computations. The first is that the data qubits are hosted within multi-qubit registers, enabling a higher connectivity than the usual square lattice. The second is that for certain operations, noise is completely biased towards dephasing errors (ie. Z errors). This work explores how fault-tolerant codes can be tailored to leverage these two characteristics, and in particular how magic states can be cultivated into a surface code efficiently.

\begin{figure}[!ht]
\includegraphics[width=0.45\textwidth]{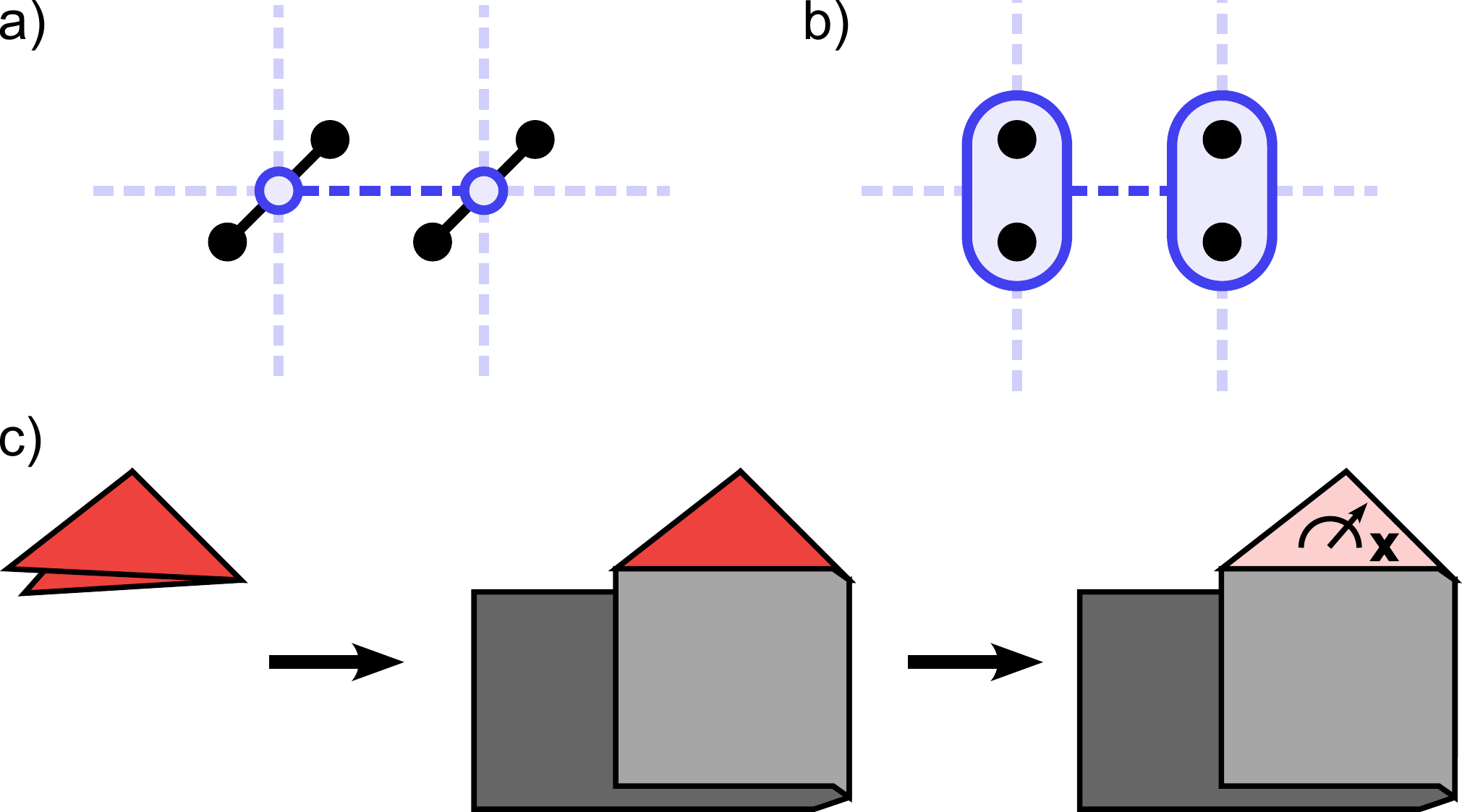}
\centering
\caption{\textbf{Overview of the multi-qubit register architecture, and the Magic Scroll cultivation it enables.} (a) Illustration of connectivity present for multi-nuclear spin registers in silicon~\cite{Edlbauer2025}. Open blue circles represent electron spin qubits, and filled black circles represent nuclear spin qubits. Two-qubit gates between nuclear spin qubits are mediated by the electrons. Each electron overlaps a few nuclear spin qubits, forming a qubit register, but neighbouring electrons are coupled (dashed lines) allowing inter-register operations. Typically, the nuclear spins are operated as the sole data qubits, with electrons facilitating multi-qubit gates (both within a register~\cite{Thorvaldson2025}, and between neighbouring registers~\cite{Edlbauer2025}). (b) Simplified version of a), where the ancillary electron qubits have been omitted, and instead registers are highlighted with blue ovals. This simplification results in the register connectivity model used in this paper. Specifically, any qubit in a register can be coupled with any qubit in a neighbouring register, as represented by the blue dashed lines between registers. (c) High-level summary of the Magic Scroll procedure for magic state cultivation studied in this work. First, a magic state is cultivated in a colour code (red triangular section), which is then merged with a partially folded surface code (grey), and then finally the colour code is measured out leaving the magic state purely in the folded surface code. The Magic Scroll is readily implemented with the register-based connectivity illustrated in parts a) and b).~\label{fig:paper_intro}}
\end{figure}

\subsection{Register Connectivity}

The first assumed characteristic is that the error correcting code is implemented on multi-qubit registers. These multi-qubit register systems are natively available for certain physical architectures, for example in nuclear spin qubits with phosphorus atom qubits in silicon~\cite{Thorvaldson2025} and for nitrogen vacancy centres~\cite{Waldherr2014, Bradley_10_qubit_NVs}. Moreover, they can be readily implemented in platforms with easily reconfigurable qubit geometries, such as neutral atom architectures~\cite{Bluvstein2024_neutral_atoms} and photonic qubit architectures~\cite{Bartolucci2023_fusion}. The physical qubit layout of a register model (based on systems with native registers) is shown in Fig.~\ref{fig:paper_intro}a, and involves a small group of data qubits connected to the same ancilla qubit. The connections between registers are facilitated by these ancilla qubits.\\

For both phosphorus atom qubits in silicon (ie. the 14|15 platform) and nitrogen vacancy centres, the ancilla qubits are electron spins, and the data qubits are nuclear spins. The coherence time of nuclear spins are orders of magnitude longer than the coherence time of electrons~\cite{Muhonen2014}, the latter of which are used to mediate interactions between the nuclear qubits (which are used as data qubits). Henceforth, we illustrate these registers in the format given in Fig.~\ref{fig:paper_intro}b, omitting the ancillary qubits which mediate interactions, similar to the notation used by Moussa~\cite{moussa_bilayer}. Nuclear data qubits within a register are effectively interchangeable in terms of their connectivity, meaning it is important to remember that any data qubit in a given register can be directly coupled (in one gate) to any data qubit in the same register, or any data qubit in an adjacent register. In this work, we only consider registers with one or two data qubits, but in practice solid-state registers have been demonstrated with three~\cite{Thorvaldson2025}, five~\cite{Edlbauer2025}, and even ten~\cite{Bradley_10_qubit_NVs} nuclear spin qubits.\\

\begin{figure}[!ht]
\includegraphics[width=0.45\textwidth]{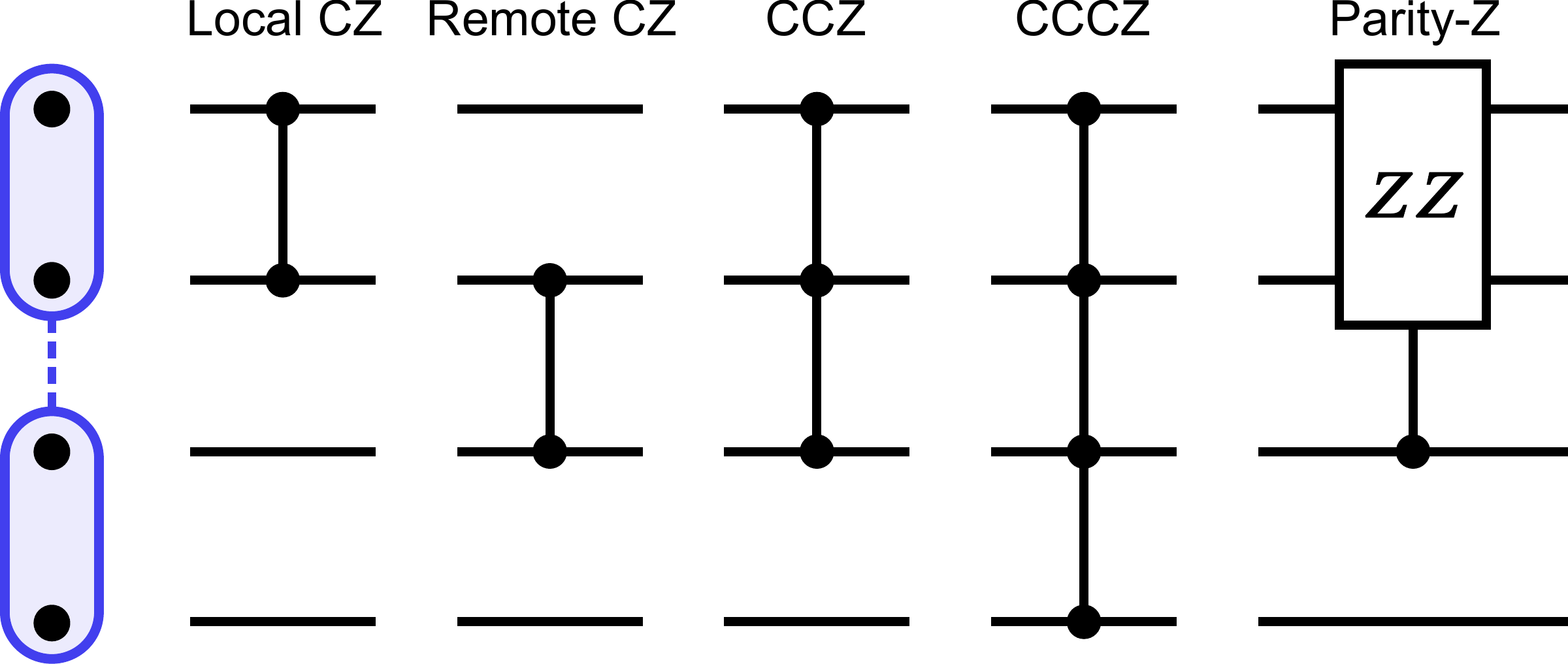}
\centering
\caption{\textbf{Overview of inter-register gates available in a two-qubit register architecture.} Based off the capabilities of multi-nuclear spin registers in silicon, we show some examples of operations that are possible between two-qubit registers. For all examples shown here, it is possible to choose anti-controls wherever a control is present, or to freely change which qubit(s) are being operated on, without incurring any extra operational overhead. Under the assumption of multi-tone driving of electrons, each of the above operations is equally expensive to perform in terms of duration, with the exception of local CZ gates which are twice as fast as the rest. Note that all gates here commute with Pauli-Z operators on any qubit, which is important for leveraging noise bias in the simulations performed here. The simulations performed in this work only use local and remote CZ gates, but for completeness we include other gates which may be of interest to future work. We assume each register can only perform one inter-register gate at a time, due to the underlying use of mediating ancilla qubits.~\label{fig:register_intro_gateset}}
\end{figure}

Examples of gates available to native register systems are shown in Fig.~\ref{fig:register_intro_gateset}. Registers not only allow any-to-any local and remote CZ gates, but also allow more complex gates, including CCZ and CCCZ gates, as well as the ability to perform Z-parity controlled Z gates (which we refer to as a parity-Z) for the same cost as a CZ gate. These gates are mediated by a single central ancilla qubit in each register, so we assume that each register can only be involved in a single multi-qubit gate within a given timestep; for example, doing a remote CZ between the upper data qubits of two neighbouring registers simultaneously with a CZ between the bottom data qubits of the same two registers would not be allowed. Instead, such an operation would require two timesteps. Despite the extended gateset of these native register systems, the QEC codes developed in this work only use CZ gates as these demonstrated the best performance, meaning that our results are applicable beyond native register systems.\\

To complete the gateset, we assume that Z-basis initialisation and measurement can be performed on any data qubit within a register, but that these operations cannot be applied simultaneously to multiple qubits within a register. This is similar to multi-qubit gates, where operations are mediated by the ancilla electron. Finally, we assume any arbitrary single-qubit gate can be applied to any data qubit within a register, and since these are not mediated by the ancilla qubit, can be applied simultaneously across multiple data qubits in the same register.

\subsection{Error Model}

Inspired by the extremely biased noise seen in the 14|15 platform~\cite{silicon_electron_lifetimes, Thorvaldson2025, Muhonen2014, silicon_electron_lifetimes}, we modify the usual single-parameter circuit-level noise (eg. as used by Gidney \textit{et. al.}~\cite{gidney_cultivation}) to include noise bias for certain operations. Not all operations are modified - for example, initialisation is modelled as usual, with some probability $p$ to have an X error after initialisation. Similarly, there is a $p$ chance to have an X error before measurement. Single-qubit gates are also the same, with a weight-$p$ depolarising error after each single qubit gate (that is, a $p/3$ equally weighted chance to have a Pauli X, Y or Z applied to that qubit).\\

\begin{table}[!ht]
    \centering
    \resizebox{0.5 \textwidth}{!}{
    \begin{tabular}{|c|c|}\hline
        \textbf{Operation} & \textbf{Error Channel} (total error $p$) \\\hline
        Reset & $X(p)$ \\\hline
        Measure & $X(p)$ \\\hline
        1-qubit gate (any) & $X(p/3), Y(p/3), Z(p/3)$ \\\hline
        Idle (M, R) & $Z(p)$ \\\hline
        Idle (1Q) & $Z(p)$ \\\hline
        CZ (inter-register) & $ZI(p/3), IZ(p/3), ZZ(p/3)$ \\\hline
        CZ (local) & $ZI(p/3), IZ(p/3), ZZ(p/3)$ \\\hline
        Parity-Z (inter-register) & $ZZI(p/3), IIZ(p/3), ZZZ(p/3)$ \\\hline
    \end{tabular}
    }
    \caption{\textbf{Summary of the biased error model used in this work.} More details for derivation of the error model for multi-qubit gates (ie. remote CZ and parity-Z gates) can be found in Appendix~\ref{sec:supp_pauli_twirling}. All error channels are performed after an ideal gate, except for Measure gates where the error is beforehand. $X(p)$ represents a $p$ probability for a Pauli-$X$ error occurring. For idling, 1Q is short for any one-qubit gate. For CZ and parity-Z, we list the data qubit(s) before the ancillas (eg. a $ZZI$ error describes a Pauli-Z on both qubits involved in the parity control).}
    ~\label{tab:error_model}
\end{table}

Idling errors are, however, treated differently. We assume that the timescale of performing multi-qubit gates is significantly shorter than single-qubit gates and the measure/reset operation times (due to the increased coupling of electron spins to magnetic drives compared to nuclear spins, see eg.~\cite{Thorvaldson2025}). As a consequence, no errors are applied to idling qubits during a timestep in which multi-qubit gates occur. Indeed, single-qubit gates are often an order of magnitude slower than their two-qubit counterparts for multi-nuclear registers in silicon~\cite{Thorvaldson2025}. Where idling is relevant (ie. timesteps where single-qubit gates or measure/reset operations occur), we assume idling qubits only experience maximally biased Z errors~\footnote{Because idling errors are maximally biased, we also ignore any idling errors that affect qubits definitely in the $|0>$ or $|1>$ state, as they will not be affected by Z errors.}, based on observed noise biases in solid state systems~\cite{Muhonen2014}. In short, our error model assumes maximum dephasing bias when a qubit operation commutes with Pauli-Z. Some systems can present even stronger forms of noise bias, for example that presented by Ruiz \textit{et. al.}~\cite{lowcost_biased_magicstates}, but in this work we limit ourselves to models motivated by solid state spin qubits.\\

The implementation of multi-qubit gates between nuclear spin qubits does not involve the use of NMR, and this, combined with the fact that nuclear spin qubits have extremely long bit-flip lifetimes, means that any of the gates in Fig.~\ref{fig:register_intro_gateset} will not cause any X or Y errors to any qubit, instead only applying errors involving products of Pauli-Z operators. A full justification of this multi-qubit gate error model is given in Appendix~\ref{sec:supp_pauli_twirling}. When extrapolating this performance to inter-register operations, we assume that all valid Pauli error channels are equally likely. We summarise the biased error model we use in this work in Table~\ref{tab:error_model}, detailing the various Pauli error channels associated with register operations, along with relative probabilities for each Pauli error to occur.

\section{Colour Codes~\label{sec:color_codes}}

\begin{figure*}[!ht]
\includegraphics[width=0.95\textwidth]{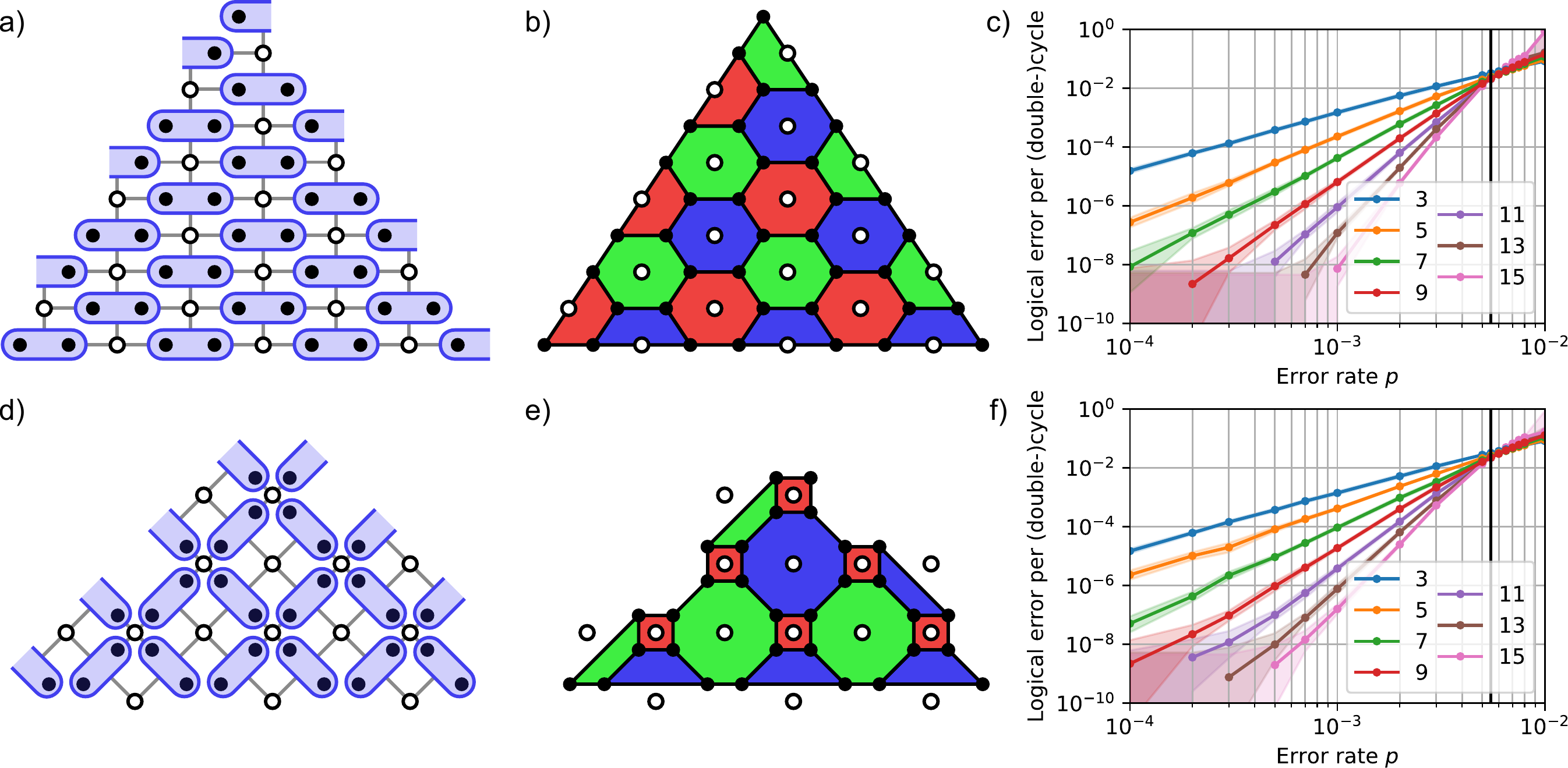}
\centering
\caption{\textbf{Implementation and performance of 6.6.6 and 4.8.8 colour codes using one- and two-qubit registers under biased noise.} (a) Layout of alternating two-qubit and one-qubit registers with square connectivity capable of implementing a 6.6.6 colour code. The register outlines have been omitted for the single-qubit registers. Open circles represent 6.6.6 code ancilla qubits, while filled circuits represent 6.6.6 code data qubits. (b) 6.6.6 colour code stabilisers corresponding to the register layout in a). Each coloured face is stabilised with both X and Z type stabilisers. (c) Simulated threshold diagram of the 6.6.6 colour code shown in a-b using the noise model given in Table~\ref{tab:error_model}. Time domain walls have been added to the stabilisation sequence, as described in the main text, to ensure equal X- and Z-logical protection under biased noise. A crossing threshold of $0.55 \%$ is marked with a vertical black line. Shaded regions indicate errors within a factor of 1000 of the maximum likelihood error. (d-f) Figures corresponding to parts a-c, but for the more compact 4.8.8 colour code. The threshold marked in f is also $0.55 \%$. Larger versions of c and f can be found in Appendix~\ref{sec:supp_col_performance}. Both implementations demonstrate logical protection on par with state of the art colour code performance, showing that square grids of registers can implement colour codes directly without flag qubits, while maintaining high performance. While both codes show similar thresholds, the 6.6.6 colour code shows better sub-threshold scaling. The 4.8.8 colour code is used later in this work as part of the Magic Scroll cultivation procedure.~\label{fig:colour_code}}
\end{figure*}

The surface code is the primary focus of many efforts to achieve fault-tolerant logical qubits due to its ability to correct both phase and bit flip errors, its use of local connectivity, and its comparatively high code threshold~\cite{Fowler_2012, Horsman_2012, pymatching, Higgott_2023, google_surface_code_paper}. Despite this, other fault-tolerant codes have recently been explored as alternatives to the surface code~\cite{Bravyi2024_LDPC, 3d_surface_codes, Bartolucci2023_fusion}. One such example is the family of 2D colour codes, which are defined on weight-three three-colourable graphs, where each vertex of the graph represents a data qubit, and each face represents both an X-type and Z-type stabiliser~\cite{colour_code_summary}. There are three such graphs which offer regular tiling of the plane, 6.6.6, 4.8.8 and 4.6.12, the former two of which are illustrated in Fig.~\ref{fig:colour_code}b and e respectively.\\

Colour codes are of particular interest as they are implementable with local connectivity on a 2D plane and feature coincident X-type and Z-type stabilisers. This latter feature in turn allows for all Clifford operations to be implemented transversely, a property which allows for the efficient creation of magic states~\cite{gidney_cultivation, Chamberland2020}.\\

One of the main drawbacks of colour codes is that they are challenging to decode efficiently. Because each data qubit in a 2D colour code is involved in at most three X-type (and three Z-type) stabilisers, single-qubit errors on data qubits will cause three stabilisers to flip. As these base errors have weights higher than two, techniques using matching graphs are not directly applicable~\cite{pymatching}. Instead, more expensive decoding schemes are required which can match arbitrary weight errors, i.e., to solve a hypergraph matching problem~\cite{gidney2023_chromobius, colour_code_unbiased_performance, vibe_decoding}.\\

Another issue with colour codes is that their stabilisers are higher weight than the surface code. This is of particular importance when considering hook errors, where a flip of the ancilla qubit in the middle of the stabiliser measurement causes multiple data qubit flips, which in turn has the potential to reduce the code distance. As a result of this, under the standard depolarising noise model, colour codes originally had an error threshold of $0.08 \%$~\cite{colour_code_summary, wang2009graphicalalgorithmsthresholderror}, compared to the threshold of $\sim 0.7-1.0\%$ for surface codes~\cite{Fowler_2012, pymatching, Higgott_2023}. By optimising stabiliser measurement circuits and decoding, recent developments have allowed the colour code threshold to climb as high as $\sim 0.4 - 0.7 \%$~\cite{gidney2023_chromobius, colour_code_unbiased_performance}. The colour code can however still be affected by hook errors.\\

Using the high connectivity register model studied here, it is possible to equalise the required connectivities for data and ancilla qubits in the colour code. For example, 6.6.6 colour codes typically require weight three and six connectivity for data and ancilla qubits respectively, but by placing data qubits on two-qubit registers, all registers require weight four connectivity instead. Explicit mappings of how this is achieved are shown in Fig.~\ref{fig:colour_code}a and d for 6.6.6 and 4.8.8 colour codes, corresponding to the colour code stabilisers shown in Fig.~\ref{fig:colour_code}b and e respectively. Each ancilla qubit is housed in a 1-qubit register, and so the register outline is omitted for clarity. Also note that these qubits are ancilla qubits for the colour code, and should not be confused with the omitted, intermediary electron qubits illustrated in Fig.~\ref{fig:paper_intro}a.\\

The stabilisers of these register-based colour codes are measured in two steps: first, all Z-stabilisers are measured, and then in the second step all X-stabilisers are measured (the ``time multiplexing'' method shown in~\cite{gidney2023_chromobius}). Due to the biased nature of the noise in the 14|15 platform (see Table~\ref{tab:error_model}), this method of stabilisation has advantages over methods designed for unbiased noise, such as the superdense cycle~\cite{gidney2023_chromobius}. When Z-stabilisers are measured, data qubits do not receive any Hadamard operations, and instead they experience purely idling and CZ gates, only accumulating dephasing errors which do not spread to other qubits. When measuring X-stabilisers, the data qubits need Hadamard gates, synchronised to occur with the ancilla qubits directly after reset and before measurement. This means any dephasing errors during the sequence of CZ operations (in-between the two Hadamard layers) will not spread to other qubits.\\

The primary benefit of performing stabilisation this way is that hook errors are completely mitigated. A hook error involves an error which occurs part-way through a stabiliser measurement (ie. not at the start or the end), which propagates to some but not all of the data qubits involved in the stabiliser. For example, an X error which occurs to the ancilla qubit halfway through a Z-measurement will propagate to two data qubits, potentially reducing the code distance. However, in our error model, only Z errors affect qubits during the middle CZ section of the circuit, which commute with the CZ, making these higher-weight hook error mechanisms impossible at $O(p)$. Because two separate steps of measurement are needed to measure all stabilisers of the colour code in this fashion, we define a ``(single-)cycle'' as a single measurement of either Z or X stabilisers, and a ``(double-)cycle'' as the combined process of measuring both Z and X stabilisers once. Assuming that measurements dominate circuit runtime, when comparing our results to the existing literature, we directly compare in terms of (single-)cycles.\\

Our error model in Table~\ref{tab:error_model} is biased towards Z errors, such that performing the usual colour code stabilisers will make the colour code protect against X-logical errors better than Z-logical errors (see Appendix~\ref{sec:supp_col_performance}). To avoid this asymmetry for systems with biased noise, we instead introduce domain walls to the colour code (ie. we apply Hadamard gates to contiguous regions of data qubits), analogous to the procedure presented by Tiurev \textit{et. al.}~\cite{Tiurev_domain_wall_colour_code}. However, instead of inserting spatial domain walls, we introduce domain walls in the time dimension - which can easily be achieved by applying Hadamard gates to all data qubits exactly once per (single-)cycle, rather than none for Z and twice for X (see Appendix~\ref{sec:supp_col_stabilisation} for circuit definition). This protects against logical X and Z errors equally (see Appendix~\ref{sec:supp_col_performance}), and also equalises the threshold for both X and Z protection.\\

The resultant performance of register-based colour codes using biased noise is shown in Fig.~\ref{fig:colour_code}c and f for 6.6.6 and 4.8.8 colour codes respectively, using the Chromobius decoder~\cite{gidney2023_chromobius}. From Fig.~\ref{fig:colour_code}c, we find a crossing threshold of $\sim 0.55 \%$ for the 6.6.6 colour code, and likewise in Fig.~\ref{fig:colour_code}e we find a crossing threshold of $\sim 0.55 \%$ for the 4.8.8 colour code. Further scaling performance can be found in Appendix~\ref{sec:supp_col_scaling}. These thresholds are considerably higher than the thresholds under unbiased noise, reported as $\sim 0.08 \%$ for time-multiplexed measurement~\cite{colour_code_summary}, as well as flagged weight optimisation ($0.36 \%$)~\cite{colour_code_unbiased_performance}. The performance is on par with superdense ($\sim 0.4 - 0.5 \%$) and middle-out measurements ($\sim 0.6-0.7 \%$)~\cite{gidney2023_chromobius}. These exceptionally high thresholds demonstrate how biased noise can significantly improve colour code performance, mainly as a result of the first-order mitigation of stabiliser hook errors.

\section{Bilayer Surface Codes~\label{sec:bilayer_codes}}

\begin{figure*}[!ht]
\includegraphics[width=0.7\textwidth]{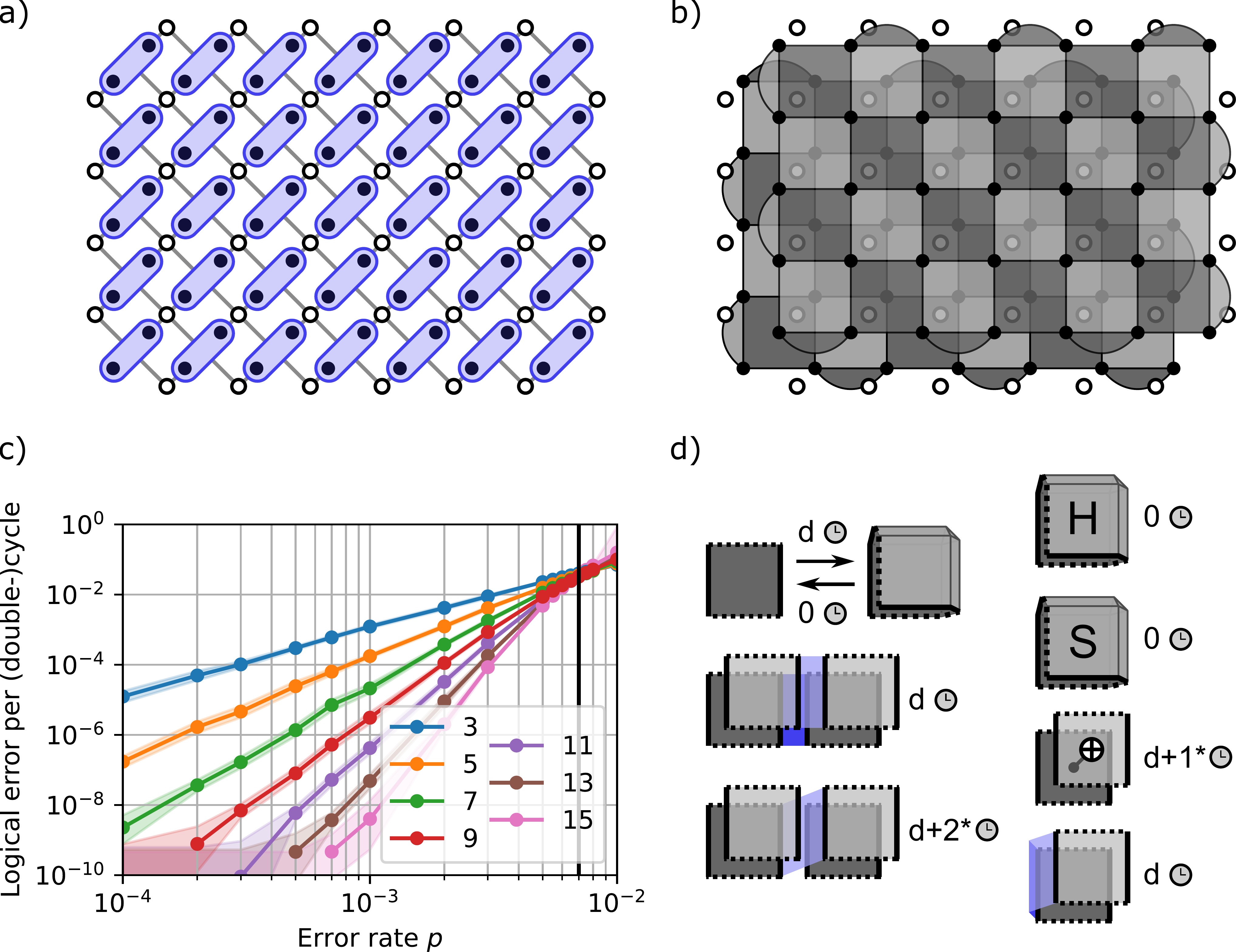}
\centering
\caption{\textbf{Implementation and performance of bilayer surface codes using 1- and 2-qubit registers under biased noise.} (a) Layout of alternating grid of 2-qubit and 1-qubit registers used for implementing a bilayer surface code, as in Fig.~\ref{fig:colour_code}a and d. Register outlines have been omitted for single-qubit registers, and the registers have square lattice connectivity. (b) Stabilisers of the bilayer surface code, hosting two surface codes, with the same register layout as per a). Note that each ancilla is responsible for at most two stabilisers, handling at most one Z-type and at most one X-type stabiliser. This allows stabilisers to be measured in two steps, first measuring all Z stabilisers in both layers, and then measuring all X stabilisers in both layers. (c) Threshold diagram of a surface code logical qubit hosted in one of the layers of a bilayer code. The (double-)cycle referenced on the y-axis involves one set of measurements of both X and Z stabilisers (ie. two rounds of measurement). Similar to the colour codes, the bilayer code here is simulated for square bilayer codes stabilised with time-domain walls to equalise logical X and Z protection. A crossing threshold of $\sim 0.7 \%$ is marked with a vertical black line. This is higher than the crossing threshold reported in Fig.~\ref{fig:colour_code}c,f of $\sim 0.55 \%$, showing a similar difference to previous work with circuit-level noise on surface and colour codes~\cite{pymatching, gidney2023_chromobius}. (d) Summary of non-standard logical operations available to bilayer surface codes, illustrated as logical qubit patches in the style of Litinski~\cite{Litinski_game_of_surface_codes}, where the clock symbol (\clock) indicates the number of (double-)cycles needed to perform the operation. A full description of the operations available for bilayer codes can be found in Appendix~\ref{sec:supp_bilayer_operations}. Note that the transversal CNOT can be $d$ (double-)cycles instead if one of the logical qubits is measured directly after the CNOT, and also that $d$ (double-)cycles of the transversal CNOT can be combined with a preceding $d$-cycle translation of one (or both) of the involved logical qubits. Similarly, the diagonal lattice surgery can be performed in $d+1$ cycles if one of the \textit{uninvolved} logical qubits is measured out after the surgery or translated before the surgery, and in $d$ cycles if one of the uninvolved qubits is both translated and measured in this way, or if one of the uninvolved patches is unused.~\label{fig:bilayer_code}}
\end{figure*}

In addition to facilitating efficient stabilisation of 6.6.6 and 4.8.8 colour codes, the extra connectivity provided by a register architecture also allows for non-standard surface codes to be constructed. In particular, we consider the construction of bilayer surface codes, where logical surface code qubits are not confined purely to a 2D plane, but instead are hosted on two 2D planes and can locally move between planes as well as laterally. Effectively, this is the same as cubic lattice logical connectivity, but where one of the dimensions of the lattice is capped at size two. Such bilayer codes have previously been studied theoretically by Moussa for qudit-based systems~\cite{moussa_bilayer}, but the same idea is directly applicable to the qubit register approach studied here. As illustrated in Fig.~\ref{fig:bilayer_code}d and explained in detail in Appendix~\ref{sec:supp_bilayer_operations}, the transversal Clifford gates studied by Moussa are also applicable here to logical qubits defined on a bilayer surface code.\\

To define the bilayer surface code, we again start with a square lattice of alternating two-qubit and one-qubit registers, with the one-qubit registers assigned as ancilla registers, as per Fig.~\ref{fig:bilayer_code}a. We construct layer one of the bilayer surface code by arbitrarily choosing a qubit from each two-qubit register, and using these in conjunction with the ancilla registers. Similarly, layer two is constructed using the remaining data qubits (ie. the qubits belonging to the two-qubit registers which were not chosen for layer one). We define edge stabilisers of these codes such that the X and Z boundaries are coincident, but the two-qubit edge stabilisers are offset between the two layers. We (arbitrarily) label one layer as being in the ``foreground'' and the other as being in the ``background'', to assist with visualisation and explanation of bilayer code operations. The full set of stabilisers for an example $5 \times 7$ bilayer code is shown in Fig.~\ref{fig:bilayer_code}b, corresponding to the same register layout as in Fig.~\ref{fig:bilayer_code}a.\\

To stabilise the bilayer code, similarly to stabilisation of the colour code, we alternate between measurements of the Z stabilisers and the X stabilisers. In addition, as with the colour code, we introduce time domain walls when stabilising the bilayer code so that X and Z logical errors are equally protected, somewhat analogous to the spatial domain walls used in the XZZX code~\cite{BonillaAtaides2021_XZZX}. As is evident from the illustration in Fig.~\ref{fig:bilayer_code}b, each ancilla qubit is responsible for at most one Z stabiliser and at most one X stabiliser, allowing each ancilla to participate in at most one stabiliser measurement each round. This therefore ensures no ancilla is ``overloaded'', needing to measure two separate stabilisers simultaneously. The cost of this strategy is that it takes two sets of measurements to fully stabilise both layers of the bilayer code, similar to the colour code, which we again refer to as a (double-)cycle. The full stabilisation sequence can be found in Appendix~\ref{sec:supp_bilayer_stabilisation}. The register connectivity model assumes that only one inter-register gate can occur at a time (see Fig.~\ref{fig:register_intro_gateset}), and only one qubit per register can be initialised/measured at a time. Under these assumptions, this (double-)cycle stabilisation scheme is optimal in terms of number of measurement timesteps and inter-register gates. In other platforms where measurements of multiple qubits in a register may be made simultaneously, such as superconducting qubits~\cite{Bravyi2024_LDPC, yoder2025tourgrossmodularquantum} and/or neutral atom arrays~\cite{Bluvstein2024_neutral_atoms}, it might be possible to measure all stabilisers in a single cycle.\\

Bilayer codes exhibit useful properties which could outweigh their higher stabilisation time-cost. Firstly, they are highly connected, with lattice surgery possible between layers in the same patch (eg. between the two qubits shown in Fig.~\ref{fig:bilayer_code}b), between either layer in a patch and/or either layer in a neighbouring patch, as illustrated in Fig.~\ref{fig:bilayer_code}d (see Appendix~\ref{sec:supp_bilayer_operations} for implementation details). With another logical qubit in the same patch and four pairs of neighbouring logical qubits, each logical qubit in the bilayer code has nine potential neighbours it can perform lattice surgery with. In addition, bilayer codes support efficient transversal $d+1$ (double-)cycle CNOT gates, and can be folded (in $d$ (double-)cycles) to facilitate instant Hadamard gates and $S$ gates, as detailed in Appendix~\ref{sec:supp_bilayer_operations}.\\

The performance of an individual logical qubit measured as part of a bilayer surface code is shown in Fig.~\ref{fig:bilayer_code}c, with additional results shown in Appendix~\ref{sec:supp_bilayer_performance}. Notably, these results show that the bilayer code has error threshold of $\sim 0.7 \%$, at the lower end of the commonly accepted single-layer surface code threshold of $\sim 0.7 - 1.0\%$, despite having a higher circuit depth to measure a full cycle of stabilisers. The high-threshold operation of bilayer codes, along with their high connectivity, makes them promising for fast logical operation and for solving routing problems between distant logical qubits. In addition, by performing lattice surgery along one matching edge of two bilayer qubits, they facilitate folded surface codes. Folded surface codes can allow for transversal Clifford gates~\cite{moussa_bilayer}, which allows for faster logical operations. Moreover, folded surface codes can be used to improve the performance of magic state distillation, as explored in the following section.

\section{The Magic Scroll~\label{sec:magic_scroll}}

Any quantum circuit involving purely Clifford operations can be efficiently computed with a classical computer in polynomial time~\cite{Aaronson_2004, Gidney_2021_stim}. As such, the advantage of quantum computation stems from the ability to involve non-Clifford operations in the quantum circuit~\cite{Howard_2014}. Unfortunately, both colour codes and surface codes cannot fault-tolerantly implement non-Clifford operations, and indeed this is impossible for any two-dimensional topological stabiliser code~\cite{bryan_nogo}. Instead, a common approach is to produce resource ``magic'' states which can then be consumed by a two-dimensional fault-tolerant code via Clifford operations. Magic states are any state which cannot be expressed as an eigenstate of Pauli products, ie. any state which cannot be obtained via Clifford operations acting on qubit(s) in the initial state $\ket{0}$. Common magic states involve the $\ket{T}$ state ($\ket{T} = T \ket{+}$), and the $\ket{CCZ}$ state ($\ket{CCZ} = CCZ \ket{+}^{\otimes 3}$). Traditionally, production of high-fidelity magic states has been significantly more expensive than the operations which consume magic states, and so making magic state production more efficient has been a significant area of research~\cite{gidney_cultivation, vaknin2025magicstatecultivationsurface, Chamberland2020, Litinski_distillation}.\\

One significant recent milestone was the switch from magic state distillation to magic state cultivation, as presented by Gidney \textit{et. al.}~\cite{gidney_cultivation}. Magic state cultivation involves hosting a magic $\ket{T}$ state on a colour code which is made progressively larger, using heavy postselection to ensure the magic state is maintained with higher and higher code distance. Finally, the magic state is escaped to a larger hybrid colour-surface code which is able to support the magic state without significant postselection. By only scaling up the number of qubits involved when the previous rounds have passed postselection, the expected volume (ie. expected qubit-rounds) is reduced, eventually making the quantum volume on-par with native fault-tolerant operations such as the CNOT gate~\cite{gidney_cultivation}.\\

Incorporating both colour codes and bilayer surface codes, we present a modified procedure of magic state cultivation based on the qubit-register models considered here. The cultivation process proceeds similarly to that by Gidney \textit{et. al.}~\cite{gidney_cultivation}:

\begin{enumerate}
    \item Start with a single-qubit magic $\ket{T}$ state,
    \item Progressively grow (g), stabilise (s) and double-check (dc) the magic state, improving its error while also increasing the distance of the colour code it is hosted in,
    \item Escape the magic state into a larger code which can support the high-fidelity state without postselection, and
    \item Compute the complementary gap of the final escaped state to determine whether to keep the magic state.
\end{enumerate}

However, our procedure differs from Gidney in four main areas:

\begin{itemize}
    \item Our magic state is hosted in a 4.8.8 colour code, rather than a 6.6.6 colour code, to minimise stabiliser count.
    \item We escape to a (partially folded) surface code, measuring out the colour code segment to complete the escape sequence, as opposed to leaving the magic state in a grafted code, to ensure efficient protection post-escape.
    \item All stabilisers involved with the 4.8.8 colour code are fully postselected (including during escape), allowing a normal surface code decoder to compute the complementary gap.
    \item The surface code height (distance) is limited during escape, and only grown to its full size at the end of escape, again to reduce qubit count where possible.
\end{itemize}

By combining these modifications, we find that the performance of magic state cultivation is significantly improved, as shown in Fig.~\ref{fig:gidney_style_figure}. Specifically, we find an improvement to error rates for a given volume by up to $\sim 100\times$, or alternatively we find an improvement to quantum volume for a fixed error rate by up to $\sim 3\times$. We explain these improvements in more detail in the following sections.

\subsection{Magic Scroll: Cultivation}

\begin{figure*}[!ht]
\includegraphics[width=0.75\textwidth]{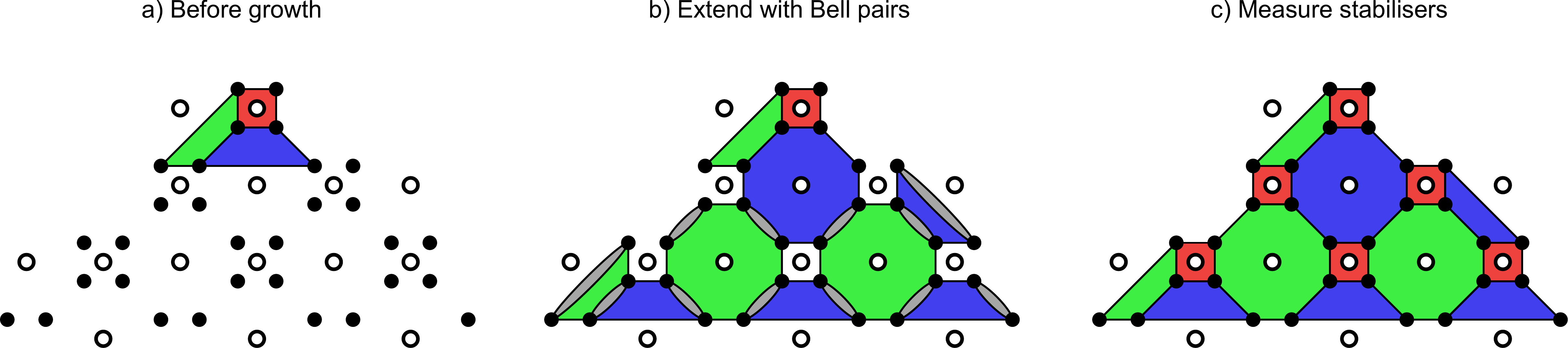}
\centering
\caption{\textbf{Method used to grow a 4.8.8 colour code from distance-3 to distance-7 using Bell pairs.} (a) Initial stabilisers of the distance-3 colour code. (b) Bell pairs are created between pairs of data qubits (marked in grey), which in the absence of errors creates deterministic values for the green and purple stabilisers (ie. all octagonal stabilisers), as shown. Note that only the leftmost green and rightmost purple stabilisers require Bell pairs between data qubits in different registers, with all other Bell pairs being created locally within a register (see Fig.~\ref{fig:colour_code}c for register layout). (c) After measurement of all stabilisers of the distance-7 colour code, the logical state hosted within the original distance-3 code has been transferred to the full distance-7 code. Note that now the red (ie. square) stabilisers have determined values, while the individual Bell pairs have lost their stabilisers due to the red stabiliser measurement.~\label{fig:colour_code_grow}}
\end{figure*}

The first section of the Magic Scroll procedure is to create a low-error magic state hosted in a 4.8.8 colour code. For simplicity, instead of injecting a distance-1 magic state into a distance-3 colour code (as is undertaken in~\cite{gidney_cultivation}), we opt to create a single-qubit magic state and grow this effective distance-1 colour code into a distance-3 colour code. This allows us to use the same growth procedure as in the rest of the cultivation component.\\

\begin{figure*}[!ht]
\includegraphics[width=0.7\textwidth]{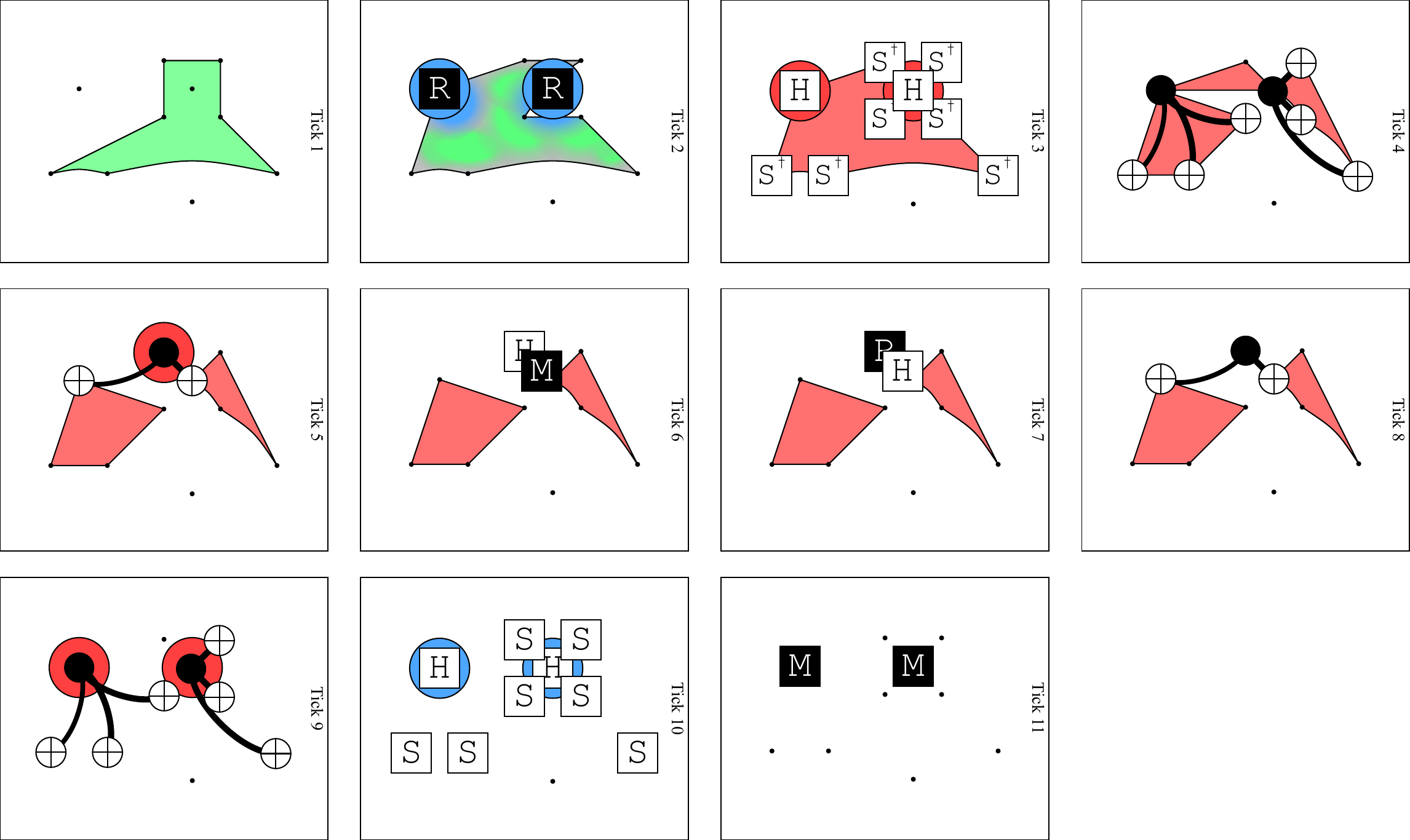}
\centering
\caption{\textbf{Detector-slice diagram of method used to double-check a magic state encoded in a distance-3 colour code.} In tick 2, two ancillas are initialised, which will serve as flag qubits at the end of the double-check. In tick 3, single-qubit gates are applied such that the original $Y$ observable is mapped to a transversal X parity. In ticks 4-6, this parity is folded into a single qubit using CNOT gates, which is then measured, performing the first check. Ticks 7-11 perform the inverse operation of ticks 2-6, finishing with measurement of the flag ancilla qubits in tick 11 to perform the second check. For magic states, replace all $S$ ($S^\dag$) gates with $T$ ($T^\dag$) gates, which will then double-check the $(X+Y)/2$ observable instead of the $Y$ observable. For clarity, the circuit has been notated using CNOT gates, however in simulation these are decomposed into our native gateset of $H$ and $CZ$. Please see Appendix~\ref{sec:supp_double_check} for the full decomposed circuit. $H$ refers to the Hadamard gate, $\dag$ indicates Hermitian conjugation, $M$ and $R$ indicate measurement and reset respectively, and as usual $S \equiv \sqrt{Z}$ and $T \equiv \sqrt{S}$. ~\label{fig:colour_code_d3_dc}}
\end{figure*}

\begin{figure*}[!ht]
\includegraphics[width=\textwidth]{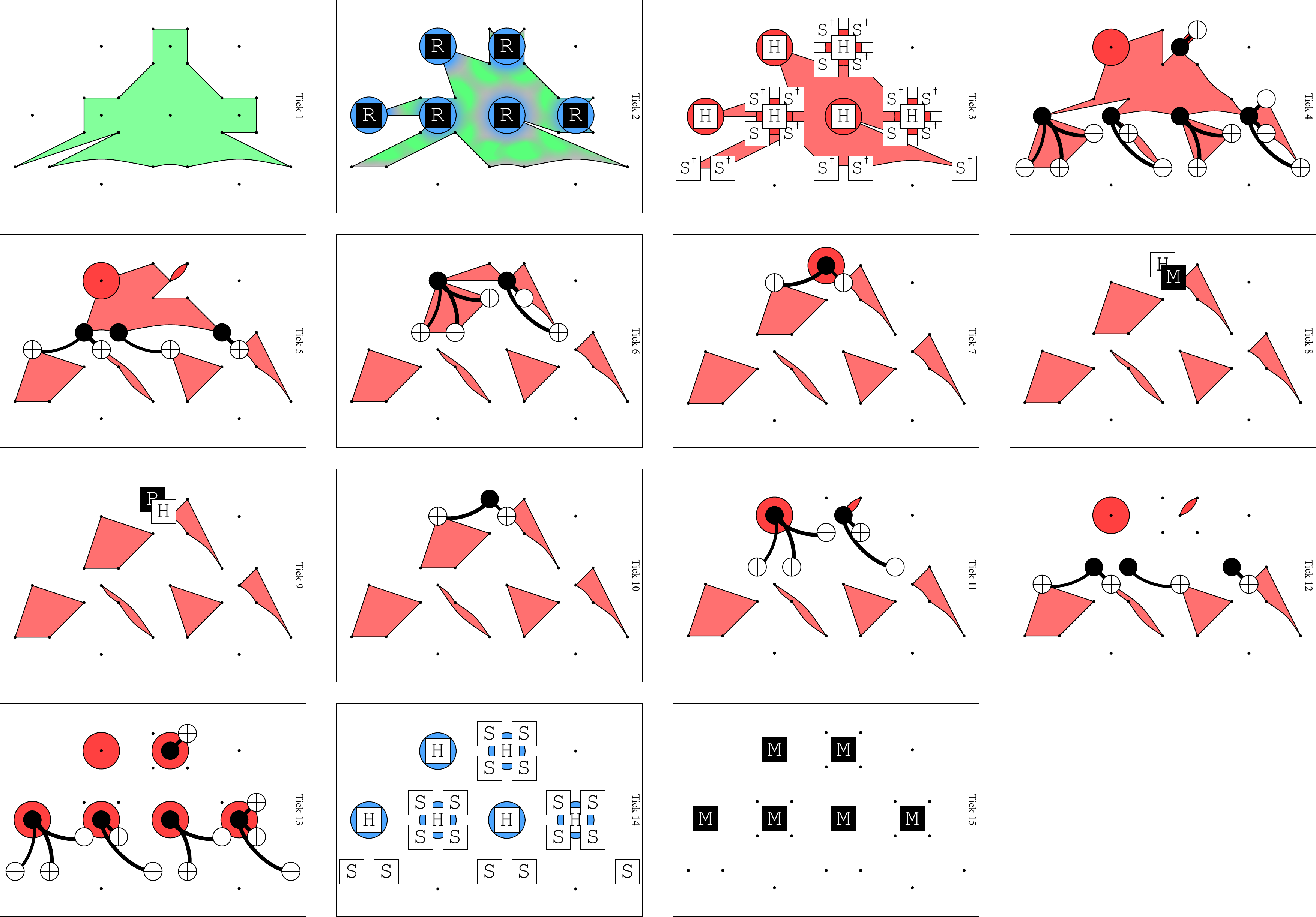}
\centering
\caption{\textbf{Detector-slice diagram of method used to double-check a magic state encoded in a distance-5 colour code.} The procedure is implemented in a similar way to Fig.~\ref{fig:colour_code_d3_dc}. In tick 2, six flag qubits are initialised. Tick 3 then applies gates to map the desired observable to a transversal $X$ parity. This parity is then folded using CNOT gates in ticks 4-8, performing the first check. The remaining ticks 9-15 perform the inverse circuit, finishing by measuring the ancilla flag qubits to perform the second check. For magic states, replace all $S$ ($S^\dag$) gates with $T$ ($T^\dag$) gates. For clarity, the circuit has been notated using CNOT gates, however in simulation these are decomposed into our native gateset of $H$ and $CZ$. Please see Appendix~\ref{sec:supp_double_check} for the full decomposed circuit.~\label{fig:colour_code_d5_dc}}
\end{figure*}

Similar to the 6.6.6 colour code, the 4.8.8 colour code can be grown by creating Bell pairs between data qubits which define most stabilisers, and then measuring the remaining undefined stabilisers. This procedure is shown in Fig.~\ref{fig:colour_code_grow} for the example of growing a distance-3 colour code to a distance-7 colour code. Errors are reduced here significantly by the fact that most Bell pairs can be created locally between the two qubits within a single register, without needing to use remote CZ gates (which have higher error) and without needing to involve a mediating ancilla qubit. Circuits for creating these Bell pairs (specifically, $(\ket{01}+\ket{10})/\sqrt{2}$), including the edge pairs which require a mediating ancilla qubit, are described in Appendix~\ref{sec:supp_bell_pair_circuit}. Importantly for cultivation, the growth process maintains the (arbitrary) quantum state hosted in the original colour code. The transversal logical X or Z operators are maintained by the Bell pairs shown in Fig.~\ref{fig:colour_code_grow}, and all other logical X or Z operators (eg. strings of X/Z operators crossing the colour code) can be obtained by multiplying by stabiliser products.\\

The second part of the cultivation process is to stabilise the code. The stabilisation process for colour codes has already been established in previous sections (see Fig.~\ref{fig:colour_code} and Appendix~\ref{sec:supp_col_stabilisation}). Stabilisation of the colour code proceeds identically to the process previously discussed, including using time-domain walls to equally protect X and Z logical errors.\\

The final component of the cultivation process is to double-check the magic state is still hosted within the colour code. The double-check procedure, if it passes postselection, decreases the error of the magic state beyond the error rates afforded by the lower distance code it was grown from. The circuit that performs this double-check leverages the fact that Clifford gates are transversal for colour codes to measure the eigenvalue of the Clifford $(X+Y)/\sqrt{2}$ operator, for which the $\ket{T}$ magic state is the $+1$ eigenstate. Similar to the approach by Gidney \textit{et. al.}, we map the parity of the transversal $(X+Y)/\sqrt{2}$ operator to the transversal X parity using by applying a $T$ gate to all data qubits, and then folding this parity to just one qubit to measure it. Simultaneously, additional flag qubits are used to ensure the unfolding procedure proceeds correctly, representing a secondary check. This procedure is based on the work by Gidney \textit{et. al.}~\cite{gidney_cultivation}.\\

One key difference compared to the process shown by Gidney \textit{et. al.} is that, due to the lower ancilla count in register-based colour codes, instead of having one flag qubit entangled with the single-qubit X parity of each data qubit, ancilla qubits are instead entangled with the two-qubit XX parity of a pair of data qubits on the same register (which can be done efficiently using a parity-Z gate or two CZ gates). The overall sequences for distance-3 and distance-5 double-checks are shown in Fig.~\ref{fig:colour_code_d3_dc} and Fig.~\ref{fig:colour_code_d5_dc} respectively. For a full decomposition of this procedure into native gates, as well as the circuit used to double-check a distance-7 colour code, please see Appendix~\ref{sec:supp_double_check}.\\

During the cultivation stage, we postselect all detector results. That is, if any stabilisers flip, or if any of the flags of the double-check procedure are triggered, then the cultivation process is fully restarted. Because the colour code is progressively grown larger, early restarts do not significantly contribute to the overall expected quantum volume as only a few qubits are in use. By repeatedly using the grow (g), stabilise (s) and double-check (dc) steps, if the procedure passes postselection, it is possible to produce very high quality magic states hosted in distance-3, distance-5 or distance-7 colour codes. However, these magic states now have such low errors that it is impossible to maintain them in such low-distance colour codes without postselection. Instead, we escape these low-error states to a much larger code capable of supporting it, as detailed in the following section.

\subsection{Magic Scroll: Escape}

\begin{figure*}[!ht]
\includegraphics[width=0.75\textwidth]{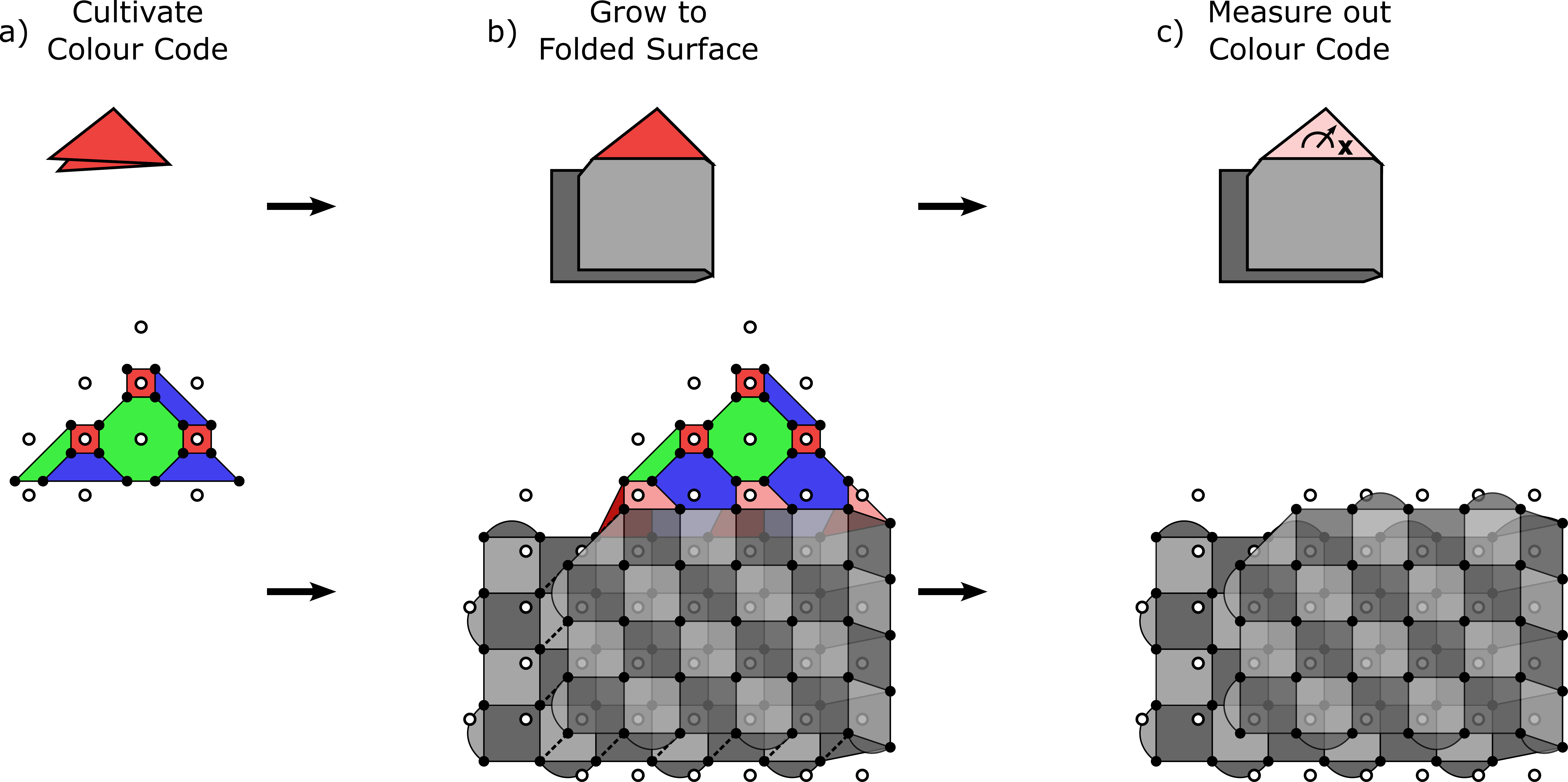}
\centering
\caption{\textbf{Summary of steps used to transfer the cultivated magic state from the colour code to a folded surface code.} Here, we escape the magic state held in the 4.8.8 colour code (cultivated as per Figs.~\ref{fig:colour_code_grow}-\ref{fig:colour_code_d5_dc}) into a partially folded surface code, constructed utilising the bilayer surface code defined in Fig.~\ref{fig:bilayer_code}. Top: High-level illustration of the steps taken to transfer the state from the colour code into the surface code. Note that in the final step, the colour code is measured in the X basis by measuring all data qubits in the X basis, leaving the magic state just in the folded surface code and allowing the upper-edge stabilisers of the folded surface code to be checked after measurement. Bottom: Full layout of the escape sequence for a distance-5 colour code escaping to a $13 \times 5$ partially folded surface code. For actual escape sequences, much larger heights are used, but we use $5$ here to keep the diagram compact. Data qubits are illustrated by filled circles, and ancillas are illustrated by open circles. Grey surfaces represent surface code stabilisers, while coloured surfaces represent colour code stabilisers. Light surfaces represent Z stabilisers and dark surfaces represent X stabilisers, with some stabiliser shading being omitted for clarity in the colour code. In the folded surface code, two qubit data registers are used (similar to the bilayer surface code in Fig.~\ref{fig:bilayer_code}), with some registers being indicated with dotted lines in (b) for clarity.~\label{fig:magic_scroll_escape}}
\end{figure*}

Other than the switch to using the 4.8.8 colour code instead of the 6.6.6 code, the cultivation stage proceeds much the same as for Gidney \textit{et. al.}~\cite{gidney_cultivation} with the main differences being the stabiliser measurement sequence and the ordering of grow, stabilise and double-check steps. The procedure for escaping the magic state into a larger code capable of sustaining it without postselection is, however, significantly different. We aim to leave the magic state in a surface code (as opposed to a half-colour half-surface grafted code per~\cite{gidney_cultivation}). As illustrated at the top of Fig.~\ref{fig:magic_scroll_escape}, this means that the final surface code portion during the escape sequence is completely separate to the colour code section. In addition, we fold the surface code (using the bilayer codes discussed earlier, see Fig.~\ref{fig:bilayer_code}) to increase the number of stabilisers joining the two codes. This also matches the topology of the colour code, as topologically the colour code is equivalent to a folded surface code~\cite{Kubica_2015_unfolding_the_colour_code}. As a final step of the escape sequence, we measure out the colour code section in the X basis, allowing us to completely transfer the magic state from the colour code to the folded surface code.\\

The qubit arrangement and stabilisers involved in the escape sequence are illustrated in the bottom half of Fig.~\ref{fig:magic_scroll_escape}. In Fig.~\ref{fig:magic_scroll_escape}a, we start with the magic state housed within the colour code at the end of the cultivation process. Next, we introduce the folded surface code section, by initialising all the bottom (wider) layer data qubits to $\ket{+}$, and the rest of the qubits (including the qubits on the fold) to $\ket{0}$. By measuring out the stabilisers shown in Fig.~\ref{fig:magic_scroll_escape}b, the colour code is grown so that the logical state is housed in the joint colour-surface code, in a similar way to how the colour code distance is grown in Fig.~\ref{fig:colour_code_grow}. This measurement procedure is also performed with time-domain walls to ensure equal protection of X and Z logical operators. Note that the foreground layer of the surface code in Fig.~\ref{fig:colour_code_grow}b is missing the top-left data qubit - this is intentional, as it facilitates merging with the colour code, while minimising the number of distance-$d$ (where $d$ is the colour code distance) error strings possible within the surface code during the escape sequence. Note that this means the distances of the final surface code in Fig.~\ref{fig:colour_code_grow}c is $13 \times 5$, even though the number of qubits involved is much closer to a $14 \times 5$ surface code.\\

For the joint colour-surface code shown in Fig.~\ref{fig:magic_scroll_escape}b, there are too many stabilisers to continue fully postselecting all stabilisers (as was done during cultivation). However, due to the low stabiliser count for the 4.8.8 colour code, we find it is possible to continue fully postselecting the coloured stabilisers in Fig.~\ref{fig:magic_scroll_escape}b (including the stabilisers along the join between the colour and surface code sections). This is significant, as the remaining folded surface code is matchable, and as such is straightforward to decode using standard matching techniques. In addition, the use of full postselection allows the merge to be performed to very low error, despite the relatively low code distance of the colour code. To further eliminate distance-$d$ error mechanisms in the surface code during escape, the two stabilisers directly below the leftmost red stabilisers on the join are postselected in the first escape cycle. The final benefit of fully postselecting the colour code and joint stabilisers in Fig.~\ref{fig:magic_scroll_escape}b is that the error rate of the Magic Scroll procedure can be easily understood by splitting it into three components: a fully postselected colour code cultivation and distance-$d$ idle; a fully postselected distance-$d$ merge operation; and a complementary-gapped rectangular surface code.\\

After the joint colour-surface code has been measured for some number of rounds (the exact number depending on the quality of the magic state desired after the escape sequence), the final step of the escape sequence is to measure out the colour code (transversely) in the X basis. A final step of postselection ensures that the measured results of the colour code data qubits match the colour code stabilisers (to ensure no measurement errors). For colour code qubits which are involved in the join between the colour and surface codes, this postselection is used to ensure the creation of X-boundary stabilisers on the top edge of the folded surface code, as seen in Fig.~\ref{fig:magic_scroll_escape}c. To enable this postselection, one round of stabilisers is measured for the lone folded surface code in Fig.~\ref{fig:magic_scroll_escape}c, with this first round being matched except for the new top X-boundary stabilisers which are postselected. In addition, after the measure-out of the colour code, in some cases we increase the height of the surface code (by growing per~\cite{Litinski_game_of_surface_codes}), allowing the escaped state to be properly maintained after the escape sequence is complete. Subsequent rounds can match the entire folded surface code, which now hosts the magic $\ket{T}$ state. For additional efficiency, we found that making the final stabiliser measurement before colour measure-out a Z-basis stabiliser measurement improved results, even though this means skipping one X-stabiliser. Finally, we combine the colour code measurement operations with the following stabilisation of the surface code to save on total rounds, and also to avoid additional idle errors on the surface code.\\

After the state has escaped into a fully matchable surface code, to ensure the quality of the escaped magic state we decide whether or not to keep the escaped magic state by using the complementary gap method~\cite{gidney2023yokedsurfacecodes}. That is, for all of the non-postselected stabilisers that were measured during the escape sequence, we decode them using MWPM methods while enforcing (via a dummy detector) that either a) there was a logical error or b) there was no logical error. By comparing the weights of the matching results for these two cases, it is possible to estimate how confident we should be in the decoder's solution. By only keeping instances with high decoder confidence in the final logical state, we significantly reduce the error rate of the kept state. We estimate that such a gap can be calculated within one (double-)cycle of stabilisation, based on the $\sim 10 \mu$s (double-)cycle times of the solid-state spin systems which inspired the register connectivity and biased error model studied here \footnote{Note that in Gidney \textit{et. al.}, which was inspired by superconducting qubits, the cycle time was assumed to be 1 $\mu$s and so instead there it was assumed that an extra 10 rounds would be needed to wait for the decoding result.}. Hence, the stabilisers in Fig.~\ref{fig:magic_scroll_escape}c are measured for two (double-)cycles, first to allow the edge stabilisers to be checked via postselection with the colour code X measurement result, and secondly to allow the complementary gap to be computed.

\subsection{Magic Scroll: Performance}

To simulate the performance of the Magic Scroll compared to the results of Gidney \textit{et. al.}~\cite{gidney_cultivation} in Fig.~\ref{fig:gidney_style_figure}, we use the proxy of cultivating a $\ket{Y} = S\ket{+}$ state instead of a $\ket{T}$ state. This amounts to replacing the $T$ ($T^\dag$) gates with $S$ ($S^\dag$) gates for the initial single-qubit seed state, as well as the double-check circuits (see Fig.~\ref{fig:colour_code_d3_dc} and Fig.~\ref{fig:colour_code_d5_dc}). We assume that the error rate of the $\ket{Y}$ state can be used to obtain a good estimate of the error rate if cultivating the $\ket{T}$ state, using the same assumption as Gidney \textit{et. al.}~\cite{gidney_cultivation}. Specifically, we assume that for the cultivation (pre-escape) stage, the errors obtained for a $\ket{Y}$ state can be doubled to obtain an estimate of the $\ket{T}$ state error. When considering the full Magic Scroll sequence including both cultivation and escape stages, because the escape stage does not use any non-Clifford operations (and therefore $\ket{Y}$ and $\ket{T}$ states should appear the same), we simulate the $\ket{Y}$ error and then add to this the no-escape $\ket{Y}$ error, which emulates the doubling of the pre-escape error.\\

There are a variety of parameters which can be varied for the Magic Scroll procedure. To optimise the efficiency of magic state production, we vary the following:

\begin{itemize}
    \item The final distance $d$ of the colour code before escaping to the surface code,
    \item The width $w$ and final height $h$ of the folded surface code left after the escape sequence, as well as the height $h'$ of the folded surface code during escape,
    \item The number of rounds $r$ over which the joint colour-surface code is measured before measuring out the colour code, and
    \item The sequence of grow (g), stabilise (s) and double-check (dc) steps used in the cultivation stage.
\end{itemize}

For a Magic Scroll which cultivates a $d=5$ colour code using two rounds of grow, stabilise, double-check, before escaping to a $w=13$, $h'=7$, $h=11$ surface code with $r=3$ joint rounds, we notate these settings as follows:

\begin{align}
    \textrm{d=5: g-s-dc-g-s-dc }\to\textrm{ 13x(7)11 r=3}~\label{eq:cultivation_label}
\end{align}

Due to the tradeoff between more spacetime volume and lower error rate, we consider multiple different settings for the Magic Scroll sequence, noted in the legend of Fig.~\ref{fig:gidney_style_figure}. We simulate the circuit using the python package stim~\cite{Gidney_2021_stim}, using pymatching to perform decoding~\cite{pymatching}. To perform efficient Y-basis measurements of the surface code, we use an error-free final round of stabiliser measurements and measurement of the Y-operator on the folded surface code, similar to the technique used in~\cite{gidney_cultivation}. The timing of this final round is included in resource cost, as it accounts for the time required to compute the complementary gap.\\

The results of our simulations are shown in Fig.~\ref{fig:gidney_style_figure}, along with the performance of Gidney \textit{et. al.} for comparison. Note that each group of the same coloured symbol represents a variety of different thresholds for the complementary gap, with the highest error point representing no complementary gap threshold. Our choices of cultivation parameters were found empirically, but were chosen to provide efficient trade-offs between volume and magic state error rate. The Magic Scroll demonstrates substantial improvements in magic state production, in particular in the vicinity of $10^4$ qubit-rounds where errors have improved by up to two orders of magnitude compared to Gidney \textit{et. al.} (cyan arrow in Fig.~\ref{fig:gidney_style_figure}). Equivalently, the expected quantum volume of creating magic states of a given error rate has dropped by up to a factor of 3 (blue arrow in Fig.~\ref{fig:gidney_style_figure}). This demonstrates the power of biased noise in improving the performance of magic state production, which is leveraged by the Magic Scroll procedure.\\

Notably, we use more double-checks and fewer stabiliser measurements during the cultivation stage as compared to~\cite{gidney_cultivation}. While the optimal circuits were found by a process of trial-and-error by iterating variants of the Magic Scroll circuit and simulating the performance (rather than a more well-defined rule/procedure), we suspect this need for more double-checks is related to our biased noise model, which despite completely eliminating some error channels, will also boost the probability of certain other errors (eg. $p/3$ chance for a ZZ error after a CZ gate per Table~\ref{tab:error_model}), potentially reducing the distance/performance of the grow and double-check steps. We believe this is why the Magic Scroll performance shown in Fig.~\ref{fig:gidney_style_figure} only shows significant improvement around errors of $10^{-8}$ and volumes of $10^4$ qubit-cycles, rather than showing improvement over the full parameter range.\\

For cultivation utilising $d=7$ colour codes, the error rate of the escaped state for high thresholds proved too low to feasibly simulate in a reasonable amount of time (we predict proper full simulations would take multiple months on the hardware we utilised for this work). As such, the $d=7$ results shown in Fig.~\ref{fig:gidney_style_figure} are partially extrapolated based on the results which were tractable, using the procedure explained in Appendix~\ref{sec:supp_magic_scroll_extrapolation}. Extrapolated results are shown in the dashed lines, and should be taken as rough estimates of the true performance (likely accurate within an order of magnitude, but could feasibly be off by a factor of 2).\\

Despite this shortcoming, overall the Magic Scroll process presents significant improvement in the efficiency of magic state creation by up to $3\times$ in expected quantum volume. This further reduces the requirements for the synthesis of low-error magic states, improving the overall runtime of quantum computations on register-based quantum platforms with biased noise by up to $3\times$ if computation is magic-state limited. In addition, we see significant improvements for the lower-volume, higher-error states ($\sim 10^{-5}$ magic state error), providing cheaper magic states for nearer-term fault-tolerant applications, as well as facilitating improved distillation which we explore in the following section.

\section{Distillation from The Magic Scroll~\label{sec:magic_scroll_distillation}}

While the Magic Scroll can produce low-error T states, down to $\sim 1 \times 10^{-10}$ error for physical gate errors of 0.1\%, these error rates may still be insufficient for all quantum algorithms (for example, algorithms with $> 10^{9}$ T gates require magic state error $< 10^{-10}$ for overall algorithm success rate to be $> 90\%$). One method for producing lower-error magic states would be to reduce physical gate errors, which is particularly effective for cultivation procedures. We show this explicitly for an error rate of $p=0.05\%$ in Appendix~\ref{sec:supp_magic_scroll_005}, in which we find that reducing the error rate from $p=0.1\%$ to $0.05\%$ improves expected volumes by almost $20\times$ and improves magic state errors by almost $100\times$. These gains are largest for the distance-7 colour code, with relative gains for distance-3 and distance-5 colour code cultivation on par with Gidney \textit{et. al.}~\cite{gidney_cultivation}.\\

Despite the significant boost in cultivation performance given by halving physical error rates, it is possible that such a reduction in error rates may be challenging. As such, it is important to consider alternative schemes which may be able to achieve lower error rates, despite the increased quantum volume costs of doing so. Distillation remains a promising candidate in this respect, as it allows quadratic or even cubic suppression of input error rates~\cite{Litinski_distillation}. By using cultivated magic states as input to distillation, it is possible to achieve error rates much lower than that possible from cultivation alone. Indeed, for the cultivation presented in Fig.~\ref{fig:gidney_style_figure}, higher error schemes are where much of the volume advantage of the Magic Scroll is seen, suggesting that using these as inputs to distillation may perform well.\\

\begin{figure}[!ht]
\includegraphics[width=0.48\textwidth]{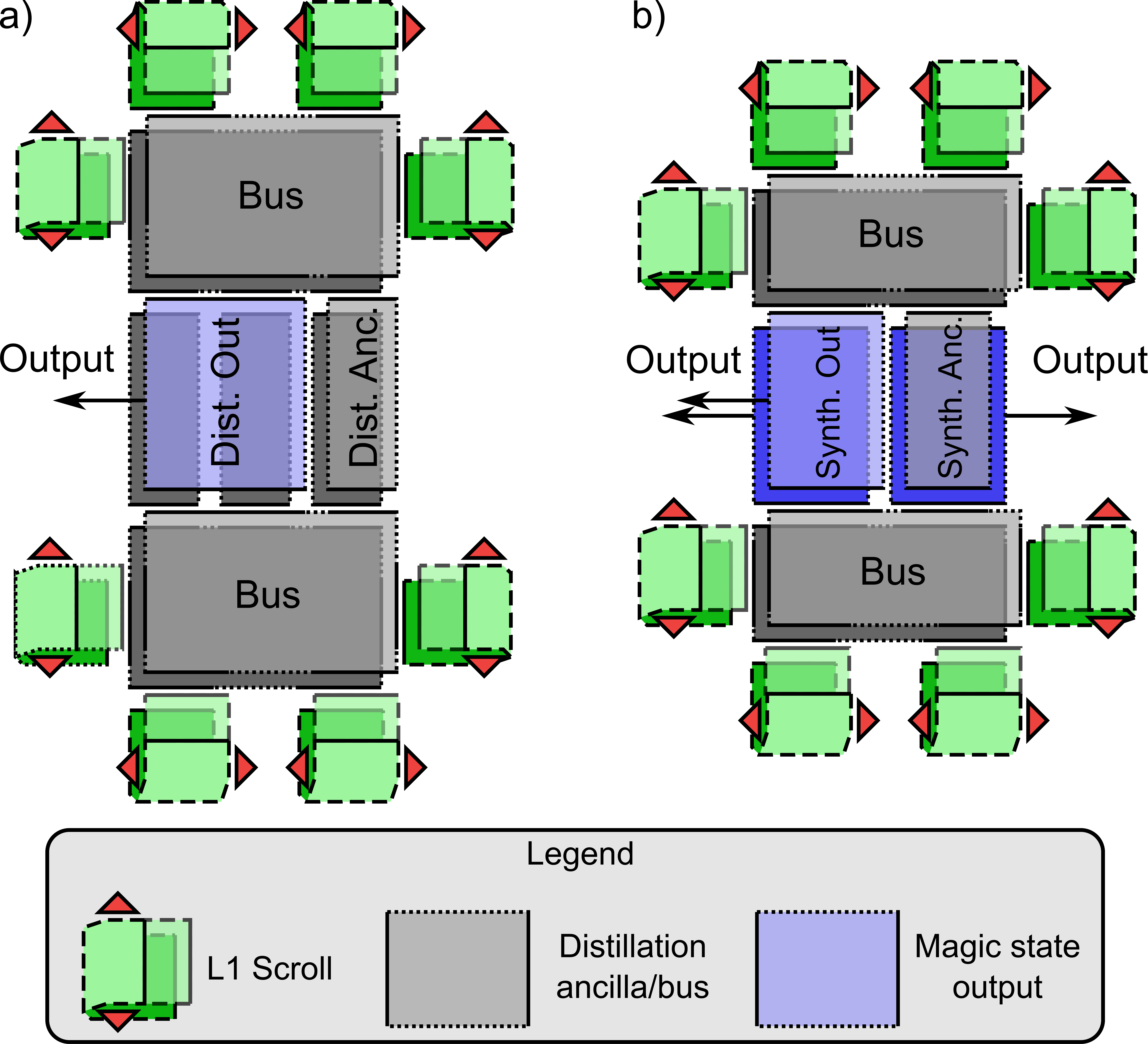}
\centering
\caption{\textbf{Layout of logical qubits when performing 15-to-1 distillation and 8-to-CCZ synthillation protocols.} a) Layout of Magic Scrolls (red representing colour code and green representing the folded surface the Scroll escapes to), output surface code patch (blue), and ancillary surface code patches/busses used in the 15-to-1 distillation protocol considered here. Note that the Magic Scrolls are so compact that they can be directly laid out around the surgery busses, without needing separate regions for level 1 distillation as per~\cite{Litinski_distillation}. When surgery is performed, the relevant bus bilayer is folded on either the left or right (opposite whichever Scroll is being merged), allowing surgery with all patches in the central block. After this surgery is complete, the surface code section of the consumed Scroll is transformed to a folded code by expanding into the transparent green sections, allowing for a transversal correction S gate to be applied if needed (see Fig.~\ref{fig:bilayer_code}d and Appendix~\ref{sec:supp_bilayer_operations}), which replaces the auto-T rotations of~\cite{Litinski_distillation}. b) Layout of the 8-to-CCZ synthillation protocol, depicted similar to (a), noting that there are now three synthillation output patches because of the three qubits involved in the CCZ state.~\label{fig:magic_scroll_distillation}}
\end{figure}

In addition to using cultivated magic states as input to distillation, we can further cheapen distillation when using the registers and biased noise models considered here. Registers allow us to use bilayer surface codes to perform the distillation operations, which slightly saves on qubits due to the reuse of ancillas (see Appendix~\ref{sec:supp_bilayer_performance}). In addition, the transversal single-qubit Clifford gates on folded bilayer codes (illustrated in Fig.~\ref{fig:bilayer_code}d and detailed in Appendix~\ref{sec:supp_bilayer_operations}, as well as by Moussa~\cite{moussa_bilayer}) allow for Clifford corrections to be applied transversely, rather than using auto-T rotations as per Litinski~\cite{Litinski_distillation}. It is important to note that these savings come at the cost of doubling the number of cycles (due to bilayer needing two (single-)cycles to measure all stabilisers), which is factored into our calculations. Overall we find the savings from using bilayer codes is worth this overhead and provides a net $3\times$ reduction in distillation volumes. Furthermore, the use of biased noise models allows for asymmetric surface code patches to be defined which equally protect X and Z errors, which can be used to further shorten bus lengths beyond those presented by Litinski~\cite{Litinski_distillation}.\\

Finally, as commented by Gidney \textit{et. al.}~\cite{gidney_cultivation}, distillation techniques often suffer in terms of volume because of their overall lack of postselection, particularly early in the distillation sequence. As such, when performing distillation, we implement complementary gapping (see~\cite{gidney2023yokedsurfacecodes}) of the surface code patches involved in the distillation circuit. To model the performance of this gapping, we modify the surface code patch error model used by Litinski~\cite{Litinski_distillation} with a single parameter, the gapping factor $f$. Specifically, if a certain surface code patch nominally performs with $p$ chance of a logical X(Z) error each cycle, then we approximate the gapped performance as

\begin{itemize}
    \item $f \cdot p$ chance to retry each cycle;
    \item $p/f$ chance of a logical X(Z) error each cycle.
\end{itemize}

\noindent which is a simplified model based on simulation results given in Appendix~\ref{sec:supp_bilayer_gapping}. This model is valid for code distances $\gtrsim 7$, and empirically provides a pessimistic approximation (ie. the actual code performs better). As this is only an approximate fit from which we extrapolate, the exact performance will depend on details of the actual implementation. Moreover, our results are already approximate since we are using the approximate noise model used by Litinski for surface code operations~\cite{Litinski_distillation}.\\

Example 15-to-1 distillation and 8-to-CCZ synthillation arrangements are shown in Fig.~\ref{fig:magic_scroll_distillation}a and b respectively. Specifically, 15-to-1 distillation produces $\ket{T}$ magic states, whereas 8-to-CCZ synthillation converts input magic $\ket{T}$ states into $\ket{CCZ}$ magic states. Both protocols are performed similarly to the level-2 protocols presented by Litinski~\cite{Litinski_distillation}, but with a few differences tailored to biased-noise registers. Firstly, as already mentioned, even the output surface code patches (purple) are asymmetric, which is possible thanks to the biased noise model. In addition, instead of level-1 distillation, we use cultivated magic states as inputs to the level-2 distillation/synthillation. These are obtained through Magic Scrolls, which are illustrated in red (colour code sections) and green (folded surface code sections). The Magic Scrolls are so compact that they can simply be placed around the lattice surgery bus sections without needing any additional routing.\\

When Magic Scrolls successfully cultivate a magic state, and the distillation procedure is ready for it, the relevant Magic Scroll is merged with one layer of the bus, with the bus then connecting its two layers on the opposite side to the Magic Scroll, forming one continuous bus which can merge with any of the central patches. Simultaneously, the folded-over section of the Magic Scroll is measured out, and grown shown as semi-transparent green) to form a folded surface code. Once the surgery is complete, the Magic Scroll section is now ready to perform a transversal S gate conditional on the outcome of the merge (which removes the need for auto-correcting ancilla patches), after which it can be immediately measured out in the X-basis while maintaining code distance, as discussed in Appendix~\ref{sec:supp_bilayer_operations}. We assume the outcome of the merge can be decoded in 50 $\mu$s, greater than the 10 $\mu$s assumed for individual Magic Scrolls due to the larger code distances used in distillation. Once this transversal S gate is complete and the Magic Scroll has been measured out, the Magic Scroll can begin cultivating again. Finally, for 15-to-1 we assume that there is sufficient time to transport the output qubits out of the factory during the first ancilla-only merge of the next distillation, and for 8-to-CCZ we similarly assume that transport-out of output qubits is negligible, using similar staggered-transport tactics to Litinski~\cite{Litinski_distillation}. The final transversal S gates are also assumed to be parallelisable, and are not factored into the expected time taken for a given distillation, but are however excluded from the available Magic Scrolls for cultivation in the following distillation until the S gates are complete.\\

{\renewcommand{\arraystretch}{1.2} 
\begin{table*}[!ht]
    \centering
    \resizebox{1\textwidth}{!}{
    \begin{tabular}{|c|c|c|c|c|c|c||c|c|c|c|}\hline
        $\mathbf{w_o}$ & $\mathbf{w_a}$ & $\mathbf{h_o}$ & $\mathbf{h_b}$ & $\mathbf{m}$ & $\mathbf{f_o}$; $\mathbf{f_a}$ & \textbf{Level 1 settings} & \textbf{Qubits} & \textbf{Cycles} & \textbf{Volume} & \textbf{Error}\\\hline
        \multicolumn{11}{|c|}{\textbf{20-to-4 (This work, $\mathbf{p=0.1\%}$)}}\\\hline
        13 & 7 & 21 & 13 & 9 & 500; 50 & d=5: g-s-dc-g-s-dc $\to$ 13$\times$7 r=3 t=12.0 & 7,860 & 57 & 450,000 & $2.5 \times 10^{-10}$\\\hline
        15 & 9 & 25 & 15 & 9 & 1,000; 100 & d=5: g-s-dc-g-s-dc $\to$ 15$\times$9 r=3 t=12.0 & 10,980 & 55 & 600,000 & $1.5 \times 10^{-11}$\\\hline
        \multicolumn{11}{|c|}{\textbf{15-to-1 (This work, $\mathbf{p=0.1\%}$)}}\\\hline
        11 & 7 & 21 & 11 & 5 & 20,000; 3 & d=3: g-s-dc $\to$ 11$\times$7 r=2 t=7.0 & 4,590 & 101 & 460,000 & $2.6 \times 10^{-10}$\\\hline
        11 & 7 & 23 & 11 & 7 & 50,000; 20 & d=3: g-s-dc $\to$ 11$\times$7 r=2 t=7.0 & 4,710 & 124 & 590,000 & $4.5 \times 10^{-11}$\\\hline
        13 & 7 & 27 & 13 & 7 & 100,000; 20 & d=3: g-s-s-dc $\to$ 11$\times$7 r=3 t=7.0 & 5,320 & 122 & 650,000 & $1.4 \times 10^{-12}$\\\hline
        13 & 7 & 29 & 13 & 7 & 1,000,000; 20 & d=3: g-s-s-dc $\to$ 11$\times$7 r=3 t=13.0 & 5,450 & 128 & 700,000 & $2.5 \times 10^{-13}$\\\hline
        15 & 7 & 31 & 15 & 7 & 5,000,000; 35 & d=3: g-s-s-dc $\to$ 13$\times$9 r=3 t=13.0 & 6,810 & 130 & 880,000 & $4.0 \times 10^{-14}$\\\hline
        15 & 9 & 33 & 15 & 9 & 10,000,000; 50 & d=3: g-s-s-dc $\to$ 13$\times$9 r=3 t=16.0 & 7,910 & 150 & 1,190,000 & $4.5 \times 10^{-15}$\\\hline
        19 & 11 & 47 & 19 & 9 & $10^9$; 100 & d=5: g-s-s-dc-g-s-s-dc $\to$ 17$\times$11 r=4 t=16.0 & 12,580 & 237 & 2,980,000 & $5.4 \times 10^{-20}$\\\hline
        \multicolumn{11}{|c|}{\textbf{8-to-CCZ (This work, $\mathbf{p=0.1\%}$)}}\\\hline
        13 & 9 & 21 & 13 & 7 & 2,000; 50 & d=5: g-s-dc-g-s-dc $\to$ 15$\times$9 r=3 t=12.0 & 6,580 & 73 & 480,000 & $2.3 \times 10^{-10}$\\\hline
        13 & 9 & 25 & 13 & 9 & 25,000; 500 & d=5: g-s-dc-g-s-dc $\to$ 15$\times$9 r=3 t=16.0 & 6,880 & 87 & 600,000 & $1.9 \times 10^{-11}$\\\hline
        15 & 9 & 29 & 15 & 11 & 50,000; 1,000 & d=5: g-s-s-dc-g-s-s-dc $\to$ 15$\times$9 r=4 t=16.0 & 8,140 & 127 & 1,030,000 & $3.1 \times 10^{-13}$\\\hline
        15 & 11 & 31 & 15 & 11 & 200,000; 4,000 & d=5: g-s-s-dc-g-s-s-dc $\to$ 17$\times$11 r=4 t=17.0 & 9,460 & 135 & 1,280,000 & $5.8 \times 10^{-14}$\\\hline
        17 & 11 & 33 & 17 & 11 & 200,000; 4,000 & d=5: g-s-s-dc-g-s-s-dc $\to$ 17$\times$11 r=4 t=20.0 & 10,730 & 139 & 1,490,000 & $6.7 \times 10^{-15}$\\\hhline{|=:=:=:=:=:=:==:=:=:=|}
        \multicolumn{11}{|c|}{\textbf{20-to-4 (Litinski, $\mathbf{p=0.1\%}$)}}\\\hline
        23 & 11 & 23 & 23 & 13 & 0; 0 & (15-to-1)$^6_{w_o=13,w_a=5,m=5}$ & 43,300 & 130 & 1,410,000 & $1.4 \times 10^{-10}$\\\hline
        27 & 13 & 27 & 27 & 15 & 0; 0 & (15-to-1)$^4_{w_o=13,w_a=5,m=5}$ & 46,800 & 157 & 1,840,000 & $2.6 \times 10^{-11}$\\\hline
        \multicolumn{11}{|c|}{\textbf{15-to-1 (Litinski, $\mathbf{p=0.1\%}$)}}\\\hline
        17 & 7 & 17 & 17 & 7 & 0; 0 & None (single level) & 4,620 & 42.6 & 197,000 & $4.5 \times 10^{-8}$\\\hline
        25 & 11 & 25 & 25 & 11 & 0; 0 & (15-to-1)$^6_{w_o=11,w_a=5,m=5}$ & 30,700 & 82.5 & 2,540,000 & $2.7 \times 10^{-12}$\\\hline
        29 & 11 & 29 & 29 & 13 & 0; 0 & (15-to-1)$^6_{w_o=13,w_a=5,m=5}$ & 39,100 & 97.5 & 3,810,000 & $3.3 \times 10^{-14}$\\\hline
        41 & 17 & 41 & 41 & 17 & 0; 0 & (15-to-1)$^6_{w_o=17,w_a=7,m=7}$ & 73,400 & 128 & 9,370,000 & $4.5 \times 10^{-20}$\\\hline
        \multicolumn{11}{|c|}{\textbf{8-to-CCZ (Litinski, $\mathbf{p=0.1\%}$)}}\\\hline
        25 & 15 & 25 & 25 & 15 & 0; 0 & (15-to-1)$^6_{w_o=13,w_a=7,m=7}$ & 47,000 & 60.0 & 2,820,000 & $5.2 \times 10^{-11}$\\\hline
    \end{tabular}
    }
    \caption{\textbf{Distillation results using the Magic Scroll, a biased-noise register model, and complementary gapping.} All settings given here are defined in the main text, noting that 0 values for complementary gap factors indicate no gapping is applied. Magic Scroll settings also indicate the threshold weight $t$ used for the Magic Scroll complementary gap, and in all cases 8 Magic Scrolls are used (arranged per Fig.~\ref{fig:magic_scroll_distillation}). Output qubit heights ($h_o$) are chosen such that an ungapped square bilayer code of height $h_o$ will not exceed the output magic state error after $h_o$ (double-)cycles, so that the magic state can be transported out and consumed properly. ``Qubits'' indicates the total number of qubits (rounded to the nearest 10) in the distillation procedure, including single rows of qubits used for lattice surgery and all qubits involved in all Magic Scrolls, assuming they are reserved for distillation even when not in use. For 20-to-4 protocols, this is still the total number of qubits in the distillery (ie. not divided by 4, which would be the number of qubits per number of magic states output). ``Cycles'' is the expected number of (single-)cycles used per distilled magic state produced, that is the number of measurements performed (ie. double the number of bilayer cycles), including retries due to complementary gapping and postselection at the end of distillation, rounded to the nearest cycle. ``Volume'' is in units of qubit-(single-)cycles per magic state produced, and is simply the product of the prior two columns (rounded to the nearest 10,000). Error indicates the average error of each magic state produced, which follow the error model given in Table~\ref{tab:error_model} using $p=0.1\%$. In addition, results are included from Litinski~\cite{Litinski_distillation} for comparison, which use distillation only, assume unbiased noise, and operate on standard square-lattice connectivity. All results from Litinski use 0.1\% physical error.}
    ~\label{tab:distillation_results}
\end{table*}
}

To characterise distillation settings, we use the following parameters:

\begin{enumerate}
    \item The width (distance) of the output logical qubits $w_o$,
    \item The width (distance) of the ancilla logical qubits in the central region $w_a$,
    \item The height of the central region $h_o$,
    \item The height of the bus $h_b$,
    \item The number of (double-)cycles $m$ used to perform lattice surgery and to stabilise the folded bilayer prior to transversal S,
    \item The degree of complementary gapping $f_o$ for gapping of the output region, as well as the bus X-type errors,
    \item The degree of complementary gapping $f_a$ for gapping of the central ancilla, Magic Scroll/folded bilayer regions, and bus measurement-type errors, and
    \item The cultivation procedure used for the level 1 Magic Scrolls (notated as per Fig.~\ref{fig:gidney_style_figure}).
\end{enumerate}

All non-cultivation operations are assumed to be performed without time domain walls in the bilayer code, except when stabilising the folded surface code prior to performing the transversal S gate. In all cases studied here, there are exactly four Magic Scrolls per bus (as per Fig.~\ref{fig:magic_scroll_distillation}), which fit on the bus without additional qubits - we found that additional Magic Scrolls did not significantly improve performance in our simulations, even when additional Magic Scrolls could fit without overhead. All simulations are performed at a physical error rate of 0.1\%, and the height of the central region $h_o$ is kept at a large enough distance so that the extrapolated error per $h_o$ (double-)cycles does not significantly exceed the distilled magic state error (allowing the state to be consumed without complementary gapping outside the region used for distillation). Results from these simulations are shown in Table~\ref{tab:distillation_results}, and the 15-to-1 distillation results are also included in Fig.~\ref{fig:gidney_style_figure}.\\

From these results, we can see that distillation performance can be significantly improved from the results presented by Litinski~\cite{Litinski_distillation}, specifically improving by a factor of $\sim 3\times$ in terms of quantum volume for the same magic state error rate. Of course, a large component of this improvement is attributed to biased noise and register connectivity, which allow for smaller overall surface code area and fewer qubits needed for a given surface code area respectively. However, the other factor which significantly improves the distillation performance is the use of complementary gapping, which improves the logical error rate of a given patch by up to $5\times10^6$ for the most extreme example in Table~\ref{tab:distillation_results}, while not significantly increasing the expected quantum volume of distillation. Specifically, we found that aiming for an overall retry chance of $\sim 10-20\%$ for a given distillation attempt yielded a good tradeoff between error reduction and expected volume increase. Complementary gapping methods like these are not specific to biased noise or register connectivity, so we expect they can also be applied to architectures implementing single-layer surface codes. Overall, these results show that biased-noise register-based architectures are not only capable of directly cultivating magic $\ket{T}$ states with errors of $\gtrsim 10^{-10}$ via the Magic Scroll procedure, but are also capable of reaching errors as low as $10^{-15}$ by using Magic Scrolls as input to distillation. These procedures can be performed with volumes reduced by up to a factor of $3-4$ compared to results using square grid connectivity and unbiased noise.

\section{Conclusion}

High qubit connectivity combined with biased noise present avenues for improving the performance of quantum error correcting codes. Here, we have shown that qubit connectivity and biased noise models physically motivated by the 14|15 platform of multi-nuclear spin registers in silicon can be leveraged to improve various aspects of fault-tolerant quantum computation. Firstly, we have shown how a square lattice of 2-qubit registers can efficiently implement QEC codes, creating 6.6.6 and 4.8.8 colour codes, as well as bilayer surface codes, with remarkably high crossing thresholds of $\sim 0.55 \%$ for the colour codes and $\sim 0.7\%$ for the bilayer code. We combined the extended operations available to both colour and bilayer codes to improve the effectiveness of magic state production by using the Magic Scroll procedure, improving expected cultivation volumes by a factor of up to $3\times$, and escaping the magic $\ket{T}$ state to a folded but otherwise standard surface code. Finally, we show that it is possible to improve the expected volume of distillation protocols by a factor of $3\times$ by reducing qubit counts using a bilayer code, using the Magic Scroll to generate input $\ket{T}$ states, and improving logical qubit performance by performing postselected complementary gapping. Overall, these results show the significant performance improvements that can be gained when leveraging both noise bias and high qubit connectivity, further reducing the requirements needed to execute useful fault-tolerant quantum algorithms.

\section*{Author Contributions}

I. D. T., S. K. G. and C. D. H. conceived of the project. I. D. T. performed the majority of simulations and analysis, with assistance from J. R. C. for colour code simulations. J. M. assisted with theoretical justification of the inter-register biased noise model. The manuscript was written by I. D. T. with input from all authors. S. K. G., C. D. H. and M. Y. S. supervised the project. No AI was used in producing this work.\\

\label{references}
\bibliographystyle{unsrtnat}
\bibliography{references}

\clearpage

\newpage
\onecolumngrid

\appendix

\onecolumngrid

\renewcommand{\thefigure}{S\arabic{figure}}%
\renewcommand{\thetable}{S\Roman{table}}%
\setcounter{figure}{0}%
\setcounter{table}{0}

\section{Magic Scroll Extrapolation Plots~\label{sec:supp_magic_scroll_extrapolation}}

\begin{figure*}[!ht]
\includegraphics[width=0.75\textwidth]{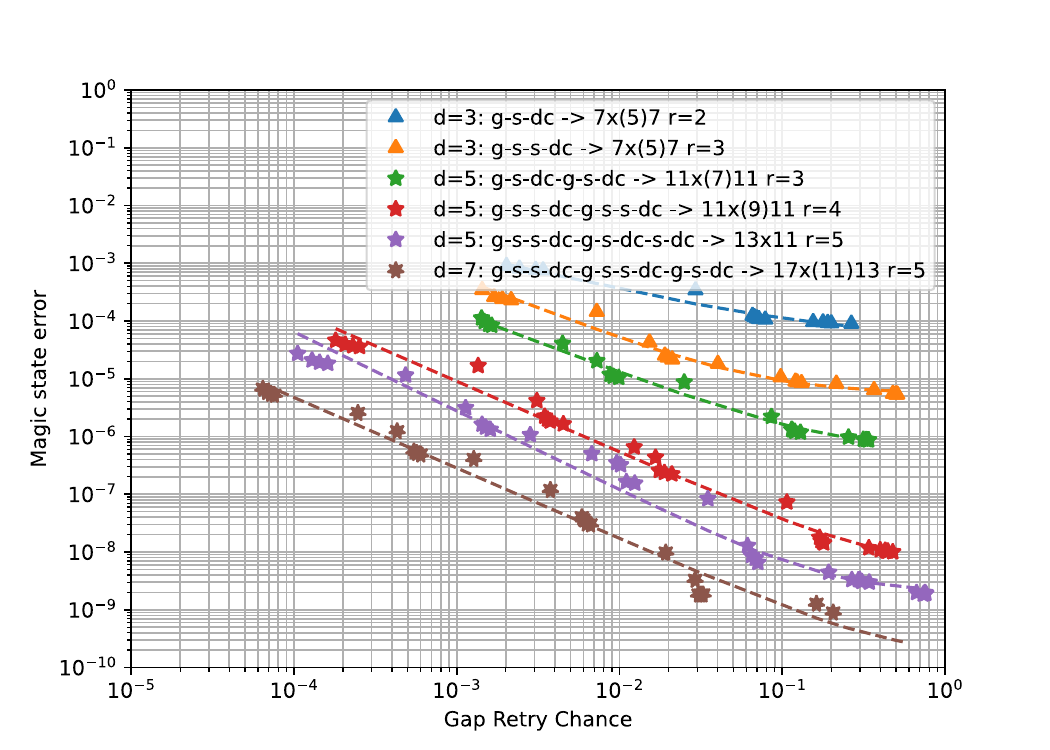}
\centering
\caption{\textbf{Extrapolation of the Magic Scroll performance, used to approximately fit/predict performance in Fig. \ref{fig:gidney_style_figure}} Instead of plotting the magic state error against the expected volume as per main text Fig. \ref{fig:gidney_style_figure}, here we plot it against the probability of the Magic Scroll procedure to retry due to failure of the complementary gap. Because the performance appears as a power law as shown here, we can use a power law fit to extrapolate the performance, which is what we do in Fig. \ref{fig:gidney_style_figure} of the main text. This allows the performance of cultivation procedures to be determined, and allows estimation of magic state errors below that which can be feasibly simulated.~\label{fig:supp_gidney_style_figure_extrapolation}}
\end{figure*}

The extrapolation method used for the Magic Scroll procedure is based on the trend seen for complementary gapped bilayer surface codes in Appendix~\ref{sec:supp_bilayer_gapping}. Specifically, we use the fact that gapped performance appears linear as a function of $\log(p)$ and $\log(P(r))$, where $p$ is the error rate and $P(r)$ is the probability of retrying due to failing the complementary gap postselection. This linear trend is however limited below by $2\epsilon_0$, where $\epsilon_0$ is the error rate of cultivating a $\ket{S}$ state without escaping - this lower limit is included when fitting the curves by first subtracting $2\epsilon_0$ from the sampled data, fitting normally, and then adding $2 \epsilon_0$ back when plotting both the sampled data and the fit. By plotting the portion of $d=7$ cultivation which is tractable to simulate in this form (with $2 \epsilon_0$ subtracted), we can extrapolate the line of best fit to predict the performance for low $\ket{T}$ state errors that would otherwise be infeasible to simulate. Note that the retry chance here is just for retries due to the complementary gap, and do not include retries due to postselected colour code stabilisers/flags. This extrapolation, with predicted no-escape errors (twice the no-escape $\ket{Y}$ state error) added, is used in main text Fig. \ref{fig:gidney_style_figure}. The performance of the Magic Scroll, plotted as a function of $\log(p)$ and $\log(P(r))$, is shown in Fig. \ref{fig:supp_gidney_style_figure_extrapolation} for reference.

\clearpage

\section{Noise Model in Multi-Nuclear Spin Registers~\label{sec:supp_pauli_twirling}}

As previously modelled \cite{Thorvaldson2025}, register-local CZ gates performed in multi-nuclear spin registers in silicon fit well to a fully dephasing-biased error channel. Specifically, the error channel for register-local CZ gates can be modelled as the following:

\begin{align}
    \begin{bmatrix}
        1 & 0 & 0 & 0\\
        0 & 1 & 0 & 0\\
        0 & 0 & 1 & 0\\
        0 & 0 & 0 & e^{i \delta}\\
    \end{bmatrix}
\end{align}
for a small angle $\delta$, sampled from a normal distribution (caused by dephasing of the mediating electron). Using a similar error model, the error channel for a register-local Parity-Z gate (using multi-toned control of the mediating electron) can be modelled as the following:

\begin{align}
    \begin{bmatrix}
        1 & 0 & 0 & 0 & 0 & 0 & 0 & 0\\
        0 & 1 & 0 & 0 & 0 & 0 & 0 & 0\\
        0 & 0 & 1 & 0 & 0 & 0 & 0 & 0\\
        0 & 0 & 0 & 1 & 0 & 0 & 0 & 0\\
        0 & 0 & 0 & 0 & 1 & 0 & 0 & 0\\
        0 & 0 & 0 & 0 & 0 & e^{i \delta} & 0 & 0\\
        0 & 0 & 0 & 0 & 0 & 0 & e^{i \delta} & 0\\
        0 & 0 & 0 & 0 & 0 & 0 & 0 & 1\\
    \end{bmatrix}
\end{align}
where, notably, only one parameter $\delta$ is used (due to the quasistatic nature of electron dephasing).\\

We convert these errors to approximate Pauli error channels via the Pauli twirling approximation (PTA) \cite{Emerson_2007, katabarwa2017dynamicalinterpretationpaulitwirling}. First we can represent the above two and three qubit unitary errors entirely in the Pauli Z basis, e.g.
\begin{align}
        \begin{bmatrix}
        1 & 0 & 0 & 0\\
        0 & 1 & 0 & 0\\
        0 & 0 & 1 & 0\\
        0 & 0 & 0 & e^{i \delta}\\
    \end{bmatrix} = \frac{1}{4}[(3 + e^{i\delta})II + (1-e^{i\delta})(ZI + IZ - ZZ)],
\end{align}
where $II = I\otimes I$ etc. The PTA converts this into a Pauli Z channel with probabilities
\begin{equation}
    \tilde{P}_{ZI} = \tilde{P}_{IZ} = \tilde{P}_{ZZ} = \frac{1}{16}|1 - e^{i\delta}|^2 = \frac{1}{8}(1 - \cos \delta), \,\, \tilde{P}_{II} = 1 - 3\tilde{P}_{ZI} = \frac{1}{8}(5 + 3 \cos \delta),
\end{equation}
where $\tilde{P}_{ZI}$ is the probability of error $Z\otimes I$, etc. To turn this into a static channel, we integrate over $\delta \sim \mathcal{N}(\sigma)$, where $\sigma$ is the standard deviation. Note that
\begin{equation}
    \frac{1}{\sqrt{2\pi \sigma^2}} \int e^{-\delta^2 / 2\sigma^2}\cos \delta d\delta = e^{-\sigma^2/2}~\label{eq:supp_gaussian_integral}
\end{equation}
We can use Eq.~\ref{eq:supp_gaussian_integral} to get final probabilities for the noise channel:
\begin{equation}
    P_{ZI} = P_{IZ} = P_{ZZ} = \frac{1}{8}(1 - e^{-\sigma^2/2}),\,\, P_{II} = \frac{1}{8}(5 + 3e^{-\sigma^2/2}).
\end{equation}

For the case of the Parity-Z gate, the result is similar:
\begin{equation}
    P_{ZZZ} = P_{ZII} = P_{IZZ} = \frac{1}{8}(1 - e^{-\sigma^2/2}),\,\,P_{III} = \frac{1}{8}(5 + 3e^{-\sigma^2/2}).
\end{equation}

When extending this error model to two-qubit gates operating between two registers, we assume that relevant information is first encoded in the electron spins on each interacting register, and then electron-electron CZ gates are performed (for example, by turning on exchange for a brief time, using the same coupling mechanism as \cite{Edlbauer2025}), after which the electrons are uncomputed. This results in a error mechanism similar to a register-local CZ on each register, but just affecting the state(s) marked, and an assumed equally-weighted dephasing-only Pauli error channel for the electron-electron CZ gate. To perform a remote CZ gate for example, one would excite the electron of each register conditional on the involved nuclear spin being $\ket{1}$, and then perform the electron-electron CZ gate, which after uncomputation leaves just a nuclear-nuclear CZ gate.\\

Presuming that the electron-electron CZ gate is the dominating source of error, the error channel remains the same as for the local CZ gate: $ZI(p/3)$, $IZ(p/3)$ and $ZZ(p/3)$. Similarly, the remote parity-Z gate remains the same as a local parity-Z gate acting on a register with $\ge 3$ qubits. This assumption, not founded on any particular physical reasoning, assumes the electron-electron CZ is the dominant source of error as it yields equally-weighted error channels, which is likely the worst case. Indeed, if we instead assume the electron rotation contributes the same error as the electron-electron CZ, we get remote CZ error channels $ZI(4p/9)$, $IZ(4p/9)$ and $ZZ(p/9)$, which significantly reduces the relative chance of correlated $ZZ$ errors and therefore is likely to improve performance. The exact error channel that should be used in future work should be informed by system performance once remote register CZ gates are experimentally characterised. For this work, we choose to use the worst realistic case.

\clearpage

\section{Colour Code Performance and Variants~\label{sec:supp_col_performance}}

\begin{figure*}[!ht]
\includegraphics[width=0.95\textwidth]{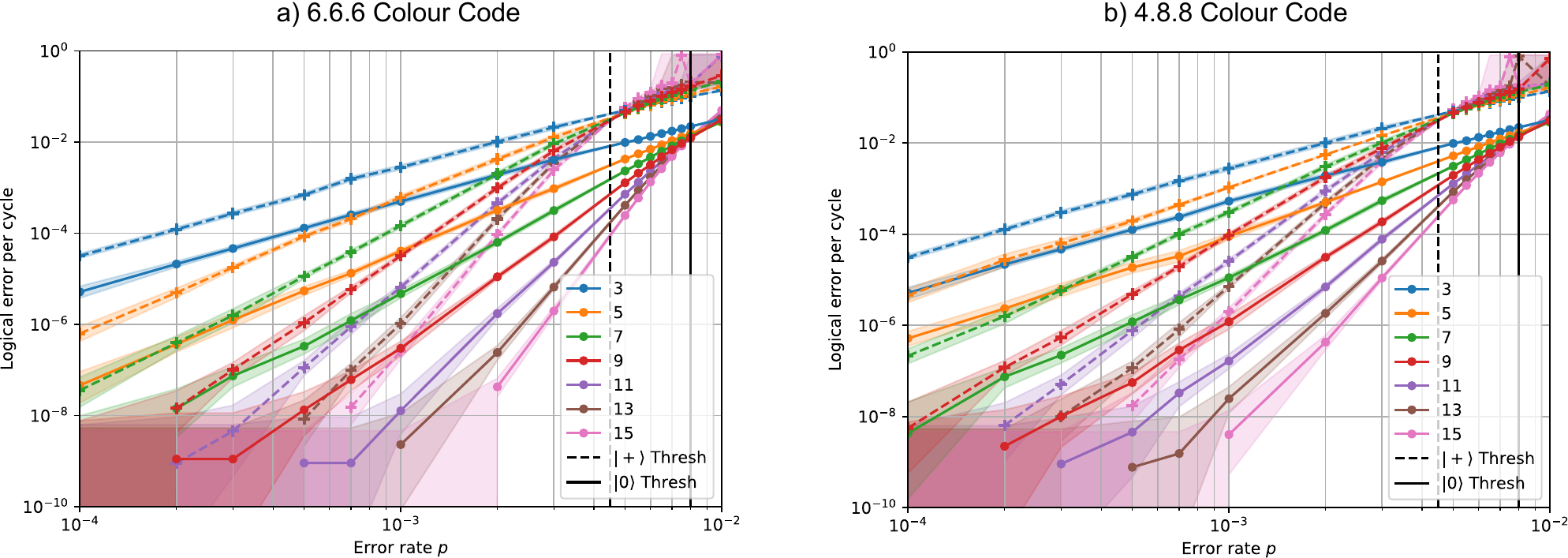}
\centering
\caption{\textbf{Performance of colour code qubits without added domain walls.} a) Performance of the 6.6.6 code, similar to Fig. \ref{fig:colour_code}c of the main text, but without the addition of time domain walls. The survival of a logical $\ket{+}$ state is shown with dashed lines, and the survival of a logical $\ket{0}$ state is shown with solid lines. The approximate thresholds are also shown, which are at $\sim 0.45$ \% and $\sim 0.8$ \% respectively. b) Performance of the 4.8.8 code, similar to (a). The marked thresholds are at $\sim 0.45$ \% and $\sim 0.8$ \% for logical $\ket{+}$ and $\ket{0}$ respectively. Because domain walls have not been added, X and Z errors are not protected equally - a consequence of the noise bias used.~\label{fig:supp_vanilla_colour}}
\end{figure*}

In Fig. \ref{fig:supp_vanilla_colour}, we show the performance of both the 6.6.6 and 4.8.8 colour codes (as defined in Fig. \ref{fig:colour_code} of the main text), but without the use of time domain walls. Because of the lack of domain walls, logical X- and Z-operators are protected at different rates and with different thresholds. Both codes have a threshold of $\sim 0.45$ \% for $\ket{+}$ survival (which is affected by logical Z errors), and $\sim 0.8$ \% for $\ket{0}$ survival (affected by logical X errors).\\

\begin{figure*}[!ht]
\includegraphics[width=0.95\textwidth]{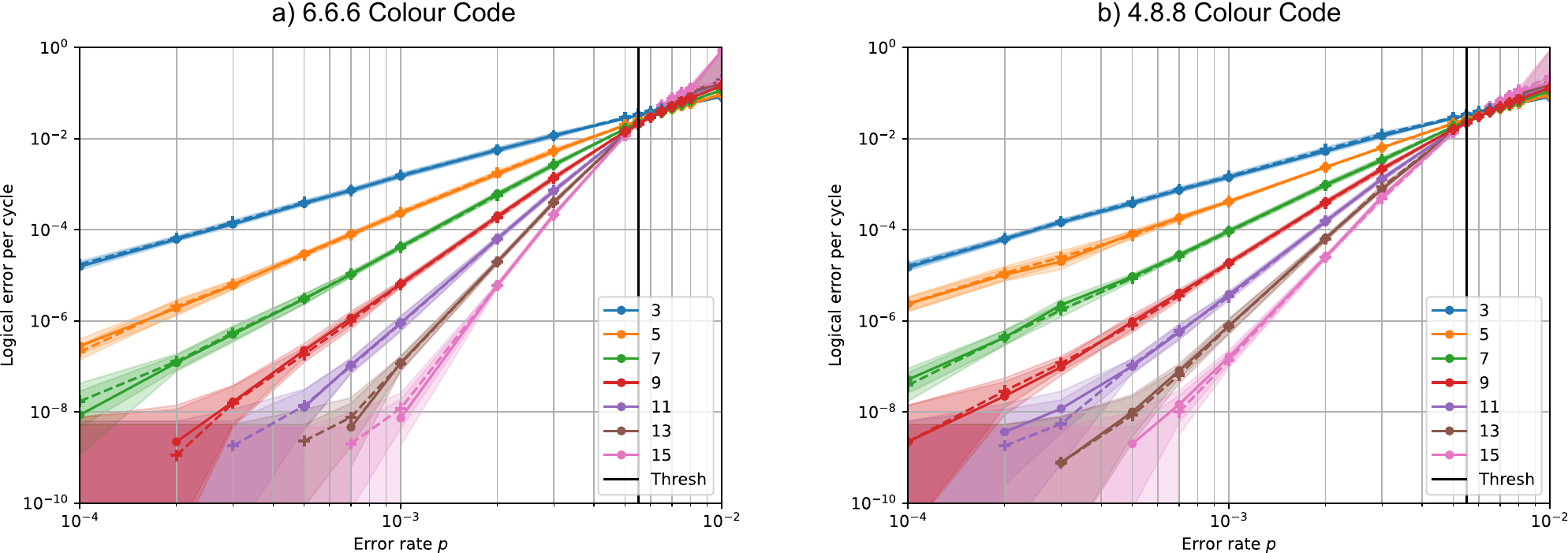}
\centering
\caption{\textbf{Performance of the time domain wall colour code, showing equal protection of logical X and Z states.} We show the performance for (a) the 6.6.6 code and (b) the 4.8.8 code, with labels otherwise as per Fig. \ref{fig:supp_vanilla_colour}. The thresholds marked are as per main text Fig. \ref{fig:colour_code}. Logical X and Z errors are protected equally, indicating that the symmetry provided by time domain walls equalises the error probability between the two logical operators.~\label{fig:supp_domainwall_colour_both}}
\end{figure*}

When time domain walls are added to the colour code, the logical $\ket{+}$ and $\ket{0}$ states are equally protected. We explicitly show this in Fig. \ref{fig:supp_domainwall_colour_both}.

\begin{figure*}[!ht]
\includegraphics[width=\textwidth]{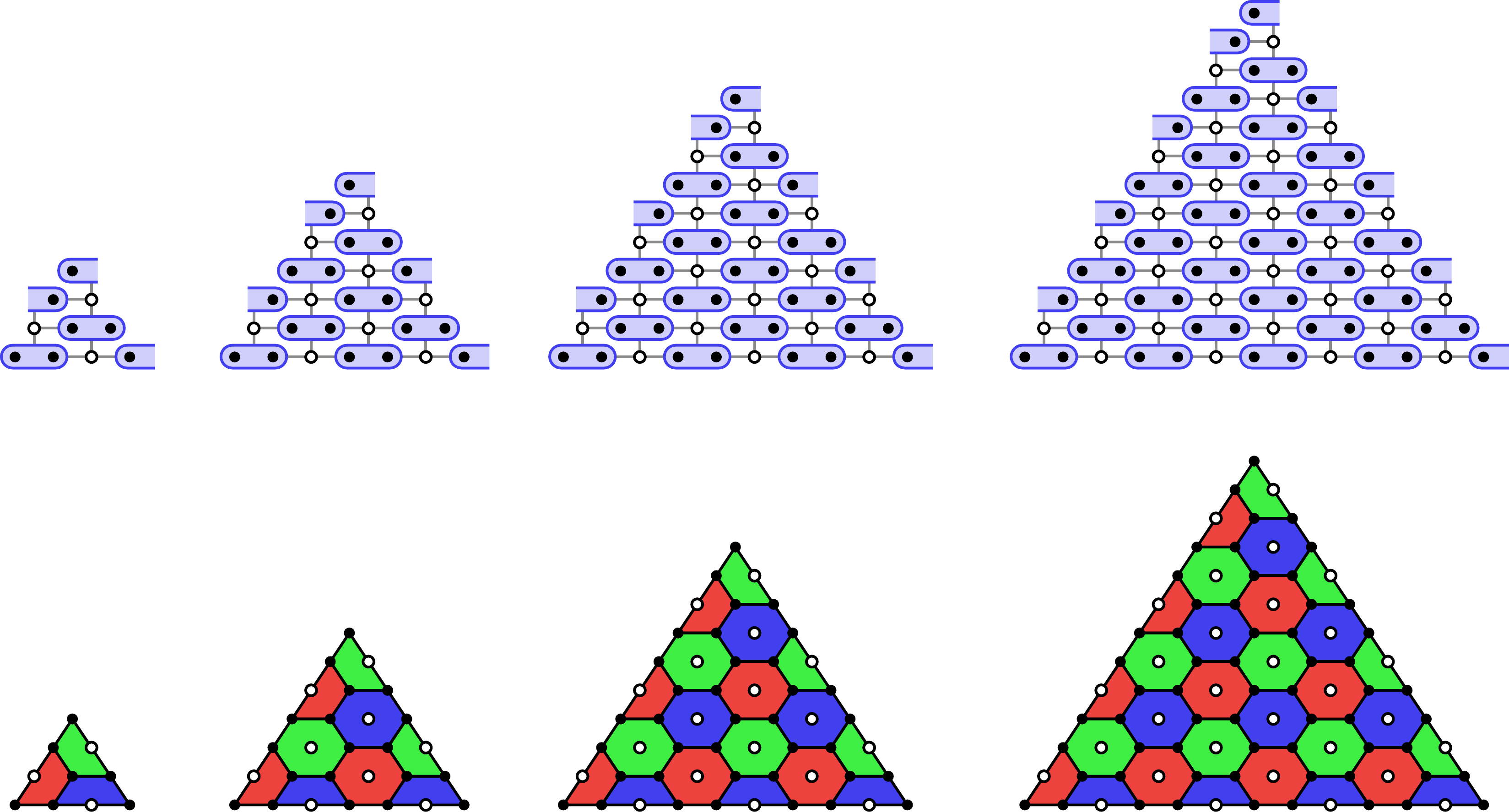}
\centering
\caption{\textbf{Mapping of various distances of 6.6.6 colour codes to registers.}~\label{fig:supp_colour_666_stab_scaling}}
\end{figure*}

\begin{figure*}[!ht]
\includegraphics[width=\textwidth]{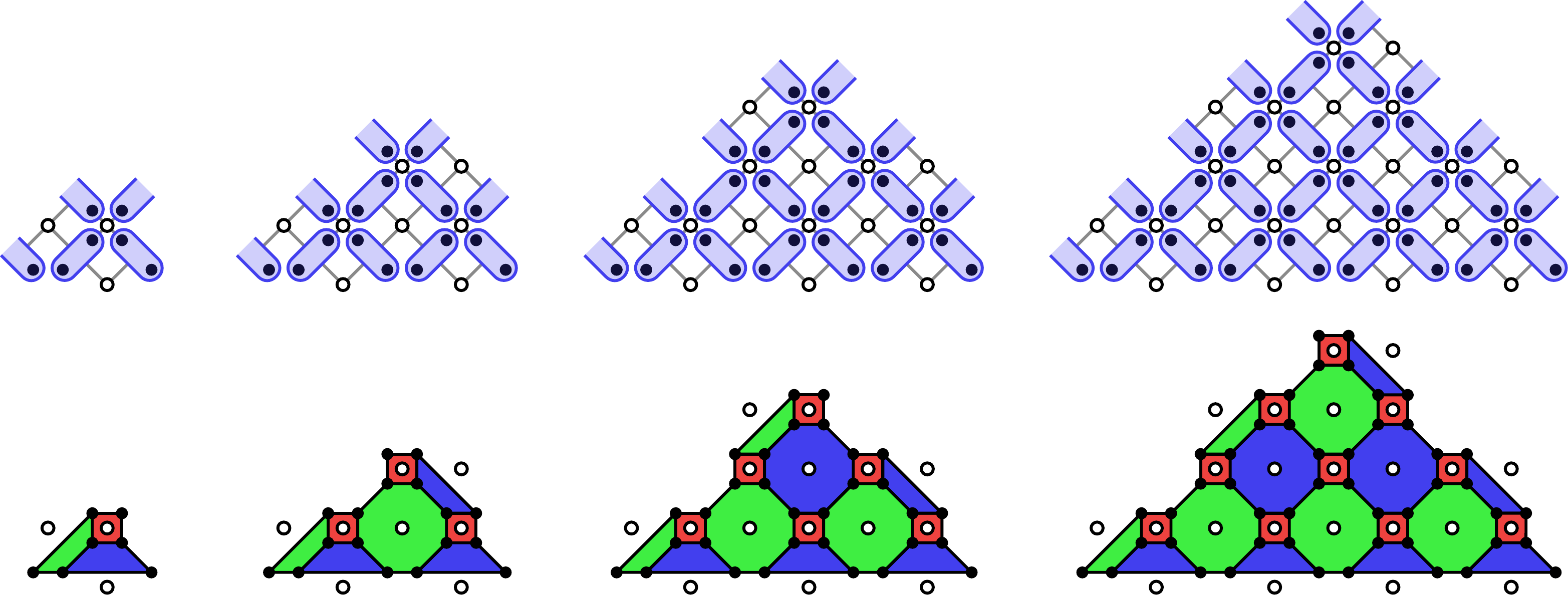}
\centering
\caption{\textbf{Mapping of various distances of 4.8.8 colour codes to registers.}~\label{fig:supp_colour_488_stab_scaling}}
\end{figure*}

To help illustrate how the colour codes map to a checkerboard-alternating grid of 1Q/2Q registers, we show the mapping of distances 3, 5, 7 and 9 to registers in Figs. \ref{fig:supp_colour_666_stab_scaling} and \ref{fig:supp_colour_488_stab_scaling} for the 6.6.6 and 4.8.8 colour codes respectively.

\clearpage

\section{Colour Code Stabilisation~\label{sec:supp_col_stabilisation}}

\begin{figure*}[!ht]
\includegraphics[width=0.9\textwidth]{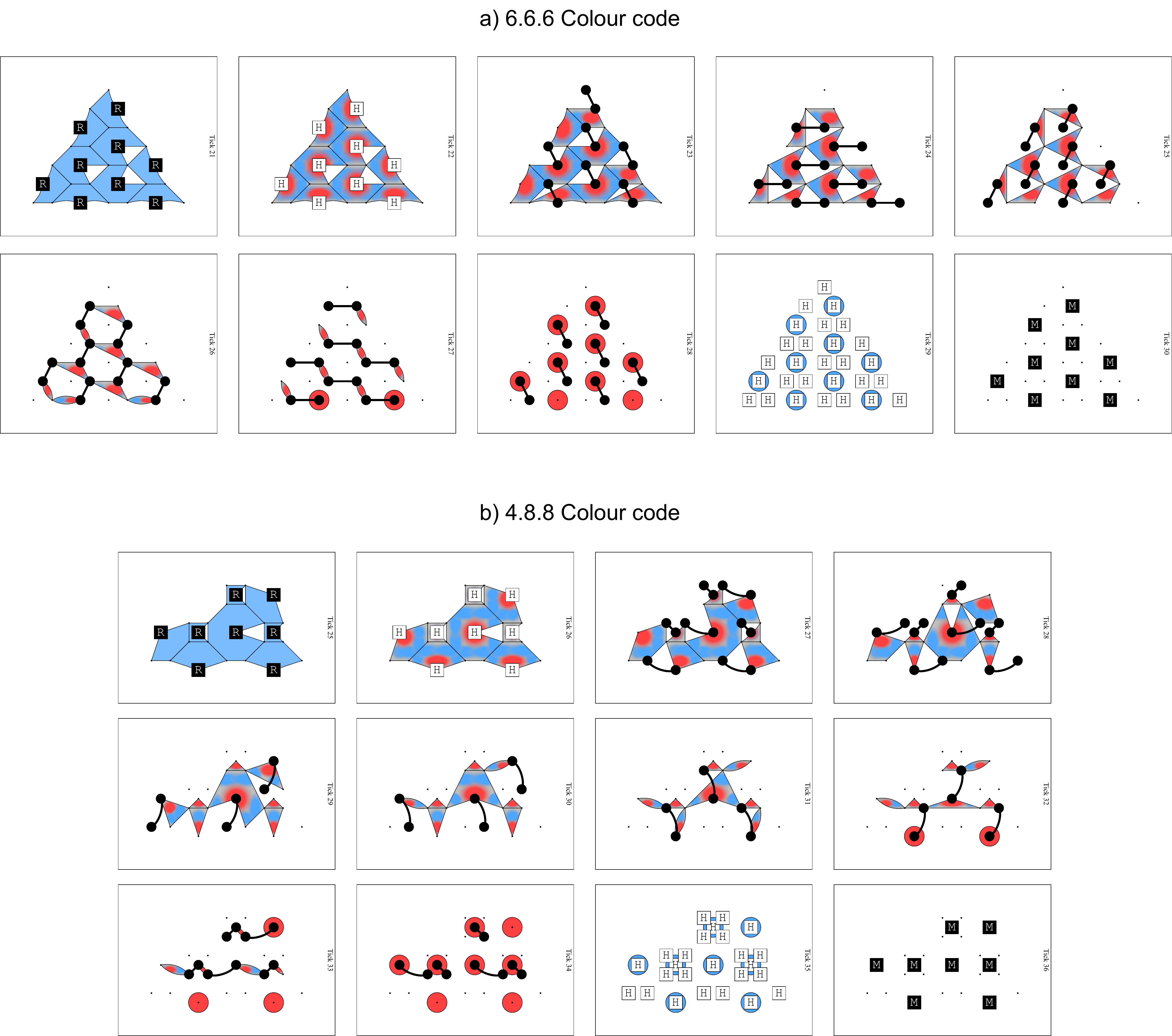}
\centering
\caption{\textbf{Measurement of a (single-)cycle of stabilisation for the 6.6.6 and 4.8.8 colour codes.} We present the single-cycle stabilisation (measuring just the Z stabilisers) for a) the 6.6.6 colour code and b) the 4.8.8 colour code, implemented with the native gates considered in this work. Note that instead of using native Parity-Z gates, we CZ gates, to avoid the correlated errors associated with Parity-Z gates. For the time domain wall code, Hadamard gates are applied to data qubits on the second last tick to swap X and Z stabilisers in preparation for measurement of X stabilisers. Here, R denotes a reset (initialise) operation, M denotes a measurement, and otherwise the figure uses standard quantum operation circuit symbols. In both (a) and (b), the first tick initialises the ancilla qubits, which are then placed into $\ket{+}$ states by the Hadamard gates in the following tick. The subsequent chain of CZ operations places the Z stabiliser value on the X projection of the ancilla qubits. In the second last tick, Hadamard gates are applied to the ancilla qubits to convert the X projection to a Z projection (preparing for measurement), and simultaneously Hadamard gates are performed on the data qubits to prepare them for X-stabiliser measurement in the next (single-)cycle. In the final tick, the ancillas are measured to determine the stabiliser values.~\label{fig:supp_colour_stabilisers}}
\end{figure*}

The exact circuits implementing (single-)cycles of stabilisation of the 6.6.6 and 4.8.8 colour codes with time-domain walls are shown in Fig. \ref{fig:supp_colour_stabilisers}. While these diagrams show measurement of the Z stabilisers, the Hadamard gates in the second last tick swap Z and X stabilisers, so that running this circuit again will yield the X stabilisers. Because these regular transversal Hadamard gates swap the logical operators, we refer to such a stabilisation sequence as having time-domain walls (in analogy to domain walls which spatially transforming qubits with Hadamard gates).

\clearpage

\section{Colour Code Scaling~\label{sec:supp_col_scaling}}

\begin{figure*}[!ht]
\includegraphics[width=0.95\textwidth]{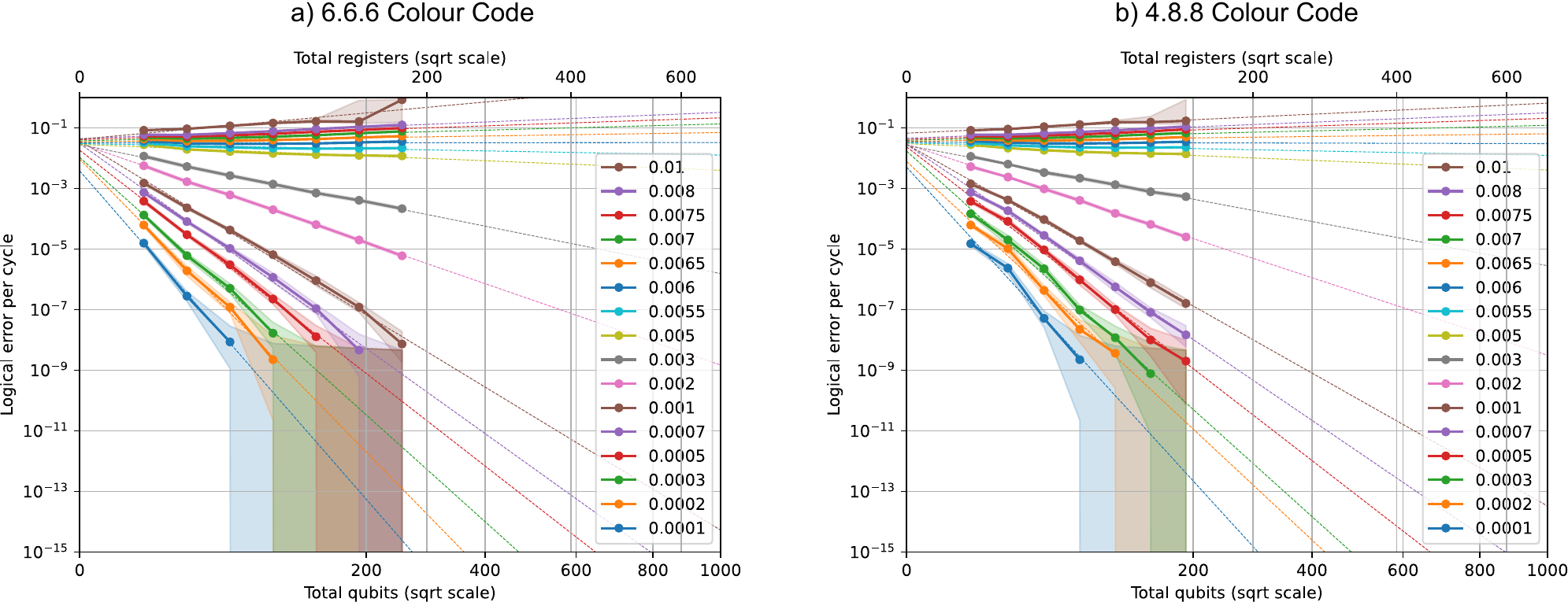}
\centering
\caption{\textbf{Scaling of time domain wall colour codes at various error rates as a function of qubit/register count.} The same data as in Fig. \ref{fig:supp_domainwall_colour_both} is plotted here, but as a function of qubits rather than error rate. Specifically, we show (a) the 6.6.6 colour code and (b) the 4.8.8 colour code. The threshold of $\sim 0.55 \%$ can be seen in both as constant error rates as a function of qubits. Data points are taken at distances of 3, 5, 7, 9, 11, 13 and 15. We also note the equivalent number of registers on the top axis. Despite the larger number of registers needed for the 6.6.6 colour code, namely $3d^2/4 + O(d)$ registers, compared to the $d^2/2$ registers used for the 4.8.8 colour code, both (a) and (b) show similar scaling as a function of the total number of registers - with the 6.6.6 colour code being slightly more efficient.~\label{fig:supp_colour_qubits_scaling}}
\end{figure*}

In Figure \ref{fig:supp_colour_qubits_scaling}, we show the same data as in Fig. \ref{fig:supp_domainwall_colour_both}, but as a function of qubits (and registers) so that it is easy to extrapolate the space impact of different physical error rates. For a distance $d$ 6.6.6 colour code, the number of data ($n_d$), ancilla ($n_a$) and total ($n$) qubits is:

\begin{align}
    n_d &= \frac{3d^2 + 1}{4}\\
    n_a &= \frac{n_d - 1}{2}\\
    n &= n_d+n_a = \frac{9d^2 - 1}{8}
\end{align}

and, noting that there are $d$ single-qubit data registers on the edge, and otherwise data registers are two-qubit registers, the total number of registers $n_r$ is:

\begin{align}
    n_r &= \frac{n_d-d}{2}+d+n_a\\
    &= \frac{3d^2+2d-1}{4}
\end{align}

Similarly, the 4.8.8 colour code has $d$ single-qubit data registers, meaning it is characterised by the following parameters:

\begin{align}
    n_d &= \frac{d^2 + 2d - 1}{2}\\
    n_a &= \frac{n_d - 1}{2}\\
    n &= n_d+n_a = \frac{3d^2 + 6d - 5}{4}\\
    n_r &= \frac{d^2 + 3d - 2}{2}
\end{align}

\clearpage

\section{Available Logical Operations for Bilayer Surface Codes~\label{sec:supp_bilayer_operations}}

As mentioned in the main text, under idle-biased noise it is possible to utilise a square grid of alternating 1- and 2-qubit registers to implement two parallel logical surface codes in a single square patch of physical qubits (see main text Fig. \ref{fig:bilayer_code}). Here, we present an efficient gateset which leverages the extra qubit connectivity afforded by this encoding, including both a complete set of local logical Clifford operations, as well as capabilities for large-scale lattice surgery (in the same vein as \cite{Litinski_game_of_surface_codes}). We note that the ``cycle'' requirements discussed here are specifically (double-)cycles of the bilayer code.

\subsection{Logical Pauli Gates}

Because bilayer surface codes consist of two surface codes, logical X, Y and Z gates can be performed to either layer logical qubit by performing them transversely (as usual for surface codes). There are no special optimisations we can make here, but we include it for completeness.

\subsection{Logical Single-Qubit Clifford Gates}

Unfortunately, surface codes do not permit local transversal Clifford gates (other than Pauli X, Y and Z), so it is not possible to perform Clifford gates directly on a surface code without using additional operations \cite{Litinski_game_of_surface_codes}. For example, surface codes implement a Hadamard gate in $3d$ cycles (where $d$ is the distance of the code), or $2d$ cycles when the Hadamard is combined with a translation of the logical qubit by at least one surface code patch \cite{Litinski_game_of_surface_codes}. However, locally transversal arbitrary Clifford gates can be implemented in both colour codes \cite{colour_code_summary} and folded surface codes \cite{moussa_bilayer}. The latter is very attractive for decoding, as folded surface codes have matchable stabilisers, and so can leverage the same decoding algorithms used for any surface code.\\

\begin{figure*}[!ht]
\includegraphics[width=0.4\textwidth]{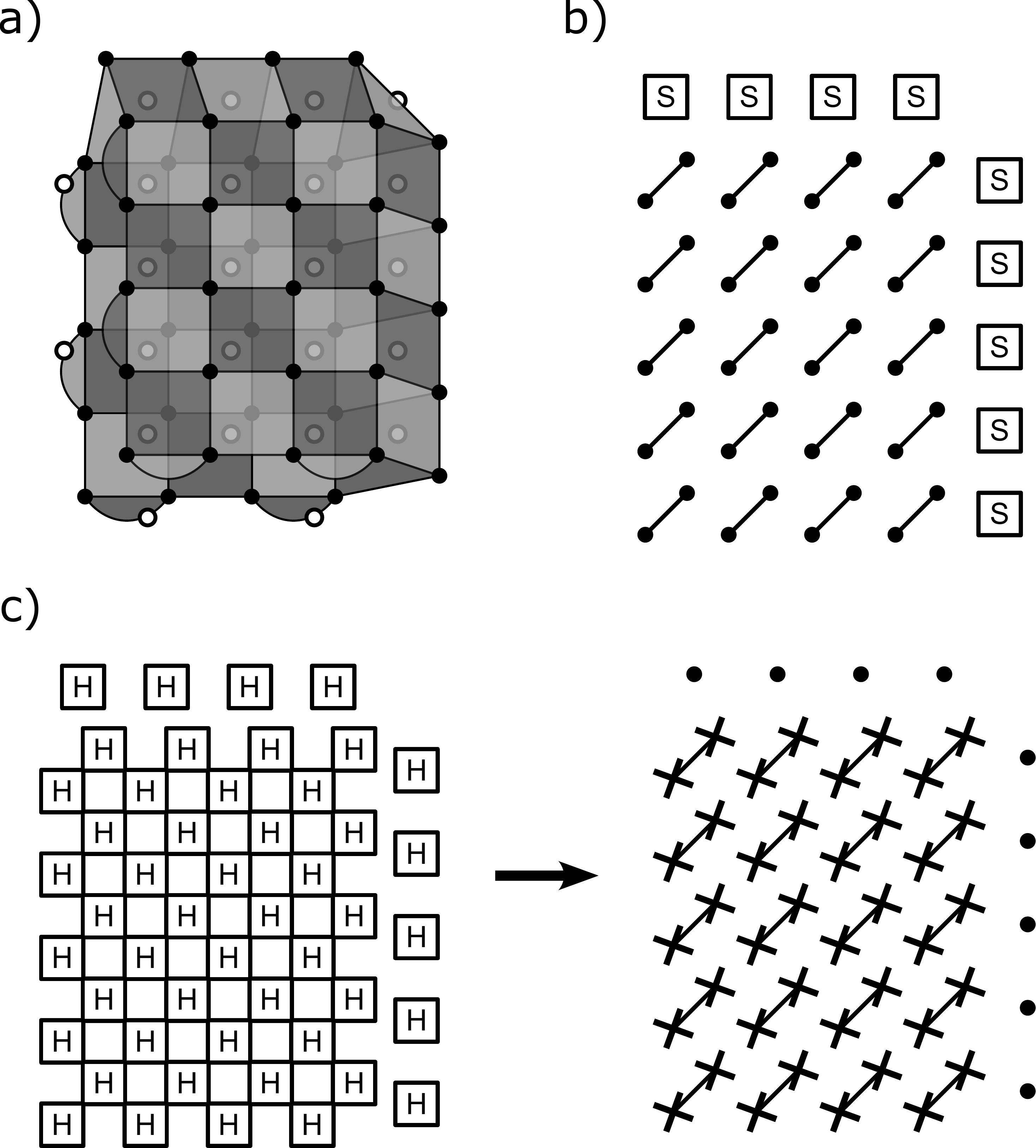}
\centering
\caption{\textbf{Transversal Clifford gates on a folded bilayer surface code.} a) Stabiliser arrangement for a folded surface code, defined using the same connectivity as Fig. \ref{fig:bilayer_code}a of the main text, implementing the code presented by Moussa \cite{moussa_bilayer}. This folded surface code is stabilised using the same technique as stabilising the bilayer surface code in the main text. b) Implementation of a transversal logical S gate on the folded surface code, performing single-qubit S gates on each fold data qubit and register-local CZ gates elsewhere. This can be decoded in $d$ cycles using the same techniques presented by Sahay \textit{et. al.} \cite{sahay2025foldtransversalsurfacecodecultivation}, or with no cycle overhead but slightly stricter assumptions per Serra-Peralta \textit{et. al.} \cite{Serra_Peralta_matching_transversal_cliffords}. c) Implementation of a transversal logical H gate on the folded surface code, first applying single-qubit H gates to all data qubits, and then performing register-local SWAP gates to all non-fold data qubits. The logical H gate can be performed with zero cycle time overhead.~\label{fig:supp_folded_bilayer_gates}}
\end{figure*}

The folded surface codes we consider here were originally developed for square lattices of qudits by Moussa \cite{moussa_bilayer}. Here, we apply this to the special case of the two-qubit registers used in this work for bilayer surface codes. The layout of a distance 5 folded surface code can be seen in Fig. \ref{fig:supp_folded_bilayer_gates}a, with X and Z boundaries existing along the unfolded edges. X (Z) logical operators can be formed, as usual, as strings of data qubit X (Z) gates connecting two of these edges. In addition, X, Y and Z logical operations can be performed as strings along the fold qubits alone, as the fold terminates on both ends at a twist defect.\\

As described by Moussa, a single-layer surface code can readily be grown into a folded surface code of the same distance using local operations. As with the majority of surface code growth operations, this growth operation requires $d$ cycles to perform to maintain the code distance. Hence, converting from a single-layer surface code to a folded surface code can be done in $d$ cycles, but the reverse operation can be performed instantly, in much the same way as the operations in Litinski \textit{et. al.} \cite{Litinski_game_of_surface_codes}. Of course, performing this conversion in a bilayer code requires the secondary layer to be free of its own logical information, so that it can be used to house the folded code. These spatial requirements are similar to single-layer surface codes, where logical Hadamard gates require one unused logical qubit patch to perform the gate \cite{Litinski_game_of_surface_codes}.\\

Once information is stored in a folded surface code, arbitrary Clifford gates can be readily performed as they are transversal. Based on the implementation described by Moussa \cite{moussa_bilayer}, we show the single-step implementation of a transversal S gate in Fig. \ref{fig:supp_folded_bilayer_gates}b, and the two-step implementation of a transversal H gate in Fig. \ref{fig:supp_folded_bilayer_gates}c. The transversal S gate involves applying single-qubit S gates to all data qubits along the fold of the folded surface code, and register-local CZ gates to all other data qubits. The transversal Hadamard gate involves first applying single-qubit H gates transversely to all data qubits of the folded code, and then performing SWAPs locally within data qubit registers to restore the stabiliser and logical operator locations to their original positions. All other Clifford gates are also transversal, and can be constructed by combining transversal S and H gates.\\

For the transversal logical H gate, each stabiliser of the original folded code is transformed into exactly one stabiliser of the final code, with top-layer Z stabilisers becoming bottom-layer X stabilisers and similarly for the other stabilisers. This means that the stabilisers remain matchable through the logical H gate, and hence the transversal H gate can be applied with zero time overhead in terms of stabiliser cycles by using standard matching algorithms. Note also that a transversal logical $\pi/2$ rotation about the $y$ axis of the Bloch sphere can also be performed by applying $R_y(\pi/2)$ to all fold qubits, and a register-local X-controlled X gate to all non-fold data qubits. This respects the same one-to-one stabiliser relationship, and so can also be performed without any modification beyond swapping the X and Z detectors when matching.\\

However, this is not true for logical S gates (Fig. \ref{fig:supp_folded_bilayer_gates}b), because the transversal S gate transforms top-layer X stabilisers in the original code into the product of top-layer X and bottom-layer Z stabilisers in the final code (similarly true for the bottom layer X stabilisers, and with Z stabilisers commuting fully with the logical S gate). Because of this, weight-3 error channels are introduced which cannot be decomposed into weight $\le 2$ channels, rendering the transversal S gate unmatchable to naive decoders. The same problem exists for decoding transversal CNOT gates between planar surface codes. While decoding methods exist which can use matching directly on such gates, these techniques require $d$ cycles of stabilisation before the transversal gate \cite{sahay2025foldtransversalsurfacecodecultivation}. However, once all nearby classical controls have been resolved, it is possible to decode transversal gates without cycle overheads by using logical observable matching (LOM) \cite{Serra_Peralta_matching_transversal_cliffords}. In addition, it is worth noting that in the case of a transversal S gate directly after logical $\ket{+}$ initialisation (before the first stabiliser measurement), as well as for an S gate directly before X-basis measurement (after the final stabiliser measurement), the problematic weight-3 error channels are actually missing one detector (before the first cycle or after the last respectively), converting these channels to weight-2 and therefore meaning that Y-basis initialisation and measurement of the folded surface code is possible to decode with standard matching algorithms. This result relies on the transversal S gate - in contrast, initialising all data qubits in the $S\ket{+}$ state followed by Z and X stabiliser measurements would not constitute a fault-tolerant Y-basis initialisation (due to $S\ket{+}$ initialisation errors causing non-decomposable weight-4 error channels). Overall, this means that S gates can generally be performed instantly without a $d$ cycle overhead, thereby enabling any single-qubit Clifford gate to be implemented without overhead.

\subsection{Transversal CNOT Gates}

\begin{figure*}[]
\includegraphics[width=0.7\textwidth]{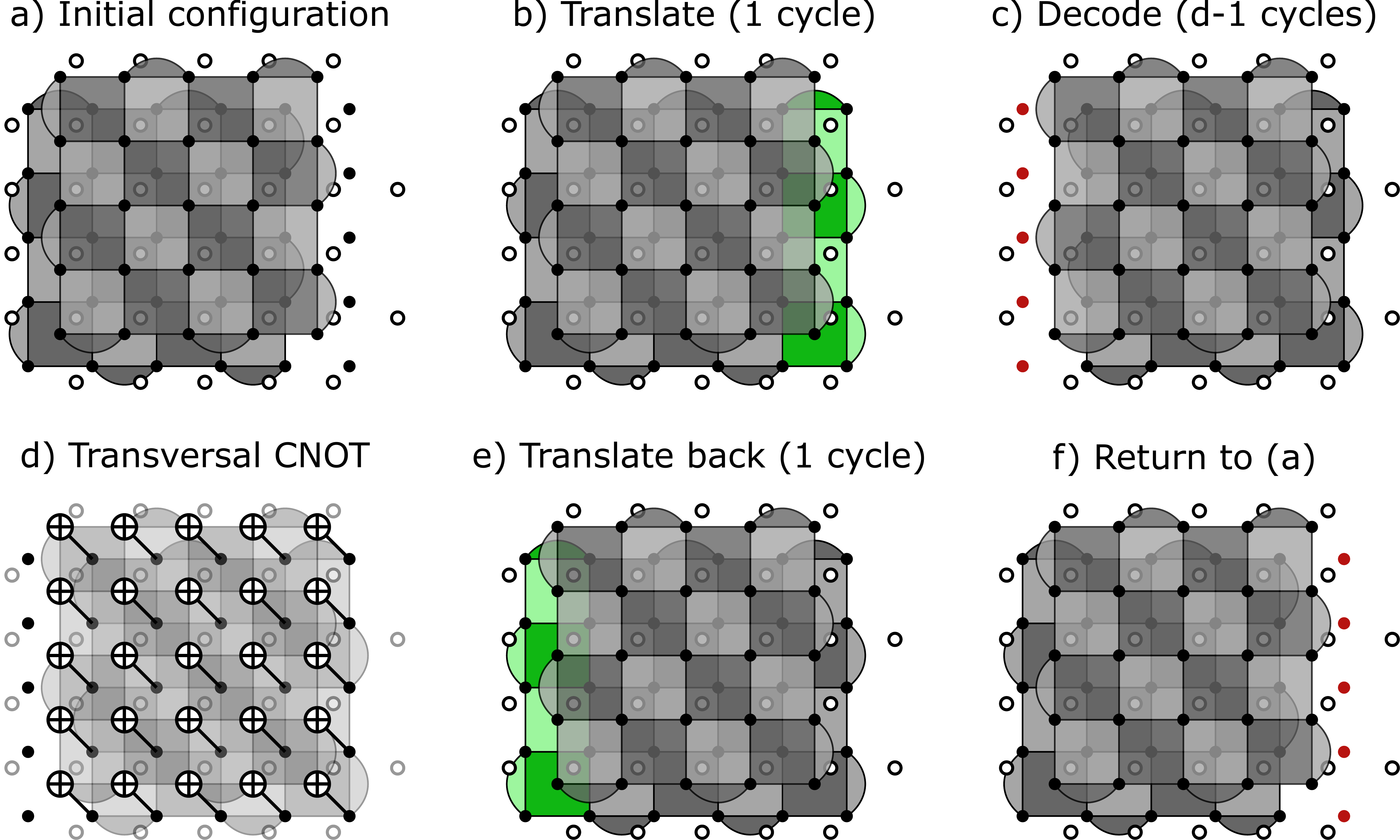}
\centering
\caption{\textbf{Transversal logical CNOT gate between the two layers of a bilayer surface code.} The sequence a-f describes, the order of steps needed to perform a transversal logical CNOT between the two layers of a bilayer code. Steps which require multiple stabiliser cycles to be measured are annotated with their cycle time overhead. To perform the CNOT, one of the layers is translated by a single data qubit column, per (a-b), to ensure the stabilisers between the two layers line up. After determining if logical errors have occurred by performing $d-1$ cycles of measurement (c), a transversal CNOT can be applied (d). Finally, we translate back to the original bilayer code setup (e-f). The entire process can be performed in $d+1$ cycles, and in $d$ cycles if either logical qubit is measured after the CNOT. In addition, the CNOT operation can be combined with a translation of one (or both) of the logical qubits for a combined duration of $d+1$ cycles, or $d$ cycles if at least one of the translated qubit(s) are measured after the CNOT. Values of new/modified stabilisers are highlighted in green, and freshly measured qubits are highlighted in red. All data qubit initialisations are to the $\ket{0}$ state and all measurements are in the Z basis. Note that the translation could equivalently have translated the X-edge rather than the Z-edge, which would instead require $\ket{+}$ state preparation and X-basis measurement.~\label{fig:supp_bilayer_CNOT}}
\end{figure*}

Bilayer surface codes also enable the application of transversal CNOT gates between two logical qubits on different layers of the same patch. Unfortunately, direct CNOT gates cannot be applied, as the stabilisers do not line up between the two patches (a feature that was intentionally picked to ensure each ancilla was responsible for at most one X and at most one Z type stabiliser, to leverage the noise bias when stabilising the bilayer code). Hence, to perform a transversal CNOT between the two layers, one of the layers needs to be offset by one row/column of data qubits to match the stabilisers. The overall strategy for implementing a transversal CNOT in this way is given in Fig. \ref{fig:supp_bilayer_CNOT}.\\

Nominally, translation of surface code qubits takes $d$ cycles \cite{Litinski_game_of_surface_codes}. However, translation by just one data qubit can be done in $1$ cycle, as the boundary defects are kept sufficiently far apart (assuming the decoding graph is split into spacelike data qubit errors and timelike ancilla measurement errors). The main exception to this is that any new stabilisers must be measured for at least $d-1$ cycles after the translation to maintain measurement-error code distance; however, these extra cycles can be absorbed into following operations. The procedure for doing this transversal CNOT gate is shown in Fig. \ref{fig:supp_bilayer_CNOT}b and e. Hence, steps b and e of Fig. \ref{fig:supp_bilayer_CNOT} can be performed in just one cycle, because there are $d-1$ cycles between them.\\

\begin{figure*}[]
\includegraphics[width=0.6\textwidth]{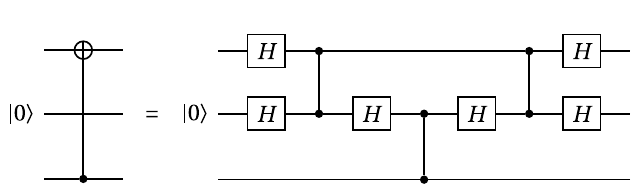}
\centering
\caption{\textbf{CNOT gate implementation mediated by an intermediary ancilla qubit.} This is used for the transversal CNOTs applied in Fig. \ref{fig:supp_bilayer_CNOT}d, because the transversal CNOTs are applied with the control qubit in a neighbouring register to the target data qubit. Hence, to keep with nearest-neighbour register interaction, we must perform this operation via the intermediary ancilla register. Note that this implementation maintains Z-noise bias for the control qubit of the CNOT. Also note that this circuit is optimised for the ancilla initially being in the $\ket{0}$ state, and does not work for an arbitrary ancilla qubit state. The final Hadamard gate to the mediating ancilla can be omitted if desired, and is simply included for symmetry so that the ancilla returns to the $\ket{0}$ state at the end of the sequence. If useful in future applications, a mediated distant CZ gate can be performed by omitting the Hadamard gates applied to the upper (target) qubit, in which case noise bias is maintained for both qubits involved in the distant CZ.~\label{fig:supp_mediated_CNOT}}
\end{figure*}

Because one of the codes has been translated, the transversal CNOT performed in Fig. \ref{fig:supp_bilayer_CNOT}d is not register-local. To keep with the square connectivity of the registers comprising the bilayer code (see Fig. \ref{fig:bilayer_code}a), this transversal CNOT needs to be performed via the ancilla registers. A circuit for this is given in Fig. \ref{fig:supp_mediated_CNOT}, which assumes the mediating ancilla is initialised in the $\ket{0}$ state. Despite being high depth, this mediated distant CNOT is still correctable (as it is transversal), and usefully also maintains Z-noise bias for the control qubit.\\

A transversal CNOT gate causes X stabilisers in the pre-CNOT control surface code to be transformed into the product of X stabilisers of both codes post-CNOT. A similar duplication occurs for the Z stabilisers of the target surface code, with both causing non-decomposable weight-3 error channels. This presents a similar challenge for the decoder as was already discussed for the transversal S gate (Fig. \ref{fig:supp_folded_bilayer_gates}b). One solution is to use the $d$-cycle overhead algorithm presented by Sahay \textit{et. al.} \cite{sahay2025foldtransversalsurfacecodecultivation}, which adds no additional cost if the CNOT is performed directly before the background layer is translated back.\\

However, it is important to note that the $d$-cycle overhead from translation (that is, the requirement that new stabilisers be measured for $d-1$ cycles after the $1$-cycle translation) is only applicable if starting from the configuration in Fig. \ref{fig:supp_bilayer_CNOT}a. If the operation before the desired transversal CNOT was to translate at least one of the involved logical qubits from some other cell (see ``Qubit movement'' in \cite{Litinski_game_of_surface_codes}), that logical qubit can be translated to the configuration shown in Fig. \ref{fig:supp_bilayer_CNOT}c rather than that shown in Fig. \ref{fig:supp_bilayer_CNOT}a. After this distant translation is done, the transversal CNOT can be done immediately by using LOM decoders per \cite{Serra_Peralta_matching_transversal_cliffords}, similarly to the transversal S gate. One cycle is still required to translate back to the usual configuration in Fig. \ref{fig:supp_bilayer_CNOT}e, meaning overall a translate + CNOT also costs $d+1$ cycles.\\

Finally, we justify the omission of a $d-1$ cycle overhead following the translate back in Fig. \ref{fig:supp_bilayer_CNOT}e. For this, we consider all possible subsequent operations, both presented by Litinski for single-layer surface codes \cite{Litinski_game_of_surface_codes} and those shown by us in main text Fig. \ref{fig:bilayer_code}d of this work. If the logical operation following the translate-back is any kind of logical idle, lattice surgery or patch-grow operation, then the $d-1$ cycle requirement is fulfilled - these operations take $d$ cycles and will continue measuring the new stabilisers highlighted in green in Fig. \ref{fig:supp_bilayer_CNOT}e for sufficient duration to maintain code distance. The following operation cannot be a corner movement, as the background surface code is a square surface code patch of distance $d$ and any corner movement would therefore reduce the code distance. The final following operation which could jeopardise the new stabilisers highlighted in Fig. \ref{fig:supp_bilayer_CNOT}e is if the background logical qubit was immediately measured out in the logical Z or X bases. However, in such a case there is no need to translate the qubit back to the normal configuration - the background qubit can be directly measured out after the transversal CNOT in Fig. \ref{fig:supp_bilayer_CNOT}d without being translated back. Hence, the full duration of the transversal CNOT can be thought of as $d+1$ cycles, with a note that if either qubit is measured out after the CNOT then the operation takes $d$ cycles with the measured logical qubit being the one translated in Fig. \ref{fig:supp_bilayer_CNOT}b.\\

The full transversal CNOT procedure shown in Fig. \ref{fig:supp_bilayer_CNOT} is annotated with these cycle time overheads. Combining all the costs, we find that a transversal CNOT gate can be performed in a total of $d+1$ cycles (not including the time taken to perform the CNOT operation in step d of Fig. \ref{fig:supp_bilayer_CNOT}), with the note that if the following operation is a measurement of either logical qubit, the duration is $d$ cycles instead. In addition, if one of the qubits is also being seperately translated before/after the transveral CNOT, this translation can absorb $d$ cycles, with the combined translation and CNOT requiring $d+1$ cycles (or $d$ if the translated qubit is then measured). Either way, the $d+1$ cycle cost is significantly cheaper than a CNOT operation executed via lattice surgery in a single-layer surface code, which is typically done in $3d$ cycles \cite{Horsman_2012}.

\subsection{Neighbouring lattice surgery}

\begin{figure*}[]
\includegraphics[width=0.4\textwidth]{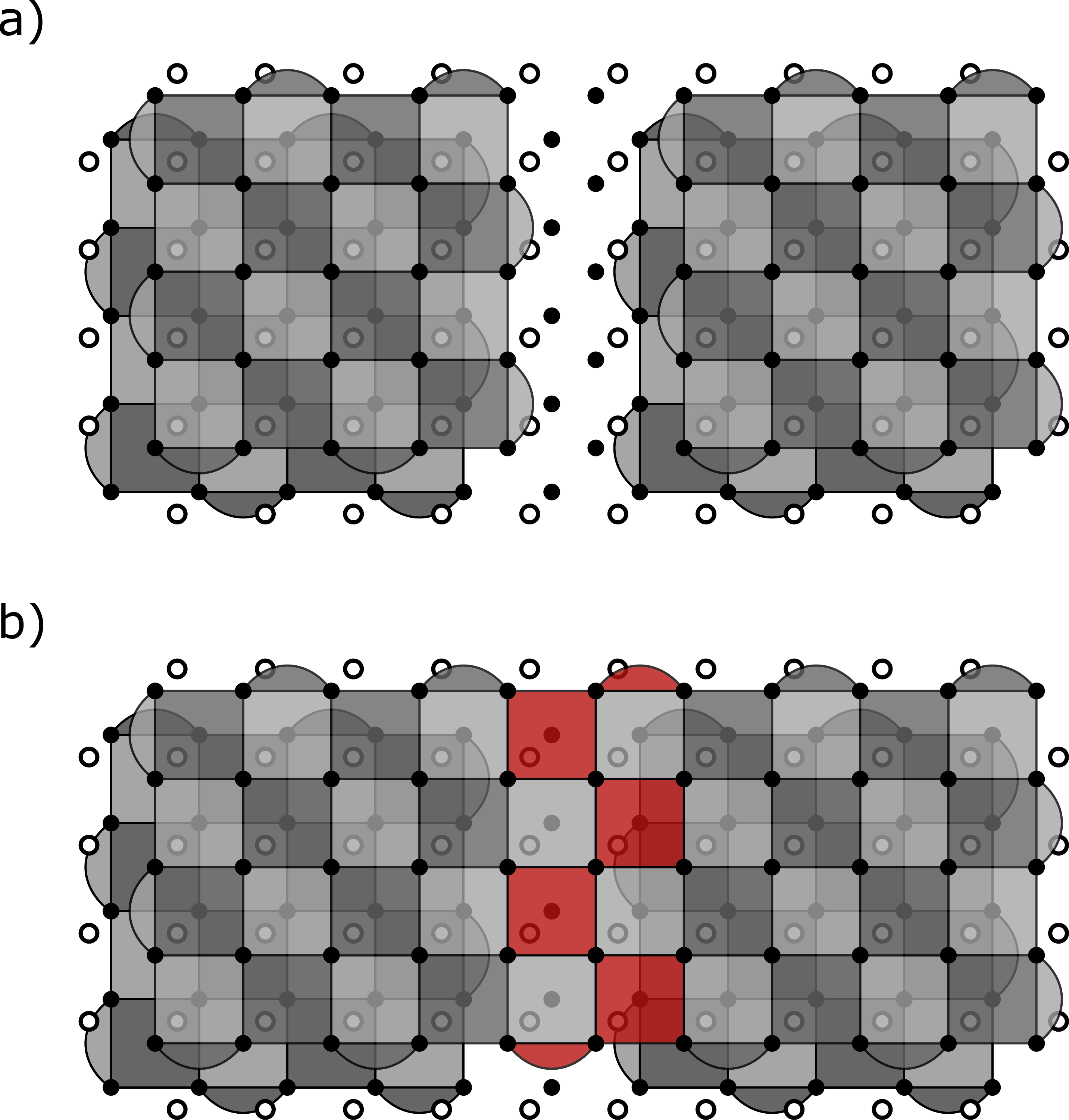}
\centering
\caption{\textbf{Example lattice surgery between the same layer of neighbouring bilayer surface codes.} a) Initial layout of the logical qubits in the bilayer code, showing two neighbouring logical patches. b) Joint stabilisers used to measure the XX logical parity between the two foreground logical qubits. The intermediary data qubits are initialised in the $\ket{0}$ state to perform the merge operation. The stabilisers highlighted in red are individually non-deterministic, but their product gives the logical XX parity between the two foreground logical qubits. As usual, the stabilisers in (b) are measured for $d$ cycles to maintain code distance, and then the intermediary qubits are measured out in the Z basis to return the logical qubits to the original configuration in (a). Note that the background layer could be merged simultaneously to the foreground layer without compromising the usual bilayer stabiliser measurement ordering.~\label{fig:supp_bilayer_neighbour_surgery}}
\end{figure*}

Surface code lattice surgery can be performed between neighbouring logical qubits in the same layer of the bilayer code. This lattice surgery is performed in much the same way as for single-layer surface codes \cite{rotated_lattice_surgery}. However, due to the fact that the edge stabilisers do not line up between the two edges being merged, an additional chain of data qubits is used in the surgery, similar to that required for merging unrotated surface codes \cite{Horsman_2012}. An example lattice surgery procedure which measures the XX logical parity is shown in Fig. \ref{fig:supp_bilayer_neighbour_surgery}. The stabilisers highlighted in red are non-deterministic, but their product reveals the XX logical parity of the two logical qubits being merged. The intermediary data qubit registers are initialised in the $\ket{0}$ state for the merge, and are measured out in the Z basis to complete the merge. As usual, the joint stabilisers are measured for $d$ cycles to maintain code distance. Note that the second layer could also be simultaneously merged in Fig. \ref{fig:supp_bilayer_local_surgery}, because the merging operation respects our bilayer stabilisation rule that each ancilla qubit is responsible for at most one X stabiliser and at most one Z stabiliser, even when both layers are merged. When measuring ZZ logical parity, the intermediary data qubits are initialised in the $\ket{+}$ state and measured in the X basis.\\

\begin{figure*}[]
\includegraphics[width=0.8\textwidth]{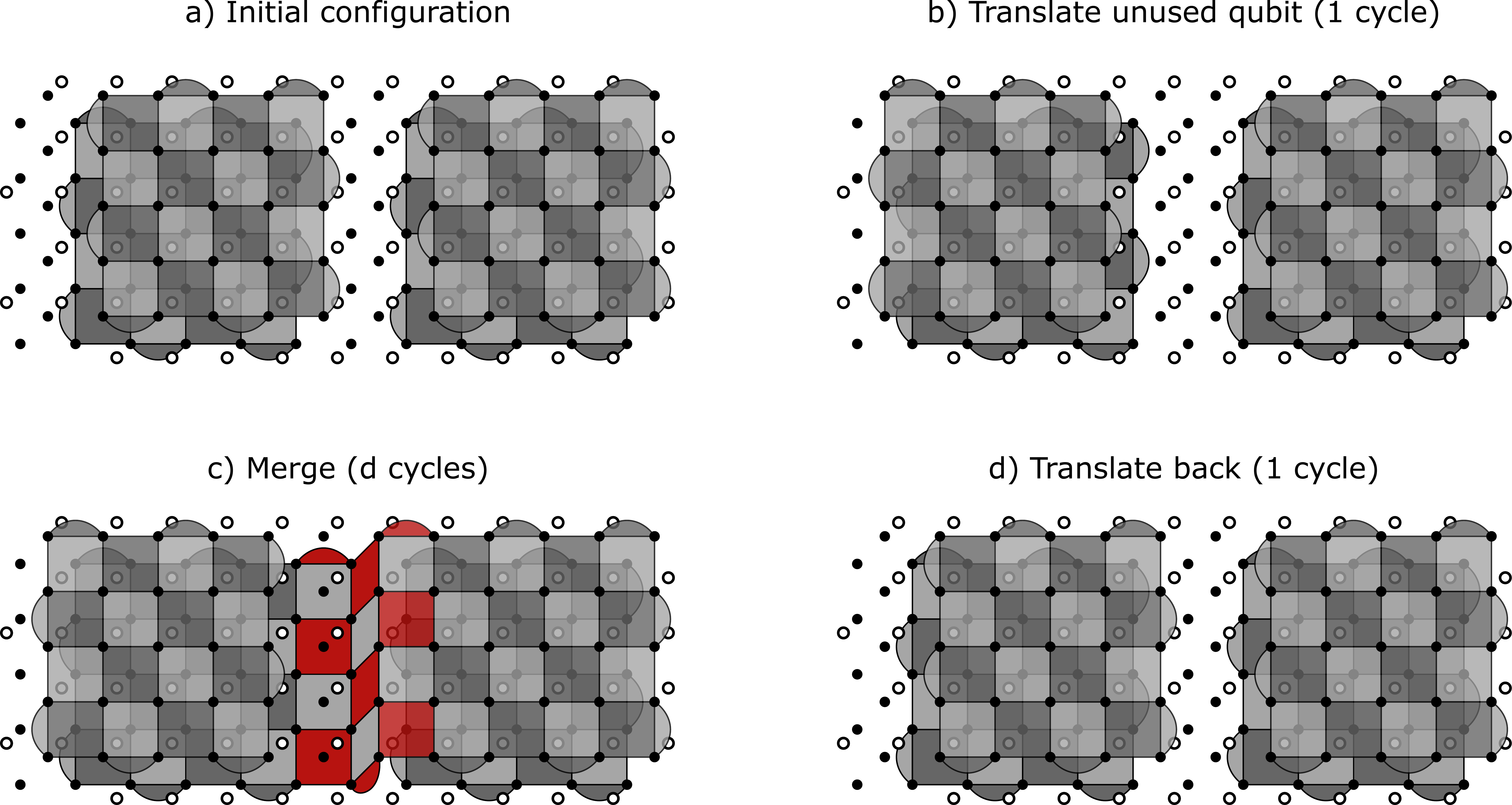}
\centering
\caption{\textbf{Example lattice surgery diagonally between different layers of neighbouring bilayer surface codes.} a) Initial layout of the logical qubits in the bilayer code, showing two neighbouring logical patches. b) One of the logical qubits not involved in the merge (in this case, the left foreground qubit) is translated out of the way to free up the edge ancillas for stabilising the joint code, similarly to Fig. \ref{fig:supp_bilayer_CNOT}b. c) Joint stabilisers are then performed to measure the XX logical parity between the left background and right foreground logical qubits. As per Fig. \ref{fig:supp_bilayer_neighbour_surgery}b, we have highlighted the nondeterministic stabilisers that are multiplied to calculate the logical XX parity. This merge otherwise occurs identically to Fig. \ref{fig:supp_bilayer_neighbour_surgery}b. d) To complete the operation, we translate the offset qubit back to its original position. Note that unlike lattice surgery between the same layer (Fig. \ref{fig:supp_bilayer_neighbour_surgery}), the diagonal lattice surgery shown here cannot be performed simultaneously (ie. you cannot also merge the foreground left qubit with the background right qubit in the operation pictured here) as this would violate the requirement that each ancilla is responsible for at most one X and Z stabiliser.~\label{fig:supp_bilayer_diag_neighbour_surgery}}
\end{figure*}

Lattice surgery between different layers of neighbouring patches is a little more complicated. The reason we cannot just use larger stabilisers which skip over the intermediary layer of qubits, in a similar way to Y-basis lattice surgery \cite{Litinski_game_of_surface_codes}, is that the performance of the bilayer codes has been optimised to avoid mid-stabiliser Hadamard gates which might cause hook errors. However, with a native gateset of 1Q rotations and CZ gates, larger stabilisers which skip a row of data qubits must involve these mid-circuit Hadamard gates to maintain nearest-neighbour connectivity and transfer the stabiliser information to one qubit for readout. Instead, we perform different-layer neighbouring lattice surgery by first translating one of the unused logical qubits out of the way (in 1 cycle as per transversal CNOT gates), and then can use the ancilla that was previously performing the edge stabilisers for this now-translated logical qubit instead to stabilise a larger connecting region, as shown in Fig. \ref{fig:supp_bilayer_diag_neighbour_surgery}.\\

The extra overhead incurred due to the translation of one of the unused logical qubits results in the diagonal lattice surgery taking $d+2$ cycles to perform. In addition, the reuse of some of both layers of intermediary data qubits means that one cannot simultaneously perform lattice surgery of both diagonals. That is, we can merge the background left qubit with the foreground right qubit (pictured in Fig. \ref{fig:supp_bilayer_diag_neighbour_surgery}), or we can merge the foreground left qubit with the background right qubit, but we cannot perform both of these operations at the same time. Similar to the discussion of transversal CNOT gates, if the preceding operation was a translation of at least one of the \textit{uninvolved} logical qubits, or if at least one of the uninvolved patches is unused, then the merge can be done in $d+1$ or $d$ cycles respectively. Also similar to transversal CNOT gates, if one of the uninvolved logical qubits is measured out directly after the merge, then the operation can be done in $d+1$ cycles ($d$ cycles if that same uninvolved logical qubit was translated before the merge).

\subsection{Local lattice surgery}

\begin{figure*}[]
\includegraphics[width=0.4\textwidth]{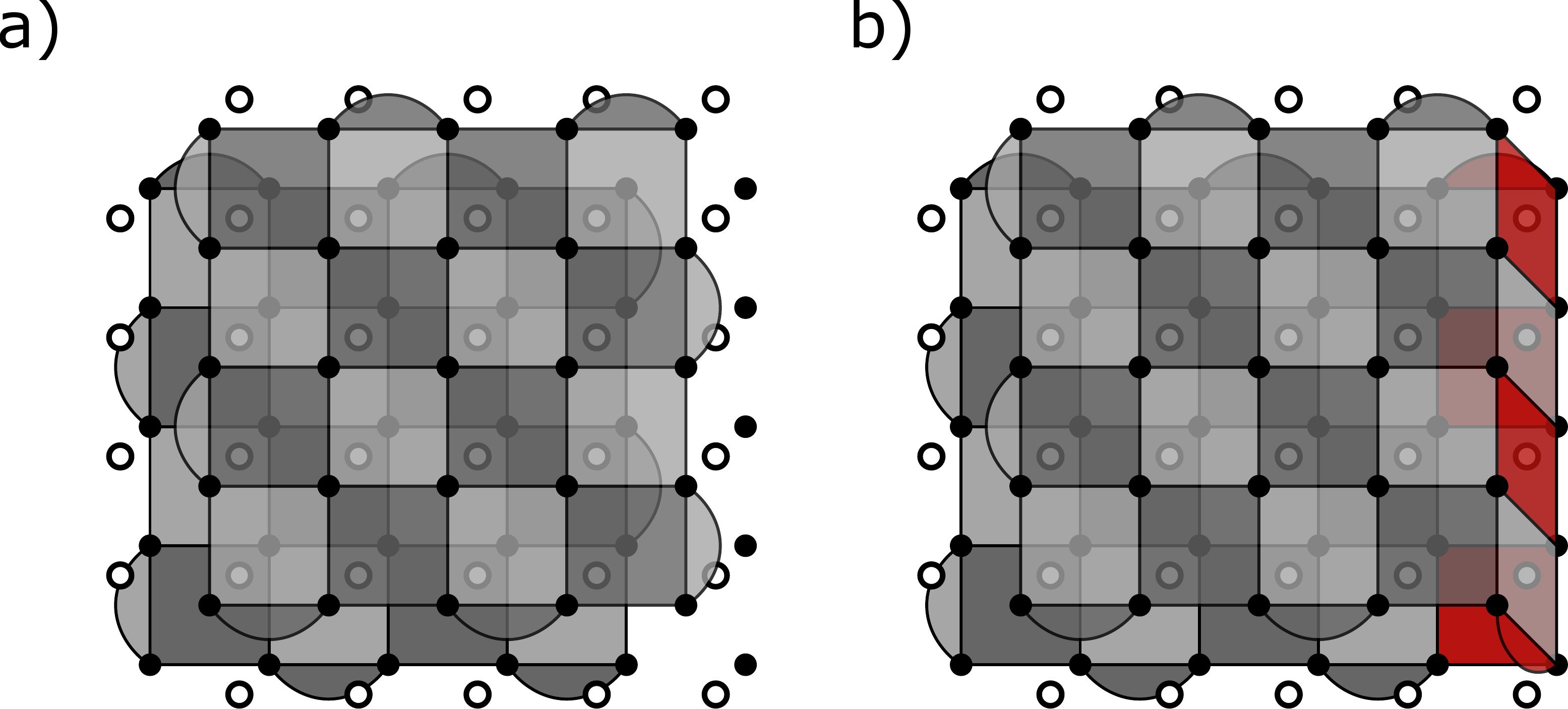}
\centering
\caption{\textbf{Example lattice surgery between the two layers of a single patch of the bilayer surface codes.} (a) represents the initial configuration of logical qubits, and (b) shows the joint stabilisers performed in the merge. The XX parity is again found by multiplying the stabilisers highlighted in red (which include X stabilisers on the background layer of the join). Again, the additional data qubits are initialised in the $\ket{0}$ state, and measured out in the Z basis after $d$ cycles to return the code to the original configuration.~\label{fig:supp_bilayer_local_surgery}}
\end{figure*}

As well as performing lattice surgery between neighbouring patches of the bilayer code, it is also possible to perform lattice surgery between the two different layers of a single patch of the bilayer code. This is shown in Fig. \ref{fig:supp_bilayer_local_surgery}. To account for the offset edge stabilisers, an extra chain of data qubits is used for the merge, similar to how same-layer codes were merged in Fig. \ref{fig:supp_bilayer_neighbour_surgery}. Again, this process takes $d$ cycles to complete.

\clearpage

\section{Bilayer Code Stabilisation~\label{sec:supp_bilayer_stabilisation}}

\begin{figure*}[!ht]
\includegraphics[width=\textwidth]{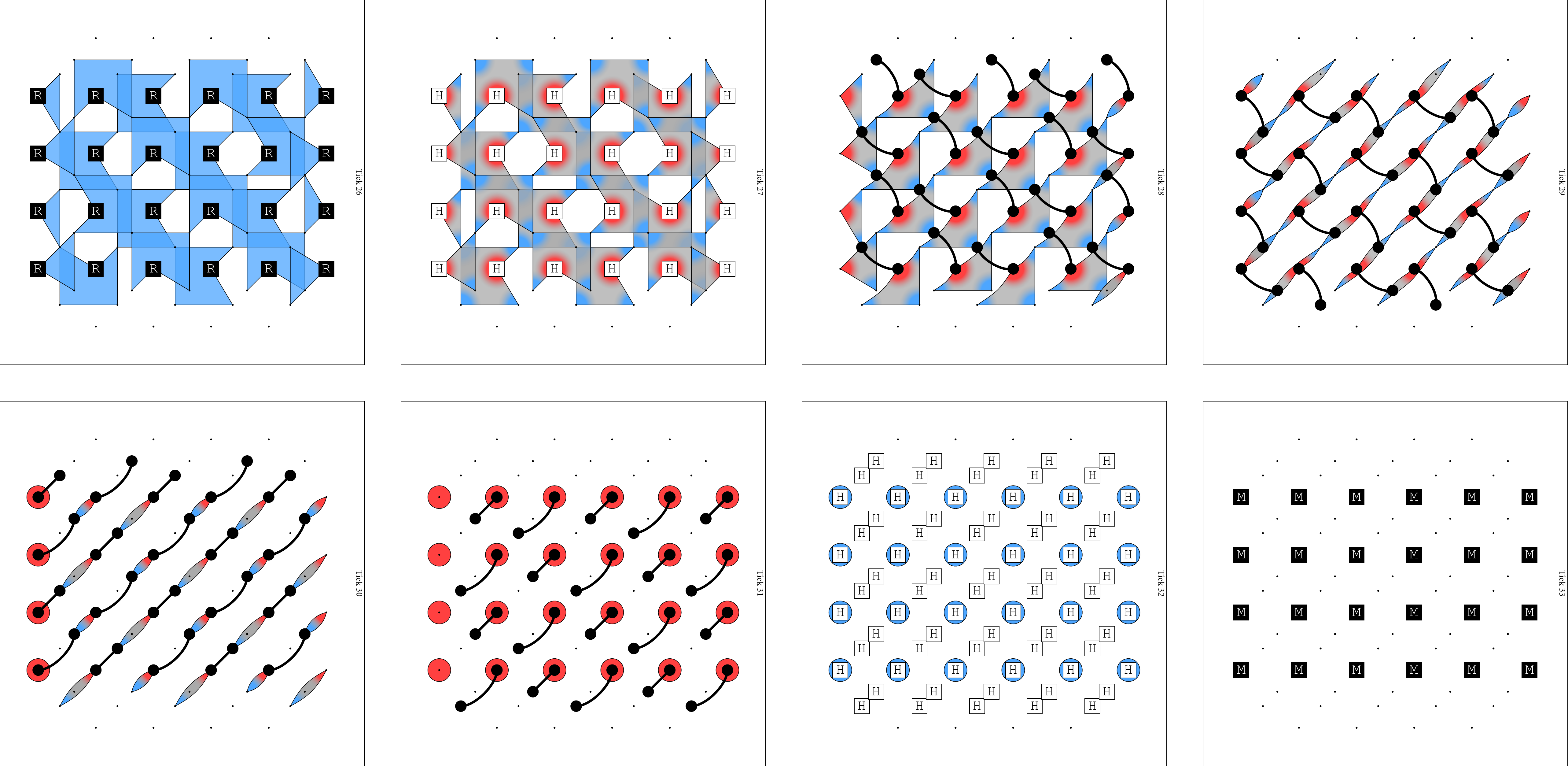}
\centering
\caption{\textbf{Measurement of the Z stabilisers for a bilayer surface code.} This circuit utilises time-domain walls to equalise protection of X and Z logical operators, which is achieved by applying data qubit Hadamard gates in the second-last tick to prepare the data qubits for X stabilisation. Similar to the colour code stabilisers given in Fig.~\ref{fig:supp_colour_stabilisers}, the first tick involves initialising all the ancilla qubits, and the second tick involves applying Hadamard gates to transfer the ancilla qubits into the $\ket{+}$ state. The following series of CZ gates transfer all the Z stabiliser values from both the foreground and background surface codes into the X projection of the ancillas. In the second-last tick, we apply Hadamard gates to transfer this information into the Z projection of the ancillas, while also simultaneously applying Hadamard gates to the data qubits to prepare for the following (single-)cycle measuring X stabilisers. The final tick measures the ancilla qubits, providing the values of all Z stabilisers. Note that unlike the colour code stabilisation in Fig.~\ref{fig:supp_colour_stabilisers}, the X stabiliser measurement is not identical to the Z measurement, instead using all ancillas responsible for X stabilisers and applying CZ gates to transfer the X stabilisers from both layers to the ancilla qubits.~\label{fig:supp_bilayer_stabilisers}}
\end{figure*}

The exact circuit implementing (single-)cycle stabilisation of the bilayer surface code is shown in Fig. \ref{fig:supp_bilayer_stabilisers} for measurement of stabilisers using time-domain walls. Similar to colour codes with time domain walls, the data qubit Hadamard gates in the second last tick swap the Z and X stabilisers, such that running a similar circuit immediately after Fig. \ref{fig:supp_bilayer_stabilisers} will measure the X stabilisers. Again, we refer to such a scheme as employing time-domain walls. However, unlike the colour codes which can reuse all ancillas, the top/bottom edge ancillas will be used for the X stabilisers, instead of the left/right edge ancillas which were used for the Z stabilisers in Fig. \ref{fig:supp_bilayer_stabilisers}. Note that when stabilising without time-domain walls, no Hadamard gates are applied to data qubits when measuring Z stabilisers, and Hadamard gates are applied to data qubits at the same time as ancilla Hadamard gates when measuring X stabilisers.

\clearpage

\section{Bilayer Code Performance, Scaling and Variants~\label{sec:supp_bilayer_performance}}

\begin{figure*}[!ht]
\includegraphics[width=\textwidth]{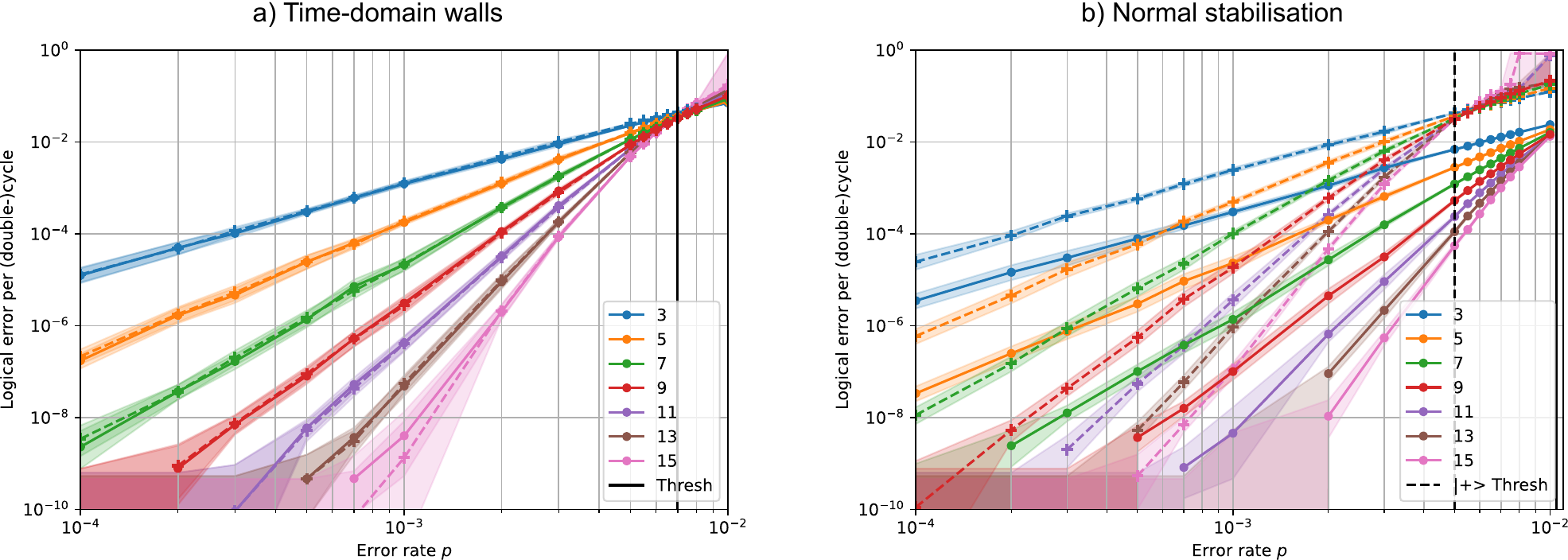}
\centering
\caption{\textbf{Threshold plots showing performance of square bilayer codes.} We show the performance of square bilayer codes at various distances as a function of physical error rate when a) stabilising with time-domain walls and b) without using time-domain walls. In both cases, we show protection of the logical $\ket{0}$ state (solid lines) and the protection of the logical $\ket{+}$ state (dashed lines). For a), we mark a crossing threshold of $0.7\%$ as per main text Fig. \ref{fig:bilayer_code}c. In b), we mark a threshold of $\sim 0.5\%$ for logical $\ket{+}$ protection and a threshold of $\sim 1.05\%$ for logical $\ket{0}$ protection. Similarly to the colour code performance shown in Figs.~\ref{fig:supp_vanilla_colour} and~\ref{fig:supp_domainwall_colour_both}, we see that adding time-domain walls in (a) equally protects X and Z logical errors, whereas stabilisation without time-domain walls in (b) shows unequal protection, a consequence of the underlying bias of the noise model.~\label{fig:supp_bilayer_threshold}}
\end{figure*}

In Fig. \ref{fig:supp_bilayer_threshold}, we show threshold plots for square bilayer patches with and without the use of time-domain walls. Similar to colour codes, when time-domain walls are used we see equal protection of the $\ket{0}$ and $\ket{+}$ states, and without these walls, we see $\ket{0}$ is better protected than $\ket{+}$ due to the noise bias.\\

In Fig. \ref{fig:supp_bilayer_scaling}, we plot the data in Fig. \ref{fig:supp_bilayer_threshold}a (ie. time-domain wall square bilayer code performance) as a function of the number of qubits rather than the physical error rate, to show how code performance can be extrapolated. For a bilayer code, the number of data qubits ($n_d$), ancilla qubits ($n_a$), total qubits ($n$) and total registers ($n_r$), all taken as average per logical qubit (noting ancillas are reused between a pair of logical qubits), is:

\begin{figure*}[!ht]
\includegraphics[width=0.5\textwidth]{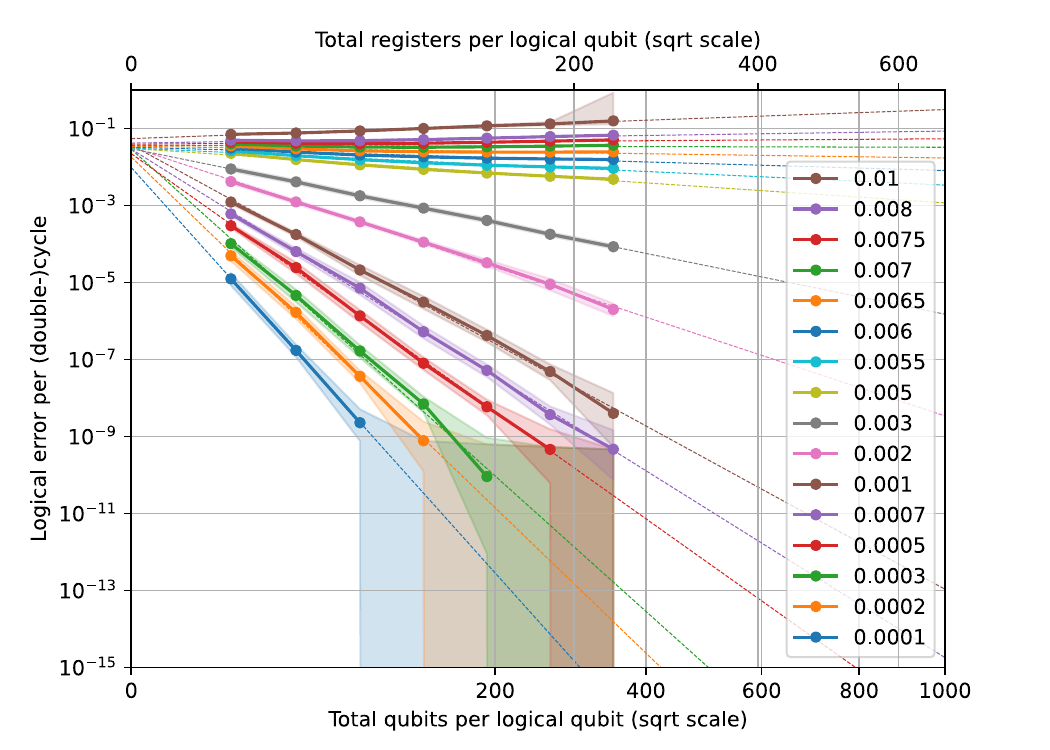}
\centering
\caption{\textbf{Scaling of time-domain wall bilayer codes as a function of qubits.} The data shown here is the same data as in Fig.~\ref{fig:supp_bilayer_threshold}a, but plotted as a function of the size of the bilayer code in qubits. The performance is sampled for square bilayer codes of distances 3, 5, 7, 9, 11, 13, and 15. When compared to the scaling shown in Fig.~\ref{fig:supp_colour_qubits_scaling}, we can see that the bilayer code scales similarly to colour codes as a function of qubits per logical qubit. As such, for the error model used in this work, larger fault-tolerant architectures based on bilayer codes are likely more efficient that ones based on colour codes, as bilayer codes are easier to decode by using matching algorithms while still having a similar qubit footprint.\label{fig:supp_bilayer_scaling}}
\end{figure*}

\begin{align}
    n_d &= d^2\\
    n_a &= \frac{(d-1)^2 + 4(d-1)}{2}\\
    &= \frac{d^2+2d-3}{2}\\
    n &= n_d+n_a = \frac{3d^2 + 2d - 3}{2}\\
    n_r &= d^2 + d - \frac{3}{2}
\end{align}

Note that this does not include any extra qubits used for the lattice surgery operations discussed in Appendix~\ref{sec:supp_bilayer_operations}, which adds $O(d)$ extra data qubits and $O(1)$ extra ancilla qubits.\\

\begin{figure*}[!ht]
\includegraphics[width=0.5\textwidth]{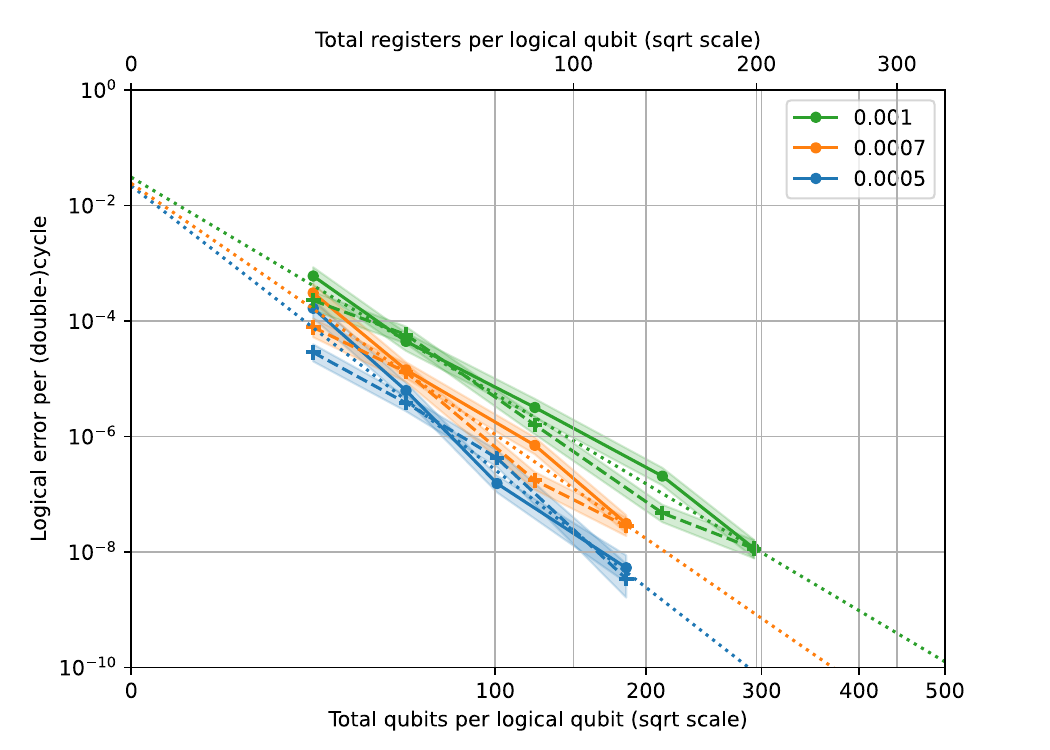}
\centering
\caption{\textbf{Scaling of performance for asymmetric rectangular bilayer patches stabilised without time domain walls.} These simulations are performed without time domain walls, only for physical error rates of 0.05\%, 0.07\% and 0.1\%, using the aspect ratios given in Tab. \ref{tab:supp_aspect_ratios}. Logical $\ket{0}$ and $\ket{+}$ states are shown with solid and dashed lines respectively. Specifically, this figure demonstrates that the chosen aspect ratios equally protect X and Z logical errors, allowing these aspect ratios to be used when simulating asymmetric surface code patches without time-domain walls (such as those in Fig.~\ref{fig:magic_scroll_distillation} of the main text.)~\label{fig:supp_bilayer_asymmetric}}
\end{figure*}

For use in distillation, we also characterise the performance of asymmetric bilayer surface code patches, which are non-square rectangular patches with different X and Z distances, such that when normal stabilisation (ie. no time-domain walls) is applied, logical X and Z protection is approximately equal. We only optimise such rectangular aspect ratios for physical error rates of 0.05\%, 0.07\% and 0.1\%, as these error rates are typically used when performing distillation simulations~\cite{Litinski_distillation}. We tabulate the approximate equal-protection aspect ratios in Table \ref{tab:supp_aspect_ratios}. To show these approximately equally protect X and Z errors, the performance of these asymmetric patches is shown in Fig. \ref{fig:supp_bilayer_asymmetric}.

{\renewcommand{\arraystretch}{1.2} 
\begin{table*}[]
    \centering
    \begin{tabular}{|c|c|c|c|c|c|}\hline
        \textbf{Physical error rate} & \multicolumn{5}{|c|}{\textbf{Aspect ratios}}\\\hline
        0.1 \% & $3 \times 5$ & $5 \times 7$ & $7 \times 11$ & $9 \times 15$ & $11 \times 17$\\\hline
        0.07 \% & $3 \times 5$ & $5 \times 7$ & $7 \times 11$ & $9 \times 13$ & \\\hline
        0.05 \% & $3 \times 5$ & $5 \times 7$ & $7 \times 9$ & $9 \times 13$ & \\\hline
    \end{tabular}
    \caption{\textbf{Patch dimensions used for simulation of non-square bilayer patches.} The aspect ratios shown here approximately equally protect logical X and Z errors when bilayer codes are stabilised without time-domain walls. These settings are used in Fig. \ref{fig:supp_bilayer_asymmetric}. For physical errors of 0.1 \%, the optimal aspect ratio seems to be close to 1.55, which is the value used when simulating performance of patches without time domain walls when performing distillation. These aspect ratios determine the performance of logical surface code qubits without time-domain walls, which is useful for designing circuits which leverage asymmetric surface code patches, such as the distillation procedures shown in Fig.~\ref{fig:magic_scroll_distillation} of the main text.}
    ~\label{tab:supp_aspect_ratios}
\end{table*}
}

\clearpage

\section{Colour Code Growth: Bell Pair Production~\label{sec:supp_bell_pair_circuit}}

\begin{figure*}[!ht]
\includegraphics[width=0.6\textwidth]{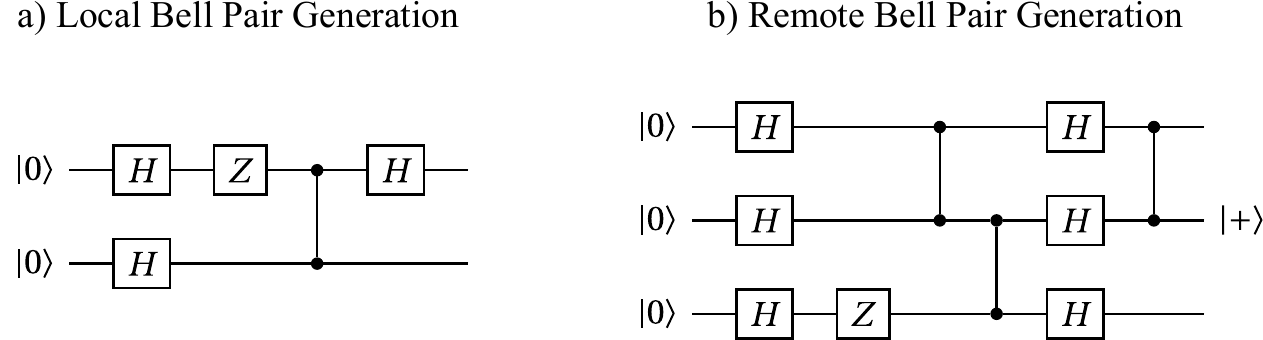}
\centering
\caption{\textbf{Generation of Bell pairs for growth of the 4.8.8 colour code.} a) Creation of a Bell pair within a single 2-qubit register. Note that if one or both qubits need to have a Hadamard gate applied (due to them participating in a domain wall), the final step of the circuit can be modified to add/remove Hadamard gates as necessary. b) Creation of a Bell pair between data qubits on different registers. The mediating ancilla is initialised in the $\ket{0}$ state, and is left in the unentangled $\ket{+}$ state at the end of the circuit allowing it to be reset without consequence. The positioning of the domain walls ensures that remote Bell pairs are only ever created with both data qubits unmodified, or with both data qubits needing to have Hadamard gates applied at the end of the circuit. In the case where both data qubits need to have Hadamard gates applied, the Z gate in the second step can be moved to the ancilla qubit, which achieves the same outcome as applying a Hadamard gate to both data qubits at the end of the circuit. ~\label{fig:supp_bell_pairs}}
\end{figure*}

In Fig. \ref{fig:colour_code_grow} of the main text, a series of $(\ket{01} + \ket{10})/\sqrt{2}$ Bell pairs were used to grow the colour code. The circuits, using the H and CZ gates native to our register model, are shown in Fig. \ref{fig:supp_bell_pairs}. These circuits can be easily modified to efficiently incorporate the Hadamard gates needed to implement the domain walls, as detailed in the caption.

\clearpage

\section{Colour Code Double-Check Circuits in Native Gates~\label{sec:supp_double_check}}

\begin{figure*}[!ht]
\includegraphics[width=\textwidth]{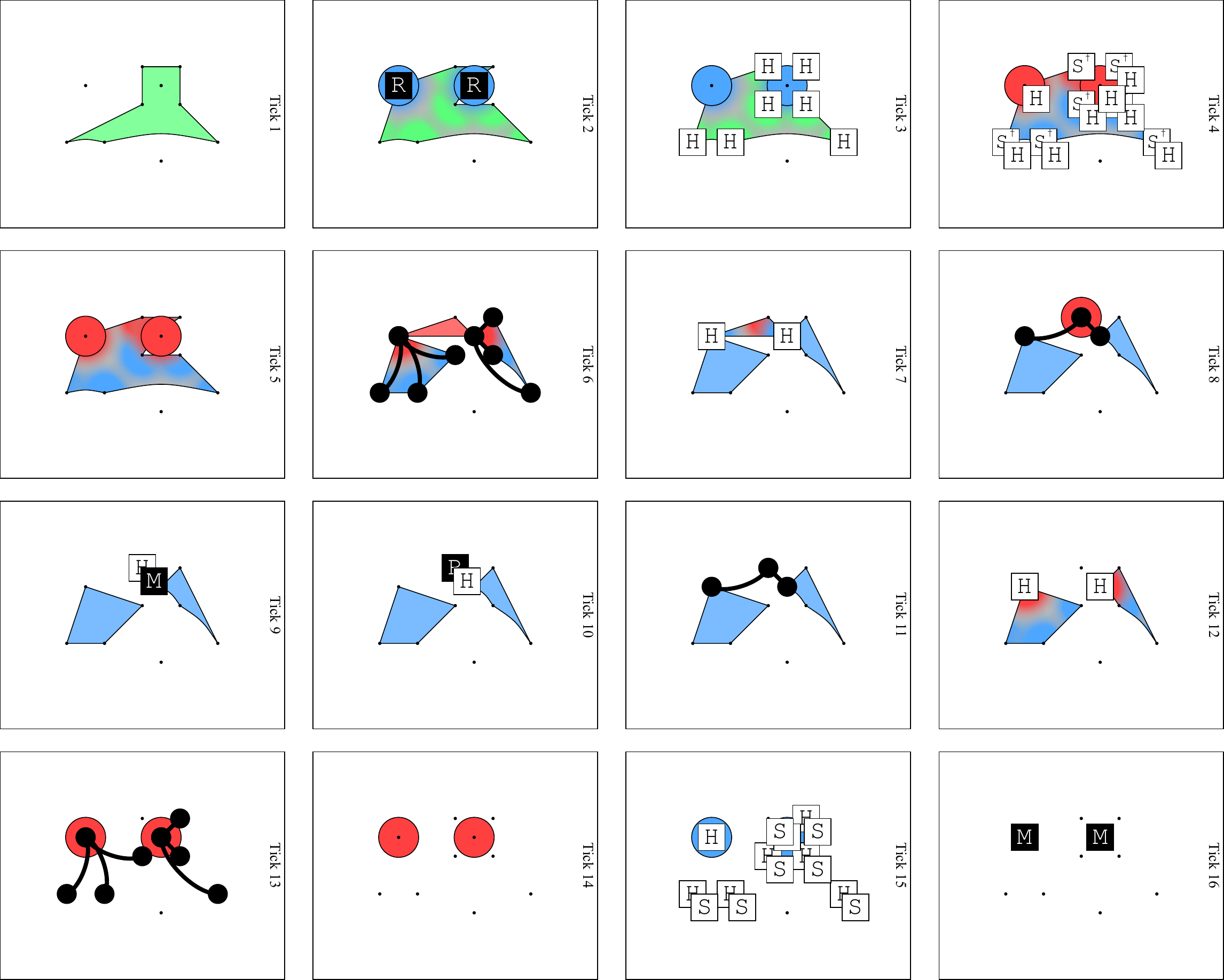}
\centering
\caption{\textbf{Detector-slice diagram of method used to double-check a magic state encoded in a distance 3 colour code using the gateset considered in this work.} This is the same circuit as Fig. \ref{fig:colour_code_d3_dc} of the main text, but decomposed into $CZ$ and single-qubit gates. There is an extra Hadamard gate added to the data qubits on the third tick, which accounts for the Hadamard which would usually be applied on the final step of stabilisation for a time-domain wall colour code. Note that some additional simplifications have been applied, mainly in the form of omitting post-CZ Hadamard gates which cancel with the reverse circuit. The full process involves initialisation of the ancilla flag qubits in tick 2 and then the aforementioned time-domain wall Hadamard gates in tick 3. Tick 4 performs the transversal $S^{\dag}$ gate transferring the relevant operator to the transversal X parity, but to reduce the total number of timesteps (and therefore improve performance) also includes the first layer of Hadamard gates in the decomposition $CNOT = H\ CZ\ H$. Ticks 5-9 perform the CNOT folding procedure, with the decomposition of the CNOT gates resulting in alternating steps between Hadamard and CZ gates, and ending in measurement of the transversal operator (and hence the first check). Ticks 10-16 reverse this process, ending in measurement of the flag ancillas and hence the second check, except that tick 14 does not include Hadamard gates to undo the action of tick 3, ensuring that the data qubtis are in the correct basis for the following (single-)cycle stabiliser measurement.~\label{fig:supp_colour_code_d3_full_dc}}
\end{figure*}

\begin{figure*}[!ht]
\includegraphics[width=\textwidth]{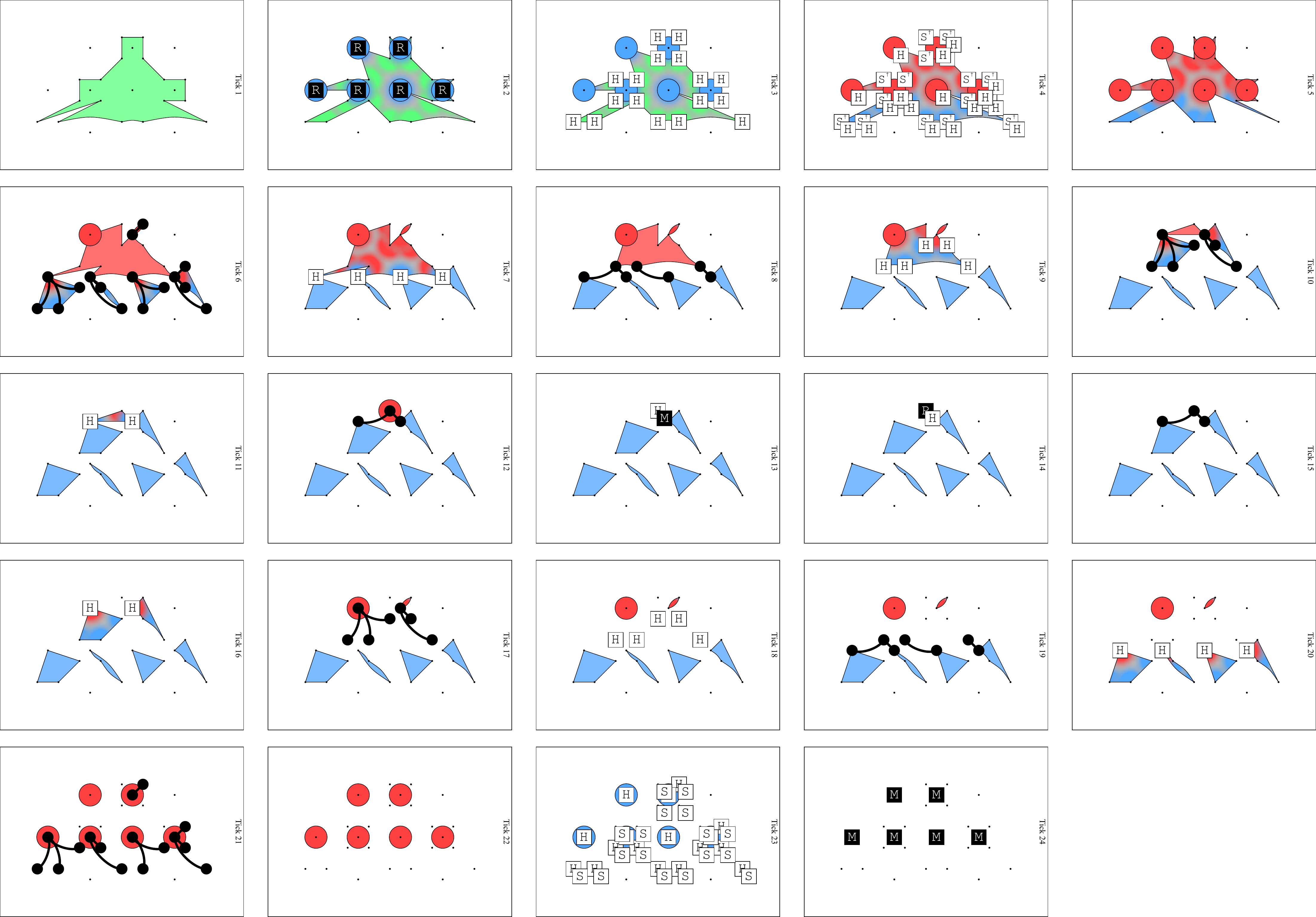}
\centering
\caption{\textbf{Detector-slice diagram of method used to double-check a magic state encoded in a distance 5 colour code using the gateset considered in this work.} This is the same circuit as Fig. \ref{fig:colour_code_d5_dc} of the main text, but decomposed into $CZ$ and single-qubit gates. The procedure used here matches the same steps described in Fig.~\ref{fig:supp_colour_code_d3_full_dc}, with ticks 5-13 folding the transversal operator, and ticks 14-24 performing the inverse of the first half of the sequence.~\label{fig:supp_colour_code_d5_full_dc}}
\end{figure*}

\begin{figure*}[!ht]
\includegraphics[width=\textwidth]{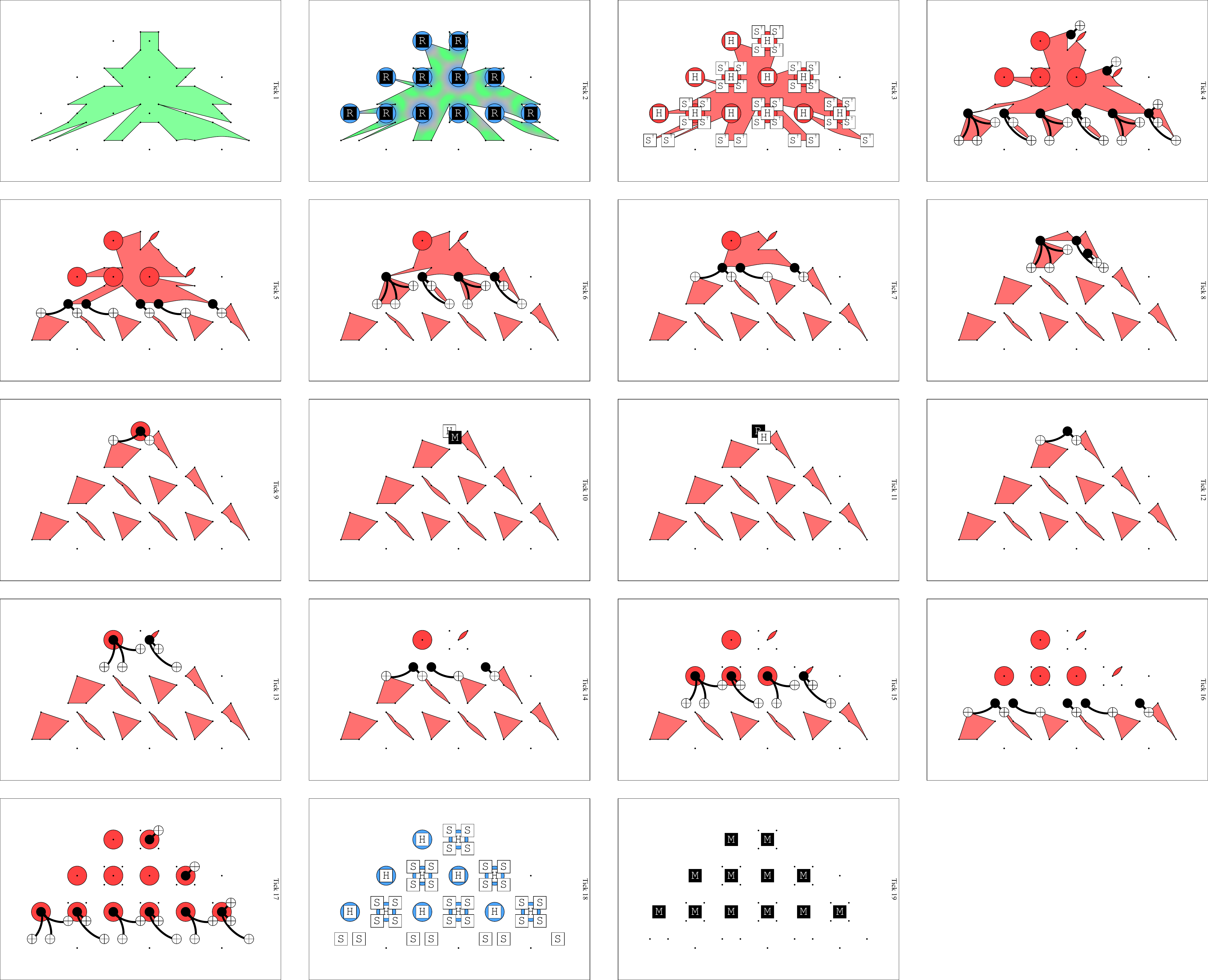}
\centering
\caption{\textbf{Detector-slice diagram of method used to double-check a magic state encoded in a distance 7 colour code.} Similar to Fig. \ref{fig:colour_code_d3_dc} and Fig. \ref{fig:colour_code_d5_dc} of the main text, this simplified circuit is used to show the method for double-checking a magic state for a distance 7 colour code. While this circuit is defined for double-checking an $\ket{S}$ state, a $\ket{T}$ state can be double-checked by replacing all $S$ ($S^\dag$) gates with $T$ ($T^\dag$) gates. The procedure used here matches the same steps described in Figs.~\ref{fig:colour_code_d3_dc} and~\ref{fig:colour_code_d5_dc} of the main text, with ticks 4-10 folding the transversal operator, and ticks 11-19 performing the inverse of the first half of the sequence.~\label{fig:supp_colour_code_d7_dc}}
\end{figure*}

\begin{figure*}[!ht]
\includegraphics[width=\textwidth]{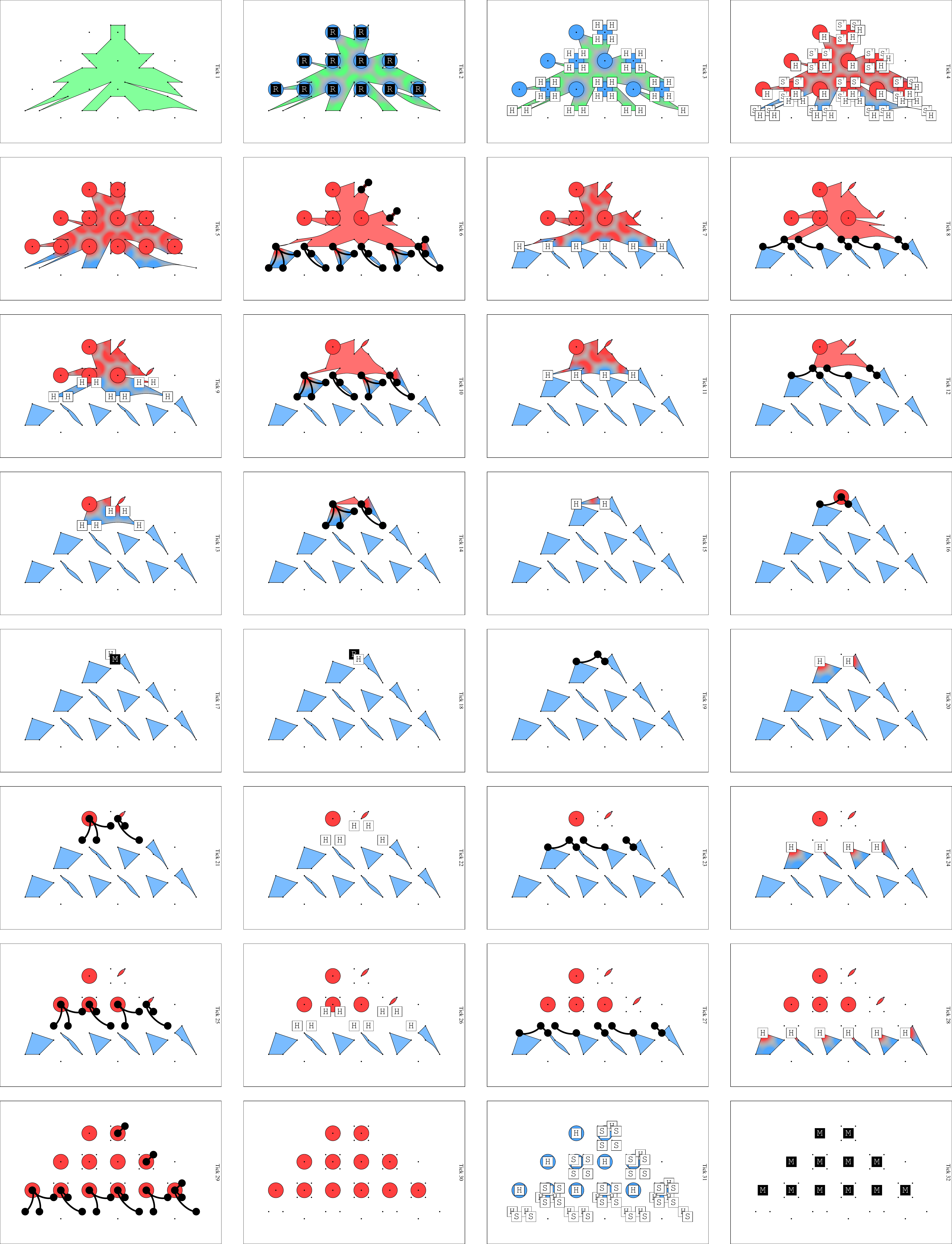}
\centering
\caption{\textbf{Detector-slice diagram of method used to double-check a magic state encoded in a distance 7 colour code using the gateset considered in this work.} This is the same circuit as Fig. \ref{fig:supp_colour_code_d7_dc}, but decomposed into $CZ$ and single-qubit gates. The procedure used here matches the same steps described in Fig.~\ref{fig:supp_colour_code_d3_full_dc}, with ticks 5-17 folding the transversal operator, and ticks 18-32 performing the inverse of the first half of the sequence.~\label{fig:supp_colour_code_d7_full_dc}}
\end{figure*}

The 4.8.8 colour code double-check circuits implemented with native gates are shown in Fig. \ref{fig:supp_colour_code_d3_full_dc}, Fig. \ref{fig:supp_colour_code_d5_full_dc} and Fig. \ref{fig:supp_colour_code_d7_full_dc} for colour distances $d=3$, $d=5$ and $d=7$ respectively. In addition, we show the simplified version of the $d=7$ double-check circuit in Fig. \ref{fig:supp_colour_code_d7_dc}.

\clearpage

\section{Magic Scroll Performance at 0.05\% error~\label{sec:supp_magic_scroll_005}}

\begin{figure*}[!ht]
\includegraphics[width=0.9\textwidth]{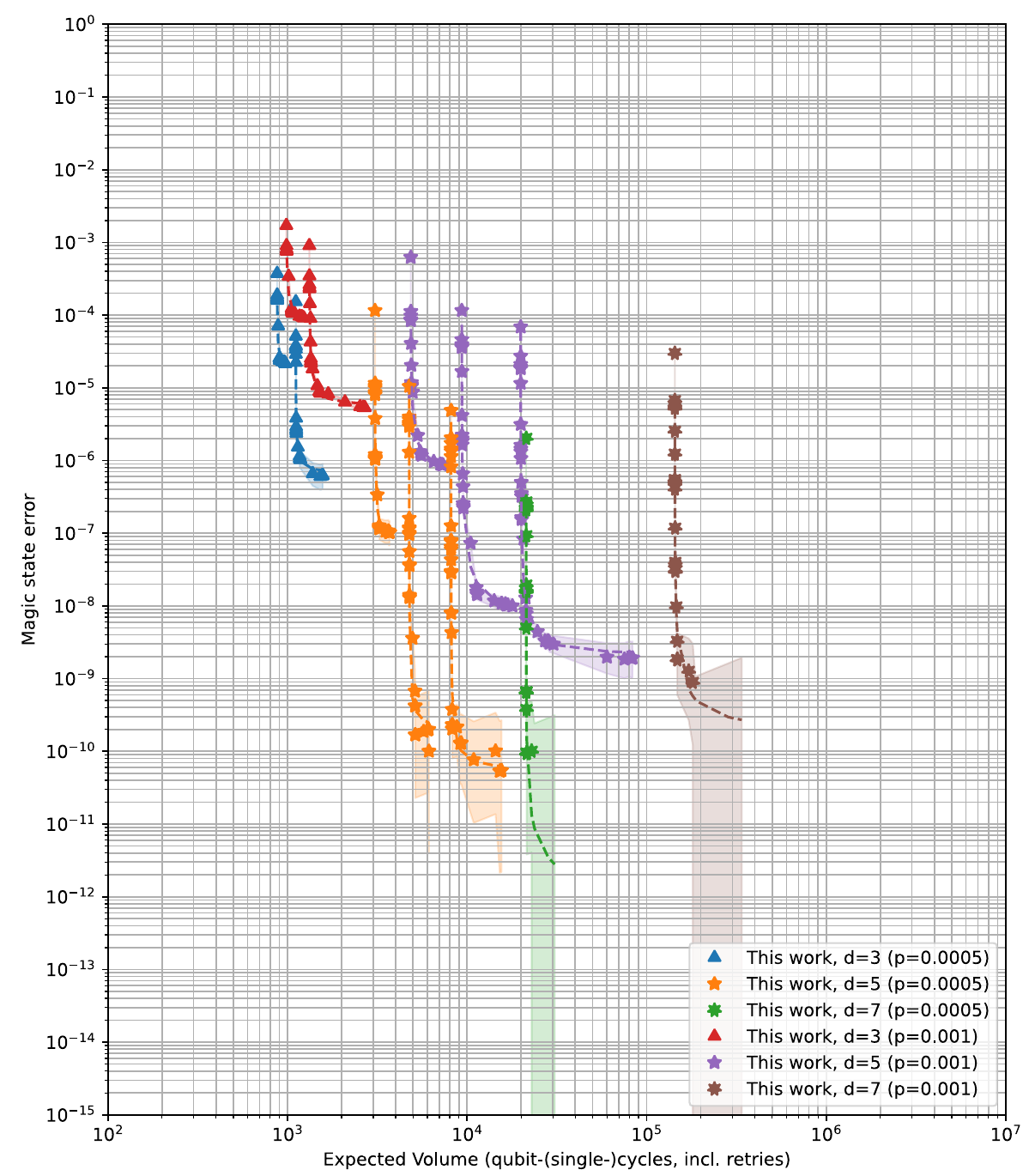}
\centering
\caption{\textbf{Magic Scroll performance at both $p=0.1\%$ and $p=0.05\%$ physical error rate.} All procedures used here are the same as were used to produce Fig. \ref{fig:gidney_style_figure} of the main text. Unlike Fig.~\ref{fig:gidney_style_figure}, this shows how the Magic Scroll procedure scales as a function of error rate. In particular, we see that reducing physical errors from $p=0.1\%$ to $p=0.05\%$ improves expected volumes by almost $20\times$ and improves magic state errors by almost $100\times$.~\label{fig:supp_gidney_style_figure_005}}
\end{figure*}

To demonstrate the scaling of the Magic Scroll as a function of physical error rate, in Fig. \ref{fig:supp_gidney_style_figure_005} we show the performance of the Magic Scroll (using the same parameters as Fig.~\ref{fig:gidney_style_figure} of the main text) at both $p=0.1\%$ and $p=0.05\%$ error.

\clearpage

\section{Bilayer Code Complementary Gapping Heuristic~\label{sec:supp_bilayer_gapping}}

As has been previously studied \cite{gidney2023yokedsurfacecodes}, postselection can be used to improve the performance of surface codes. Specifically, by postselecting on a decoded ``complementary gap'', which estimates the decoder's confidence in a solution, any passing decoding results will have a lower error rate than without such postselection. To compute the gap, one first matches the observed syndromes while forcing a logical error to occur in the decoding result, and then matches again while forcing no logical error to occur, which are both achieved by introducing a new detector which encodes whether or not a logical error occurred. By subtracting the weights of the edges from both decoding runs, one learns the difference in weight between the favoured outcome (ie. the one with lower weight) and the alternative. Because weights are related to error channel probabilities, this difference (or gap) can be interpreted as a confidence in the decoding result. By only omitting runs which have low confidence, the remaining high-confidence runs will have significantly lower logical error rates.\\

\begin{figure*}[!ht]
\includegraphics[width=0.7\textwidth]{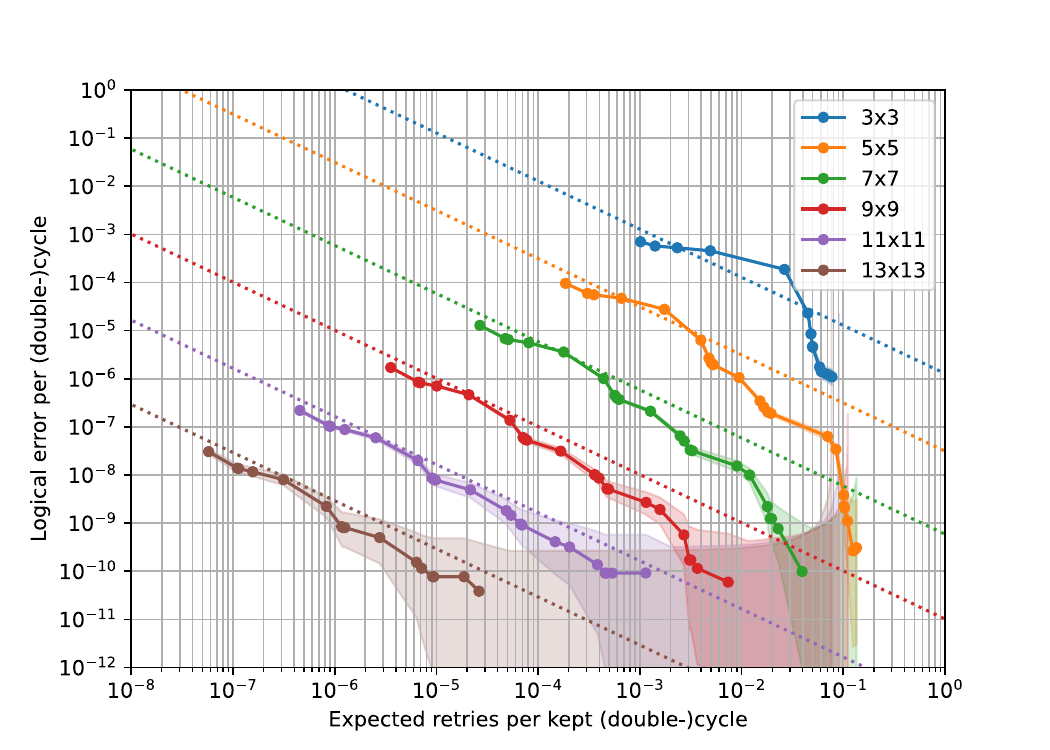}
\centering
\caption{\textbf{Performance of gapped bilayer codes, along with approximate empirical model.} Here, we simulated time-domain wall bilayer codes ($\ket{0}$ state only) with a variety of different complementary gaps. The logical error is plotted against the average chance for the gap postselection to fail, and therefore retry, each (double-)cycle. The overlaid dashed lines are not fits, but instead directly show the approximate empirical model of $\epsilon \sim p^2/P(r)$. This empirical model tends to be a pessimistic over-estimate of the error rate, and so is used for approximating the performance of distillation.~\label{fig:supp_bilayer_gapping}}
\end{figure*}

For the magic state cultivation using the Magic Scroll presented in this work (except for the extrapolated $d=7$ performance), such complementary gap performance was directly simulated. However, we also use complementary gaps to enhance the performance of distillation, for which patch sizes are much too large to directly simulate the performance (especially when aiming for logical error rates below $10^{-12}$). Instead, we develop an empirical model to approximate the performance of a gapped surface code based on lower distances for which simulation is tractable, and extrapolate this model when dealing with larger distances. The numerical simulation results used for this are shown in Fig. \ref{fig:supp_bilayer_gapping}.\\

We find that the following model is a reasonable approximation for sufficiently large distance surface codes, with ungapped logical error rates of $p$:

\begin{itemize}
    \item $pf$ chance to retry each gapped cycle, and
    \item $p/f$ chance for a logical error each gapped cycle
\end{itemize}

for some arbitrary ``gap factor'' $f > 1$ characterising the strength of gapping.\\

If we let $\epsilon$ be the gapped error probability and $P(r)$ be the retry chance per cycle, then we can combine the above two expressions to get:

\begin{align}
    \epsilon \sim \frac{p^2}{P(r)}
\end{align}

which is valid for $P(r) > p$, and otherwise $\epsilon = p$ and $P(r) = 0$. The performance of this empirical model is overlaid on Fig. \ref{fig:supp_bilayer_gapping}, showing that it is a good approximation for distances $\gtrsim 7$, and when the approximation fails tends to be an overestimate of logical error performance (meaning it is, on the whole, a pessimistic approximation). This approximation is used when modelling the performance of gapped surface code patches when distilling.

\clearpage

\end{document}